\documentclass{egpubl}
\usepackage{egsr2026_DLOnlyTrack}

\WsPaper           

\usepackage[T1]{fontenc}
\usepackage{dfadobe}

\usepackage{cite}  
\BibtexOrBiblatex
\electronicVersion
\PrintedOrElectronic
\ifpdf \usepackage[pdftex]{graphicx} \pdfcompresslevel=9
\else \usepackage[dvips]{graphicx} \fi

\usepackage{egweblnk}

\usepackage{graphicx}
\usepackage{subcaption}
\usepackage{booktabs}
\usepackage{wrapfig}
\usepackage[normalem]{ulem}
\usepackage{xcolor}

\newcommand{\new}[1]{#1}
\newcommand{\newtwo}[1]{#1}

\title[Differentiable Glare Mitigation]%
      {Glare Mitigation using a Differentiable Unified Glare Rating}

\author[L.~Beresna \& E.~Fiume]
{\parbox{\textwidth}{\centering
   Linas Beresna\orcid{0009-0004-7102-324X}
   and Eugene Fiume\orcid{0000-0001-5939-7055}
   }
   \\
{\parbox{\textwidth}{\centering
   Simon Fraser University, School of Computing Science, Canada
   }
}}

\begin{document}

\teaser{
  \includegraphics[width=\linewidth]{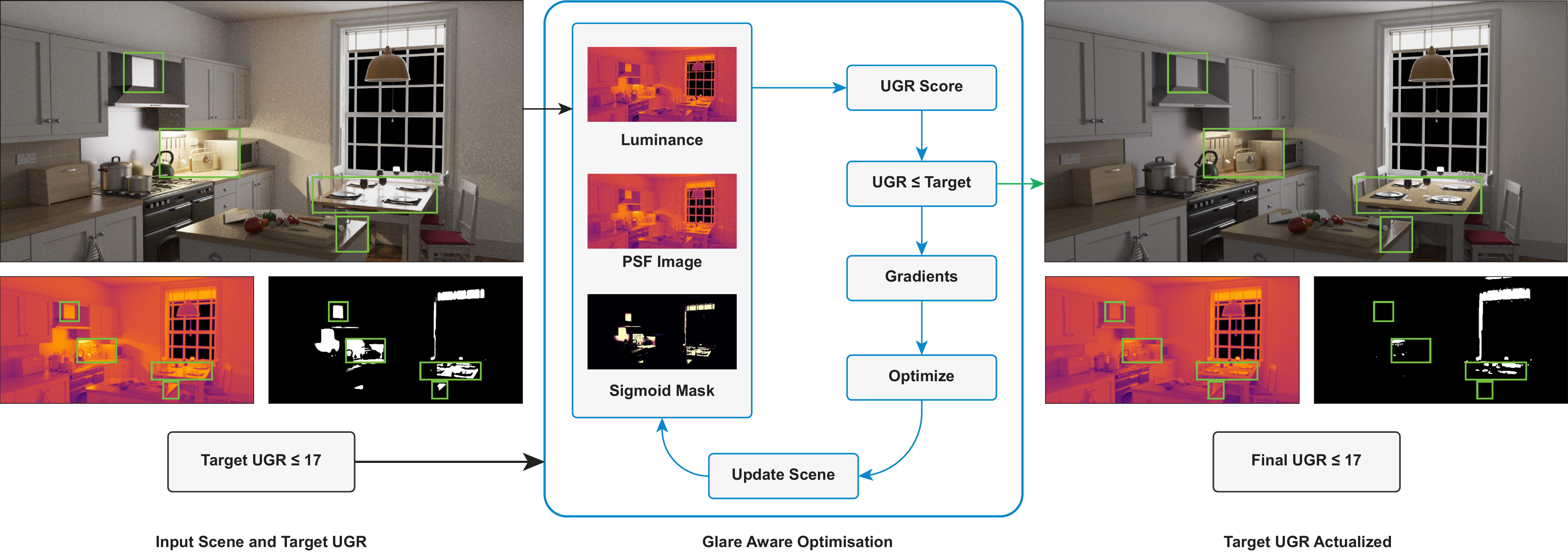}
  \centering
  \caption{Starting from an initial high glare state (left), our differentiable rendering framework jointly optimises material parameters ($\alpha, \eta$) and light source distributions to reduce the Unified Glare Rating. By utilising a point spread function and a soft sigmoid to formulate a continuous proxy, the framework smoothly guides the scene until a target ergonomic threshold (e.g., $\mathrm{UGR} \le 17$) is reached (right). Our method selectively mitigates localised glare hotspots while preserving the global illumination and aesthetic intent of the original indoor scene.}
  \label{fig:teaser}
  \vspace{5em}
}

\maketitle
\begin{abstract}
Recent research in differentiable light transport extends the utility of computer graphics algorithms beyond traditional image generation, offering powerful tools for physical inverse design. In architectural and automotive applications, visual discomfort from glare is a critical design rating, traditionally quantified by the discrete CIE Unified Glare Rating (UGR). The standard UGR formulation relies on strict binary thresholds, making it fundamentally incompatible with smooth gradient-based inverse rendering. In this paper, we introduce a continuous, fully differentiable proxy for UGR. To resolve the severe optimisation instabilities caused by Monte Carlo variance at low sample densities, we introduce a differentiable optical scattering pass that simulates the Point Spread Function (PSF) of the human eye to heal fractured evaluation masks. We replace the discrete UGR step function with a tunable sigmoid boundary, enabling gradients to flow smoothly from the psychophysical measure back to the physical scene parameters. We deploy this differentiable framework to systematically reduce glare across three radiometric domains: surface-side microgeometry roughening, boundary-side index of refraction (IOR) optimisation, and source-side emitter gobo masking. By transforming a passive perceptual evaluation into an active loss landscape, our framework provides a robust, physics-based pipeline for optimizing visual comfort in complex global illumination environments.


\end{abstract}


\section{Introduction}
\label{sec:intro}

\begin{figure}[t!]
  \centering
  \begin{subfigure}[b]{0.327\linewidth}
    \includegraphics[width=\linewidth]{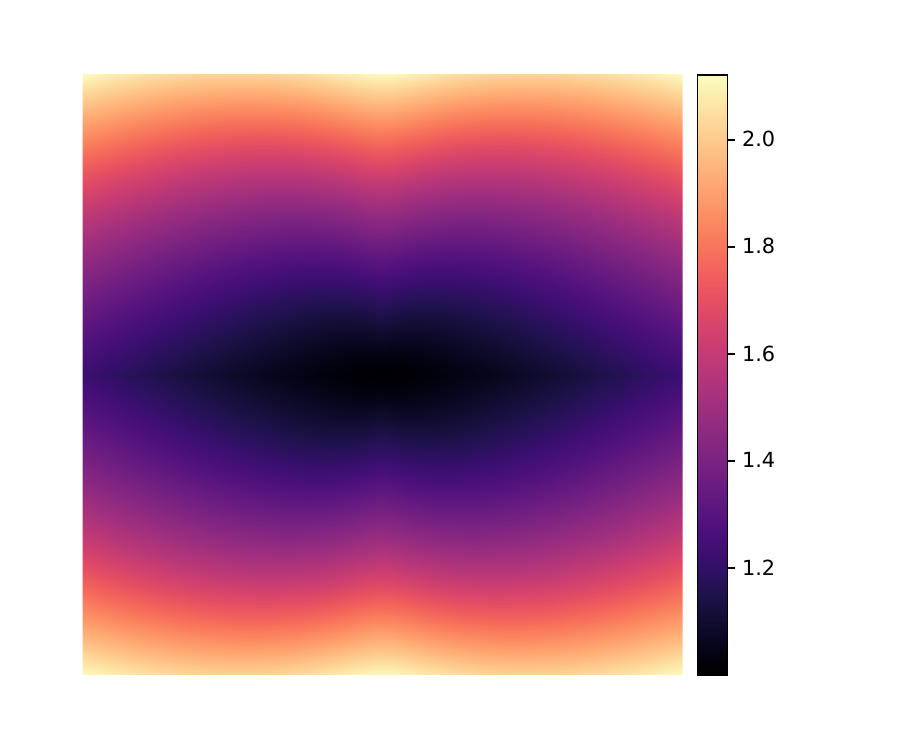}
    \caption{Position Index ($P$)}
    \label{fig:ugr_guth}
  \end{subfigure}
  \hfill
  \begin{subfigure}[b]{0.327\linewidth}
    \includegraphics[width=\linewidth]{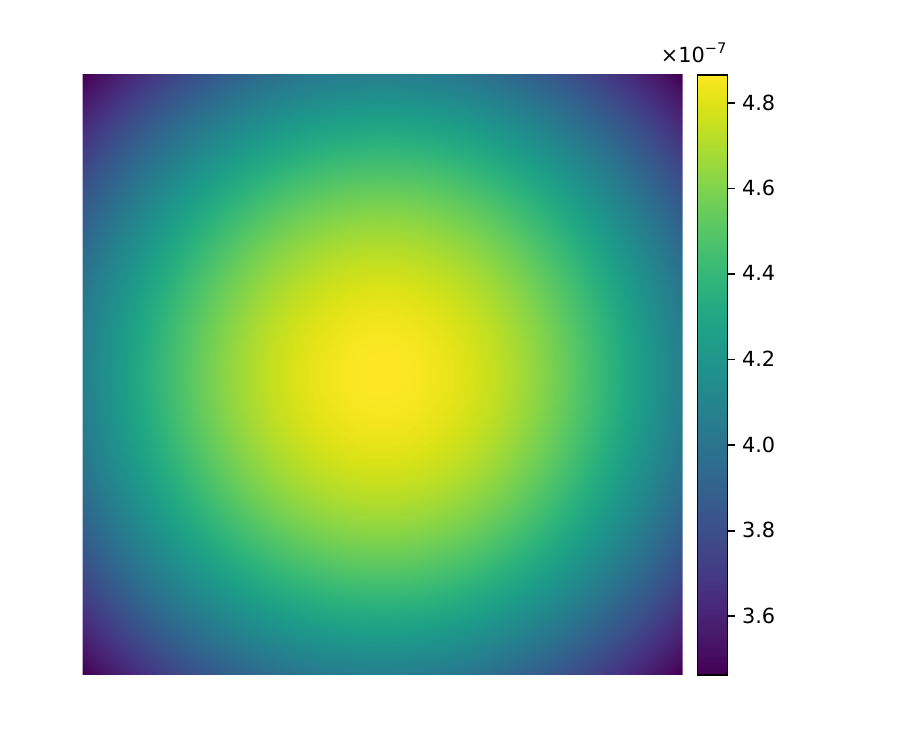}
    \caption{Solid Angle ($\omega$)}
    \label{fig:ugr_solid_angle}
  \end{subfigure}
  \hfill
  \begin{subfigure}[b]{0.327\linewidth}
    \includegraphics[width=\linewidth]{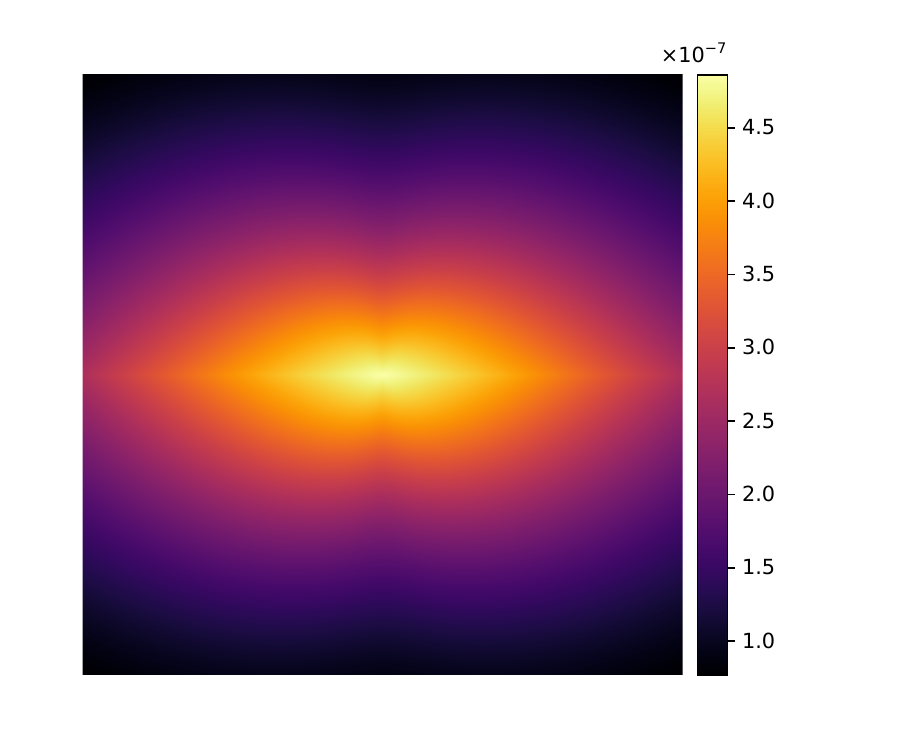}
    \caption{Weight ($\omega/P^2$)}
    \label{fig:ugr_total_weight}
  \end{subfigure}
  \caption{Spatial components of the UGR camera model.
  (a) The Guth position index $P$ maps human foveal sensitivity; dark central regions represent a value of $\approx1.0$ (highest sensitivity), while bright peripheral regions indicate lower sensitivity as $P$ increases.
  (b) The pixel solid angle $\omega$ accounts for geometric perspective distortion, with peak values at the optical center.
  (c) The combined spatial weighting $\omega/P^2$ highlights the central foveal region.}
  \label{fig:ugr_components}
  \vspace{-2em}
\end{figure}

Architectural and automotive lighting design is fundamentally constrained by the need for human visual comfort. One major artifact that compromises this comfort is glare, defined as a psychophysical perception of excessive brightness that causes annoyance, discomfort, or loss in visual performance. In industry, discomfort glare is standardly quantified by the Unified Glare Rating (UGR), originally established by the International Commission on Illumination (CIE) \cite{cie1995ugr} and formalized in modern workplace lighting standards such as EN 12464-1 \cite{en12464, energieplus_ugr}. The UGR evaluates the discomfort caused by light sources relative to the ambient environment.  From a computational perspective, it is critical to define this measure as a function of both the observer viewpoint $v$ and the physical scene description $S$. Formally, the measure $\mathrm{UGR}(v, S)$ is calculated as:
\begin{equation}
	\mathrm{UGR}(v, S) = 8 \log_{10} \left( \frac{0.25}{L_b} \sum_{i=1}^{n} \frac{L_i^2 \omega_i}{p_i^2} \right)
	\label{eq:ugr}
\end{equation}
where $L_b$ is the background adaptation luminance, $L_i$ is the luminance of the $i^{\mathrm{th}}$ glare source in the direction of $v$, $\omega_i$ is the solid angle subtended by the source, and $p_i$ is the Guth position index relative to $v$ (see Figure \ref{fig:ugr_components} for a visualization of these spatial weights in camera space). Crucially, all material and lighting parameters that we adjust during optimization lie within the scene description $S$.

Traditionally, the CIE definition of UGR is applied strictly to primary light sources shining directly at the eye. In practice, however, industry tools also evaluate reflected glare over specific spatial domains like a desk or whiteboard, typically by placing virtual evaluation planes on the objects themselves. Because our framework is built on multi-bounce path tracing, essentially all surfaces in $S$ become potential luminaires: we evaluate the UGR natively from the camera viewpoint $v$, treating any surface that reflects a harsh highlight into the observer's field of view as an apparent glare source. This naturally unifies the traditional point-to-point measure with these spatial-domain evaluations.

In standard industry workflows (e.g., utilizing lighting design software such as DIALux \cite{dialux}), evaluating this measure is typically an iterative, forward-rendering task. Designers must manually adjust light placements, material properties, or architectural layouts, re-render the scene, and re-calculate the UGR until the value falls below a targeted discomfort threshold. This trial-and-error process is highly inefficient for complex, multi-bounce light transport scenarios. Differentiable rendering offers a systematic alternative by enabling inverse design: computing the gradients of a loss function with respect to scene parameters to automatically optimize materials and lighting. However, the standard UGR formulation presents mathematical discontinuities that can prevent standard gradient-based optimization. Specifically, identifying which pixels constitute the active glare sources $L_i$ relies on a strict binary threshold (e.g., $L > 5L_b$). Mathematically, this formulation acts as a Heaviside step function. Consequently, the resulting gradients behave as Dirac delta functions, evaluating to zero almost everywhere and yielding infinite singularities precisely at the threshold boundary. This can prevent stable gradient descent and may yield unreliable convergence.

In this paper, we bridge the gap between psychophysical lighting ratings and inverse rendering by introducing a continuous, fully differentiable proxy for UGR. By replacing discrete step functions with tunable, continuous approximations, we allow gradients to flow smoothly from the human-vision value, backward through the path-traced light transport, and into the physical parameters of the scene. To our knowledge, no prior work integrates a visual discomfort rating directly into a differentiable rendering loss landscape for inverse scene design.

We deploy this differentiable UGR rating to systematically evaluate automated glare mitigation across three distinct radiometric domains: source-side emission masking, boundary-side refraction (anti-reflective coatings), and surface-side microgeometry scattering (anti-glare roughening). Through this framework, we uncover a critical regularization trap in standard industry mitigation practices. We empirically demonstrate that while spatial roughening successfully diffuses mild glare, excessive roughening can inadvertently starve the surrounding environment of ambient light. Because the UGR formulation is inversely proportional to the background luminance $L_b$, this ambient darkening paradoxically spikes the discomfort rating. We show that retaining $L_b$ within the computational graph provides a powerful self-regularizing effect that prevents this over-frosting failure state. Furthermore, we demonstrate that modifying the refractive index and source emission yields inherently stable loss landscapes, providing universally effective pathways for glare reduction.

In summary, our core contributions are:
\begin{itemize}
    \item A continuous, differentiable formulation of the CIE UGR that enables gradient-based optimization of visual discomfort.
    \item An automated inverse design framework that establishes a methodological bridge between differentiable light transport and psychovisual computing, directly optimizing physical scene parameters against empirical perceptual models.
    \item An empirical analysis of light transport mitigation strategies, demonstrating the stability of anti-reflective and emission masking versus the ambient-starvation risks of microfacet roughening under complex global illumination environments.
\end{itemize}

\section{Related Work}
\label{sec:related}
Our framework sits at the intersection of differentiable light transport, inverse material design, and psychophysical glare evaluation; in this section, we analyze the foundational techniques in these domains and highlight how our approach differs from traditional photometric optimization.

\subsection{Differentiable Rendering}
Differentiable rendering (DR) spans rasterization-based methods \cite{loper2014opendr, liu2019softras, kato2018neural, goel2020differentiable} through to physically based path tracing \cite{li2018differentiable, zhang2023projective, Loubet2019Reparameterizing}. Quantifying UGR requires strict radiometric accuracy, as glare frequently manifests through secondary light transport effects such as inter-reflections off glossy monitors, architectural glass, or metallic surfaces \cite{cie1995ugr, ward1994radiance, carlucci2015review, pierson2018review}; rasterization-based approaches are therefore insufficient. We build on physically based path-tracing DR, including recent advances in volumetric and translucent transport \cite{nimierdavid2022unbiased, deng2022reconstructing, weier2024practical}. By operating on fixed scene geometry, our framework avoids the primary visibility discontinuities that plague shape optimization, and we use Path Replay Backpropagation \cite{vicini2021PathReplay} to maintain constant memory while optimizing high-resolution spatial textures and emitter parameters.

DR is increasingly applied to source-side manipulation, from computational optics \cite{teh2025automated} to view-independent adjoint luminaire optimization for architectural targets \cite{lipp2023view}. However, because discomfort glare is inherently view-dependent, tightly coupled to the observer's retinal projections and the solid angle of each source, view-independent heuristics are insufficient. Our emitter optimization builds on these source-side principles while strictly maintaining a camera-space evaluation.

\subsection{Inverse Material Design}
Inverse material design has historically been formulated as an appearance matching problem, recovering SVBRDFs from real-world captures \cite{aittala2016reflectance, deschaintre2018single, gao2019deep, nimierdavid2021material} via pixel-wise photometric losses. Our method departs from this paradigm: rather than mimicking a target image, we optimize against a global psychophysical objective, allowing the scene to deviate radiometrically as long as it converges toward human visual comfort. Unlike post-processing neural networks that hallucinate glare removal from 2D images \cite{wu2021how, dai2022flare7k} without providing actionable design data, our gradients directly reflect physically accurate light transport.

Beyond appearance matching, inverse rendering has been applied to the functional design of optics and layered materials \cite{jakob2014comprehensive, guo2018material, tseng2021differentiable, sun2021lens}. Selectively optimizing scene parameters corresponds directly to real-world architectural interventions, for example, index-matching anti-reflective thin films and chemically etched anti-glare surfaces. Furthermore, our framework redirects this machinery toward a human-centric visual comfort rating rather than imaging fidelity.

\subsection{Psychophysical Glare Measures}
Quantifying visual discomfort relies on empirically derived psychophysical measures that aggregate complex physical light transport into a single scalar rating. Within this domain, the Daylight Glare Probability (DGP) is widely adopted for daylight-driven scenarios featuring massive, high-contrast exterior sources (e.g., the sun) \cite{wienold2006evaluation}. However, for artificially lit indoor environments featuring multiple point or area luminaires and complex multi-bounce reflections, the UGR remains the formal international standard \cite{cie1995ugr, en12464}. Because our framework targets the optimization of architectural materials, secondary specular bounces, and active indoor luminaires, UGR is the radiometrically appropriate objective.

Crucially, UGR evaluates discomfort as a ratio between localized glare sources and global adaptation luminance: naively minimizing scene brightness yields an unusable, darkened room, while flooding the scene with background luminance washes out contrast. Traditionally, these ratings have been evaluated \textit{post-hoc} using forward-simulation rendering tools, most notably the Radiance system \cite{ward1994radiance, jones2017validated}.

Recent research has focused on accelerating these evaluations using simplified screen-space heuristics or deep convolutional neural networks \cite{pierson2018review}, yet uniformly treats the measure as a read-only evaluation or forward prediction task. Such static, localized methods often fail to align with subjective perception under complex ambient lighting \cite{ekim2026perception}. In contrast, by introducing a soft, continuous proxy that natively incorporates global background adaptation ($L_b$) into the computational graph, we transform UGR from a \textit{post-process} evaluation tool into a dynamic, differentiable objective, dynamically balancing localized glare against the total ambient environment.

\section{Differentiable Glare Formulation}
\label{sec:formulation}
To utilize the Unified Glare Rating as an active objective function for inverse design, we must translate its mathematical formulation into a robust, fully differentiable computational pipeline. In this section, we detail the mathematical and stochastic challenges of standard image-space evaluation, and introduce our two-stage continuous formulation designed to resolve them.

\subsection{Image-Space Evaluation}
As introduced in Section \ref{sec:intro}, the standard UGR formula aggregates the localized discomfort produced by all individual glare sources in the visual field against the background luminance. In practical lighting design, these calculated UGR values correspond directly to subjective human discomfort. A UGR of 10 or below is generally considered imperceptible, 19 is the universally accepted maximum threshold for standard office environments, and values exceeding 28 are deemed visually intolerable \cite{cie1995ugr, en12464}.

\begin{figure}[t!]
  \centering
  \newcommand{\diagheight}{2.9cm}

  \begin{subfigure}[b]{0.33\linewidth}
    \centering
    \includegraphics[height=\diagheight]{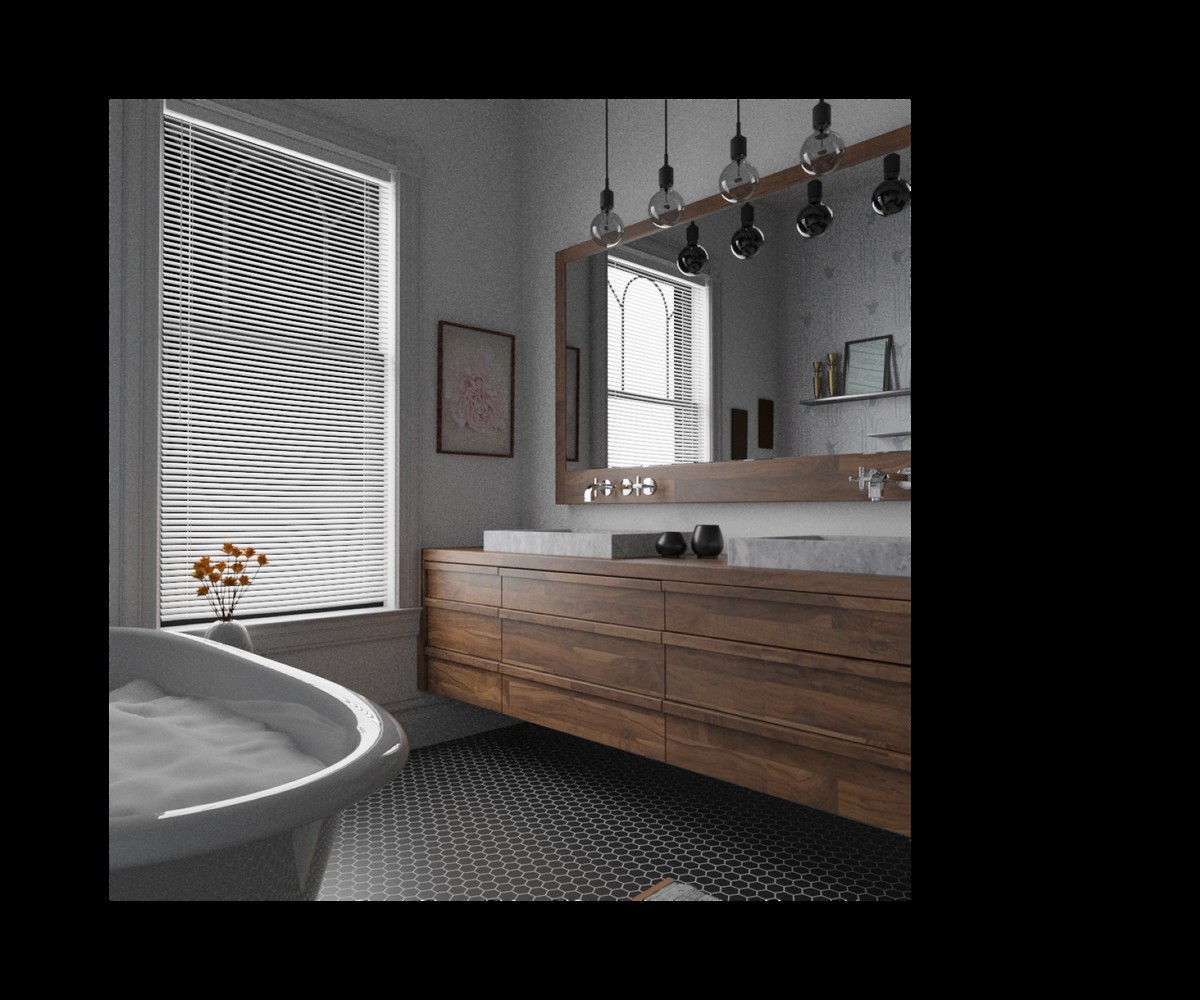}
    \caption{Render}
    \label{fig:bathroom_render}
  \end{subfigure}%
  \begin{subfigure}[b]{0.33\linewidth}
    \centering
    \includegraphics[height=\diagheight]{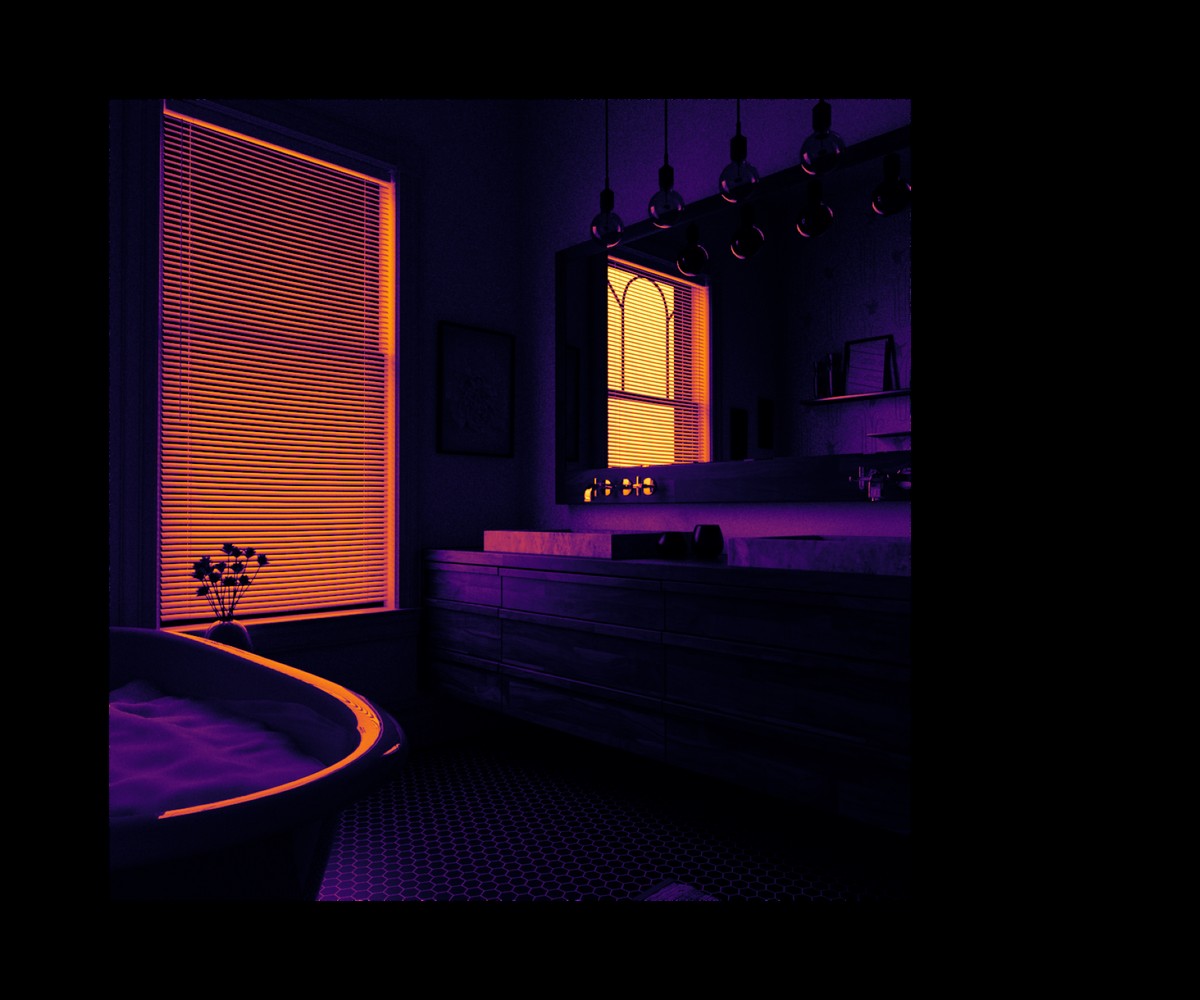}
    \caption{Luminance}
    \label{fig:bathroom_lum}
  \end{subfigure}%
  \begin{subfigure}[b]{0.33\linewidth}
    \centering
    \includegraphics[height=\diagheight]{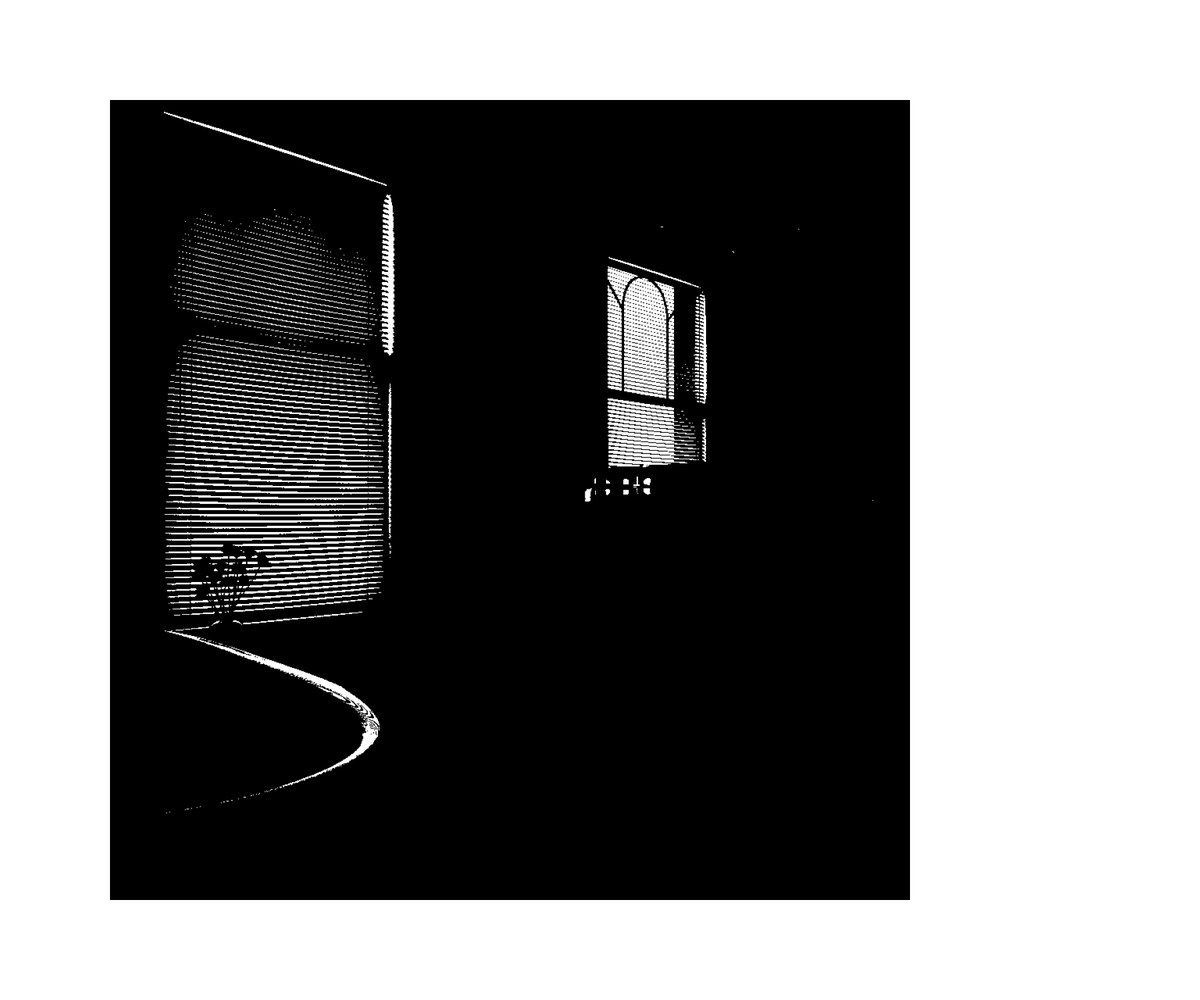}
    \caption{Glare Mask}
    \label{fig:bathroom_mask}
  \end{subfigure}

  \caption{Visualizing spatial glare from a rendered scene. (a) A forward path-traced render of a bathroom environment featuring multiple reflective surfaces. (b) The photometric luminance distribution, capturing the extreme dynamic range of the incident light. (c) The resulting spatial glare mask, isolating active glare pixels that exceed the relative background adaptation threshold ($L > 5L_b$).}
  \label{fig:bathroom_diagnostic}
\end{figure}

In the context of image-space evaluation, this formulation is calculated on a per-pixel basis using a rendered High Dynamic Range image \cite{blaszczak2013method}. The summation iterates over contiguous clusters of pixels that are classified as glare sources. The standard methodology dictates that a pixel belongs to a glare source if and only if its luminance strictly exceeds a predefined adaptation threshold. In lighting design software, this boundary is typically defined by a binary inequality condition: $L_i > 5 L_b$ \cite{wienold2006evaluation}. Throughout our method, we choose to use a hard limit of 5 times the background luminance, but any limit can be chosen. Figure \ref{fig:bathroom_diagnostic} visualizes how a rendering compares against the luminosity of the view and the glare mask.

While binary thresholding can yield converging gradients in some stochastic setups, the discrete classification creates a harsh step in the loss landscape: a pixel crossing from $4.99 L_b$ to $5.01 L_b$ is abruptly hit with the full UGR penalty rather than a proportional one, causing the optimizer to overcorrect and produce unnatural material artifacts. We explore this empirically in Section \ref{sec:results}. To traverse the landscape smoothly, we require a continuous approach that penalizes glare before it triggers a strict violation (Section \ref{sec:soft_ugr}).

\subsection{Optical Scattering}
\label{sec:physical_scattering}

While Monte Carlo path tracing computes accurate radiometric quantities, discomfort glare is fundamentally perceptual. Suppressing the stochastic variance that destabilizes optimization through high sample counts is impractical, as gradient-based inverse design requires hundreds of iterative passes; operating at low sample counts (e.g., 16 spp) is therefore necessary for practical computation times. By accommodating the eye's biological response to this variance, our pipeline operates robustly on stochastic inputs without disrupting the gradient flow.

The human eye is not a perfect optical instrument. As light passes through the ocular media, specifically the cornea, the crystalline lens, and the vitreous humor, a significant portion of the incoming light undergoes intraocular scattering \cite{spencer1995}. This biological scattering acts as a spatial low-pass filter, distributing the energy of intense localized light sources across a wider area of the retina. This phenomenon, often referred to as ``veiling luminance'' or ``bloom,'' is characterized by the Point Spread Function (PSF) of the human eye.

In the context of differentiable rendering, modeling this intraocular scattering provides a dual benefit. Physically, convolving the Monte Carlo radiance estimate with an approximation of the eye's Point Spread Function (PSF) translates incident scene radiance into perceived retinal luminance \cite{kakimoto2004}. While classical light transport defines this sensor response via the measurement equation \cite{veach1998robust}, our framework extends it into the differentiable domain, so the biological PSF becomes an active gradient pathway rather than a forward-evaluation filter.

\begin{figure}[t]
    \centering
    \setlength{\tabcolsep}{2pt}
    \renewcommand{\arraystretch}{0.5}
    
    \begin{tabular}{cccc}
        & \small $p=0$ & \small $p=1$ & \small $p=2$ \\
        
        \vspace{1mm} \\

        \rotatebox{90}{\hspace{4mm} \small Luminance} &
        \includegraphics[width=0.29\columnwidth]{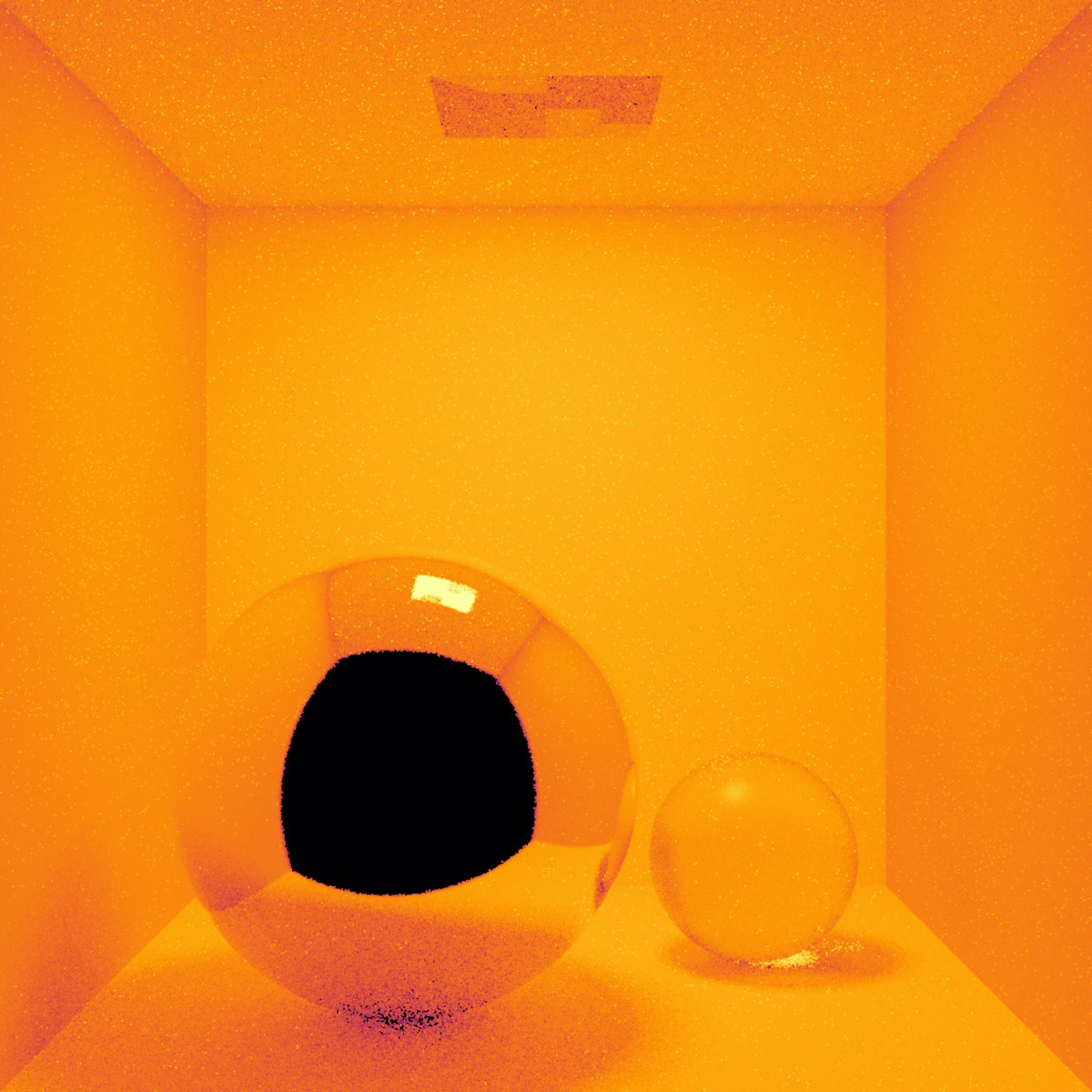} &
        \includegraphics[width=0.29\columnwidth]{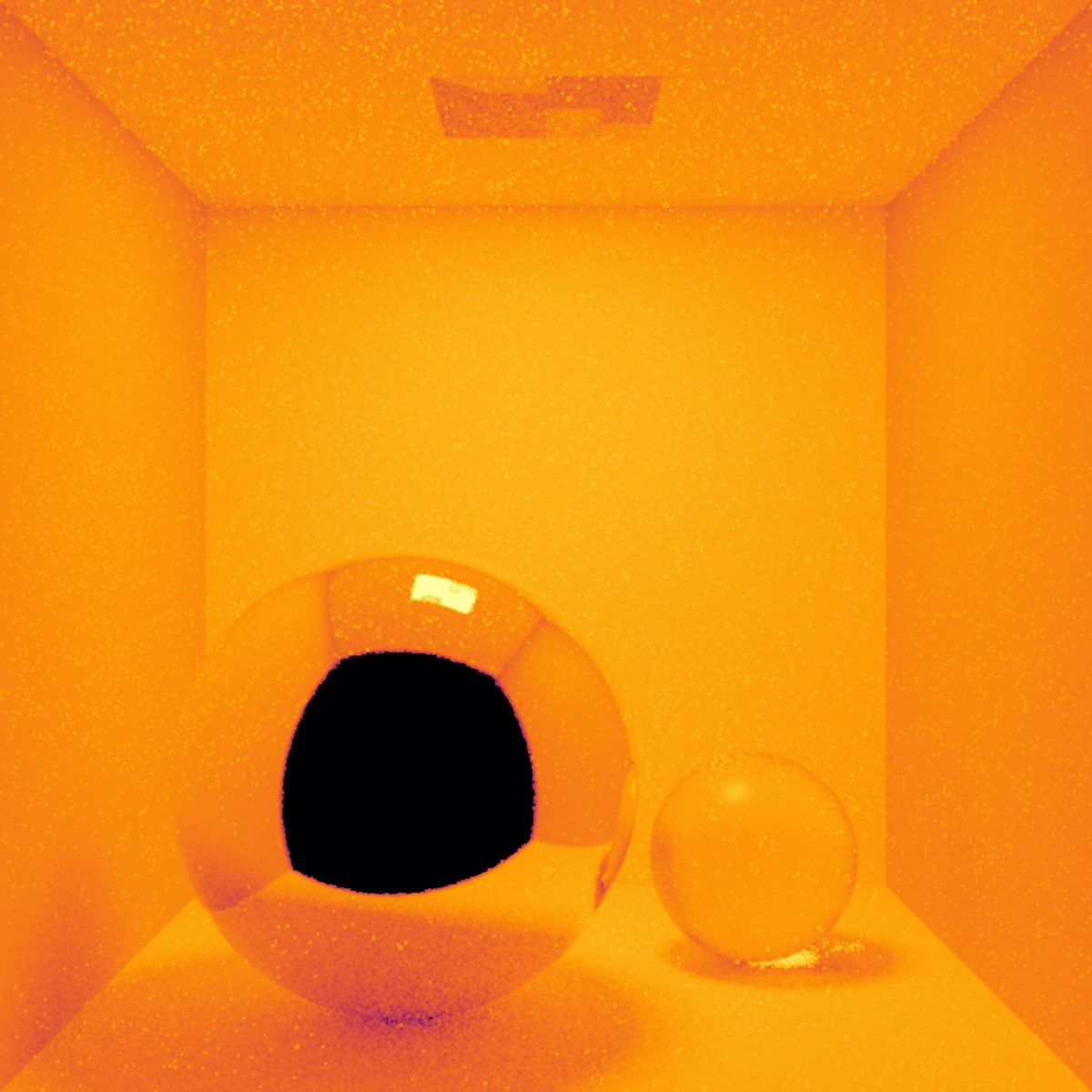} &
        \includegraphics[width=0.29\columnwidth]{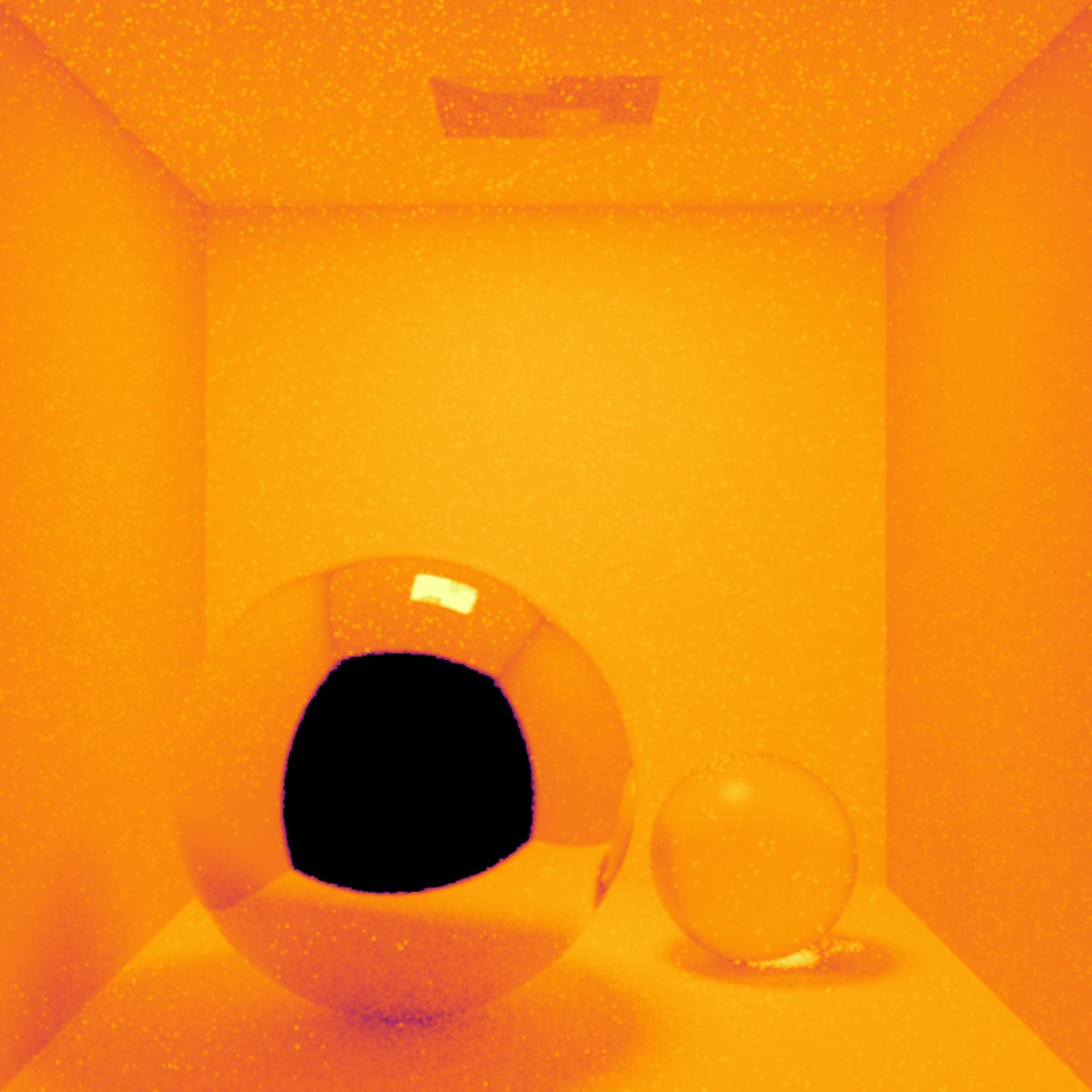} \\
        
        \vspace{1mm} \\

        \rotatebox{90}{\hspace{5mm} \small Glare Mask} &
        \includegraphics[width=0.29\columnwidth]{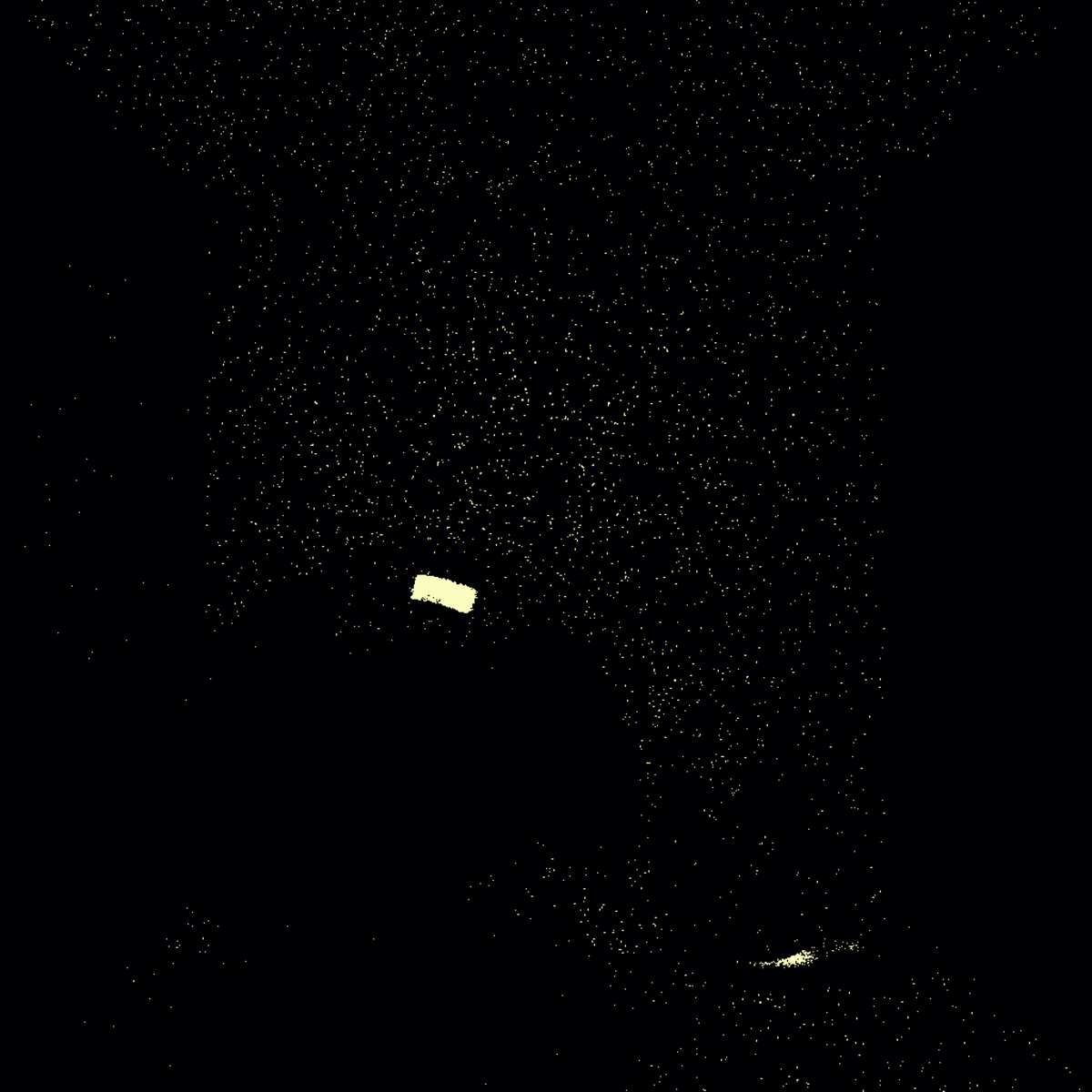} &
        \includegraphics[width=0.29\columnwidth]{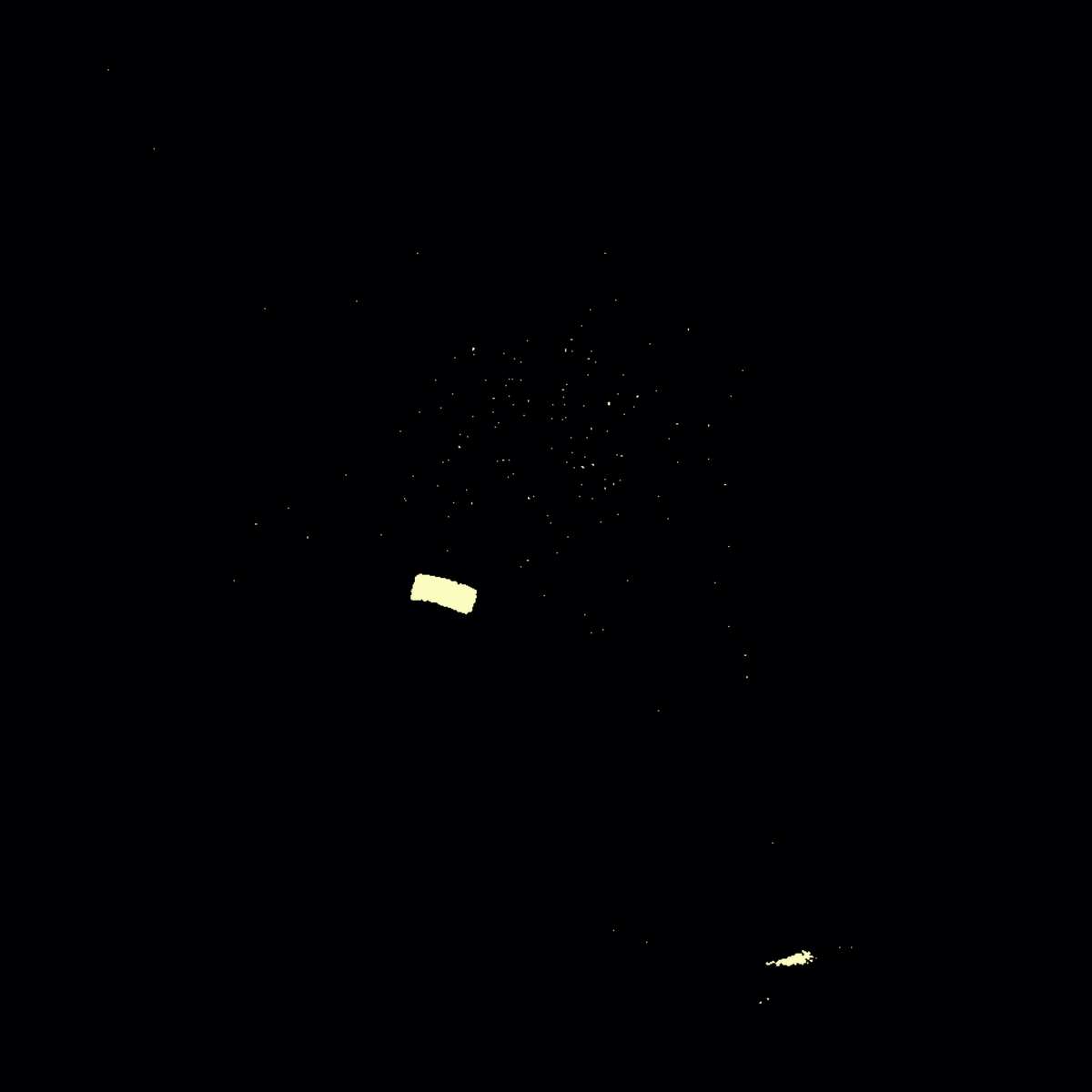} &
        \includegraphics[width=0.29\columnwidth]{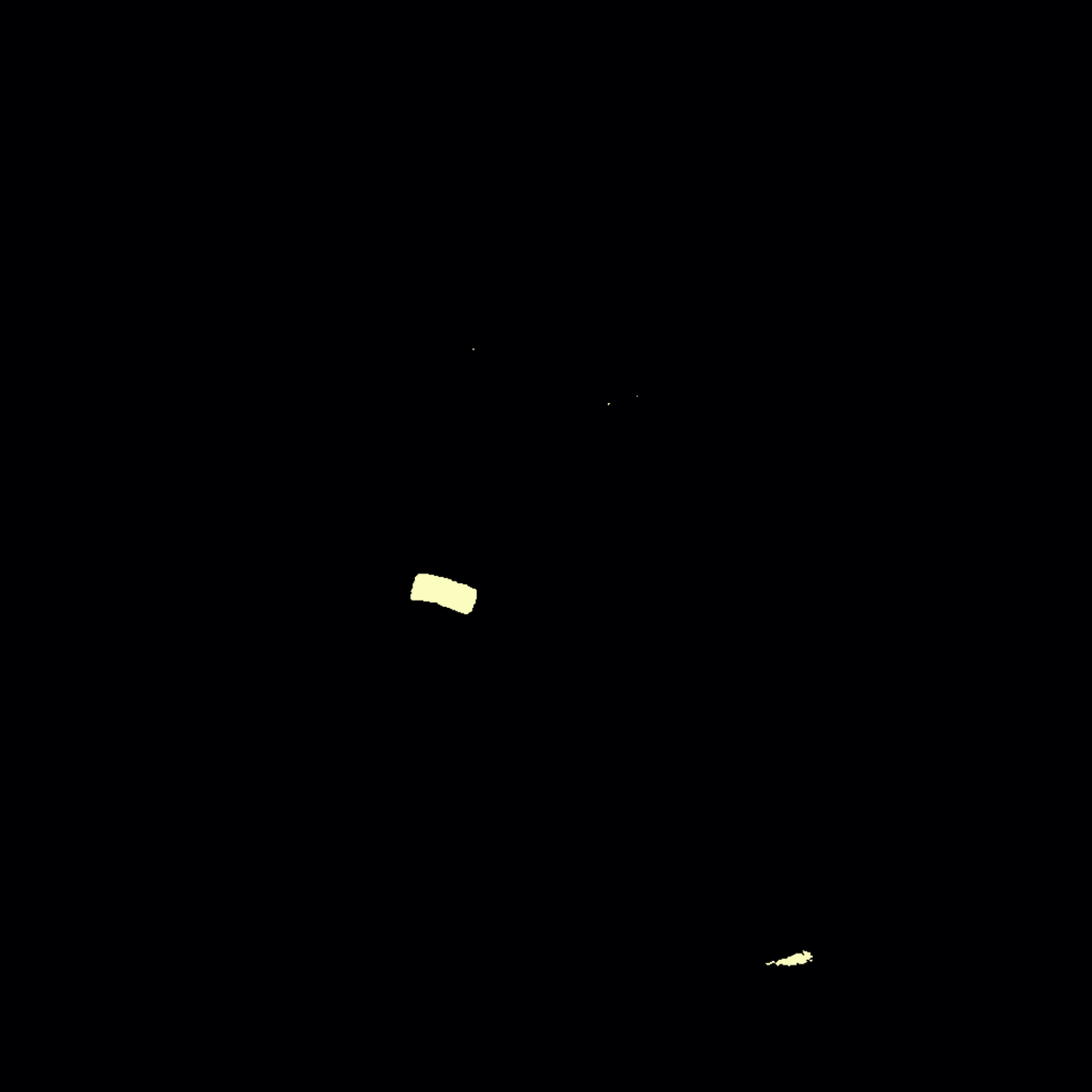} \\
        
    \end{tabular}

    \caption{(Top) As the filter strength increases ($p=0 \to 2$), the raw luminance image becomes progressively blurred, spatially distributing the energy of high-variance Monte Carlo samples. (Bottom) This energy distribution stabilizes the corresponding glare masks. Note the isolated white pixels scattered throughout the background at $p=0$; as the filter diffuses their energy, these false-positive spikes fall below the evaluation threshold and are suppressed by $p=2$. This isolates primary glare source and prevents stochastic noise from destabilizing the optimization gradients.}
    \label{fig:scattering_solution}
    \vspace{-0.5em}
\end{figure}

\new{Concretely, we implement this PSF as $p$ sequential applications of a $3{\times}3$ Gaussian kernel, with corner, edge, and center weights of $\frac{1}{16}$, $\frac{1}{8}$, and $\frac{1}{4}$ respectively, applied to the luminance image prior to mask computation. The parameter $p$ reported in Figure~\ref{fig:scattering_solution} denotes the number of scattering passes.} Mathematically, this physically-grounded scattering stabilizes the inverse rendering process, by absorbing outlier radiance values.

\begin{figure}[t]
    \centering
    
    \begin{subfigure}[b]{0.48\linewidth}
        \centering
        \includegraphics[width=\textwidth]{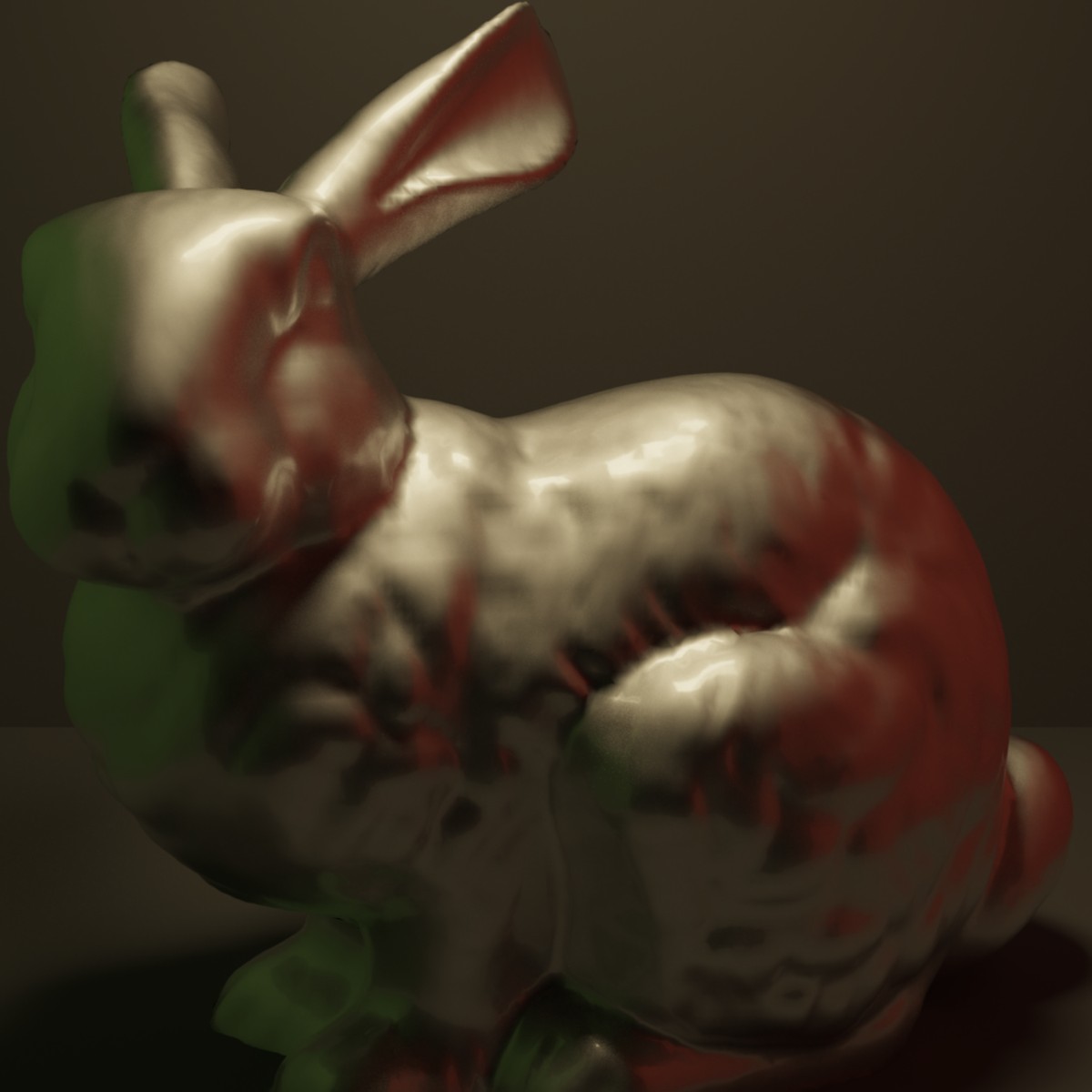}
        \caption{RGB Render}
    \end{subfigure}
    \hfill
    \begin{subfigure}[b]{0.48\linewidth}
        \centering
        \includegraphics[width=\textwidth]{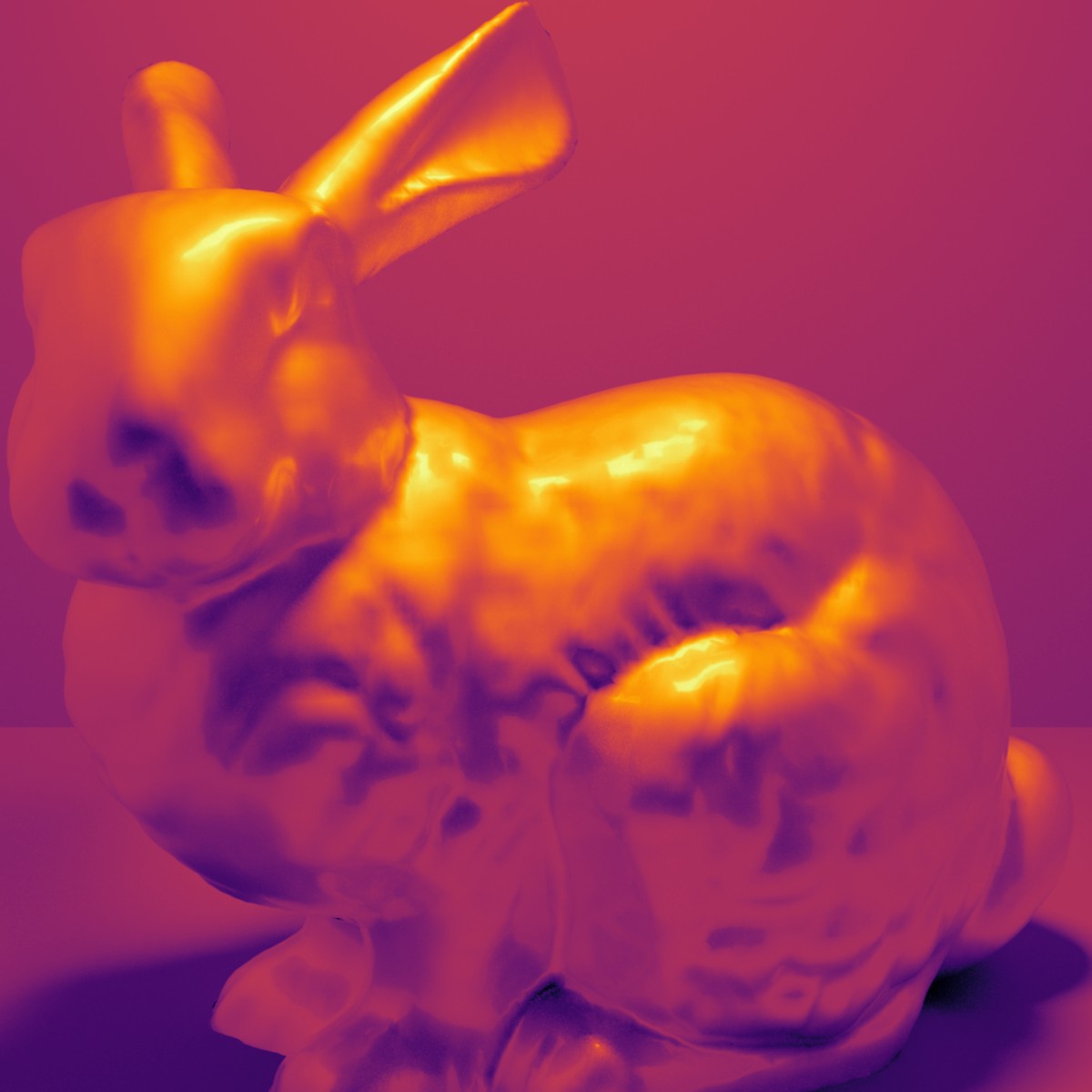}
        \caption{Perceptual Luminance}
    \end{subfigure}
    
    \vspace{0.25cm}
    
    \begin{subfigure}[b]{0.23\linewidth}
        \centering
        \includegraphics[width=\textwidth]{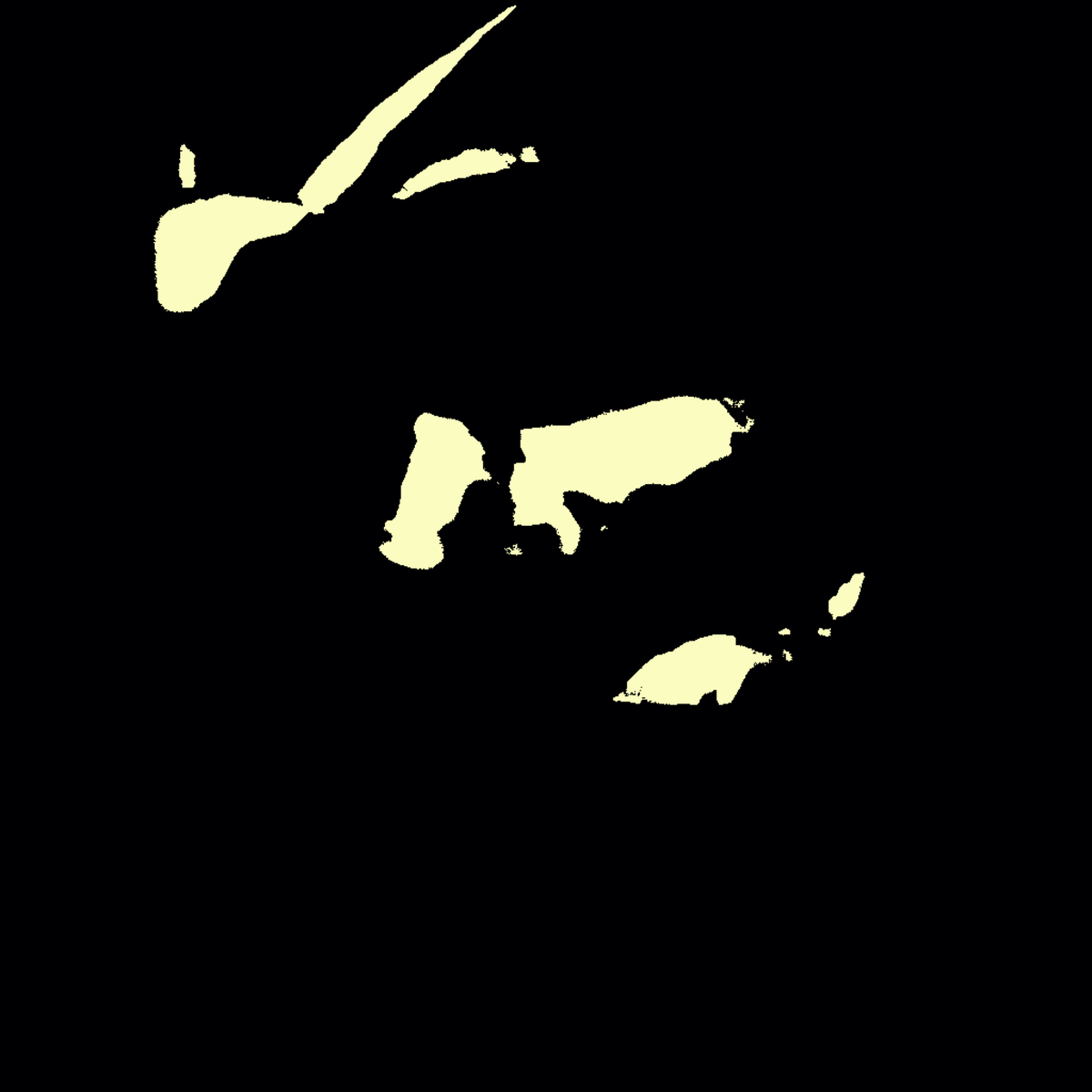}
        \caption{$k=\infty$}
    \end{subfigure}\hfill
    \begin{subfigure}[b]{0.23\linewidth}
        \centering
        \includegraphics[width=\textwidth]{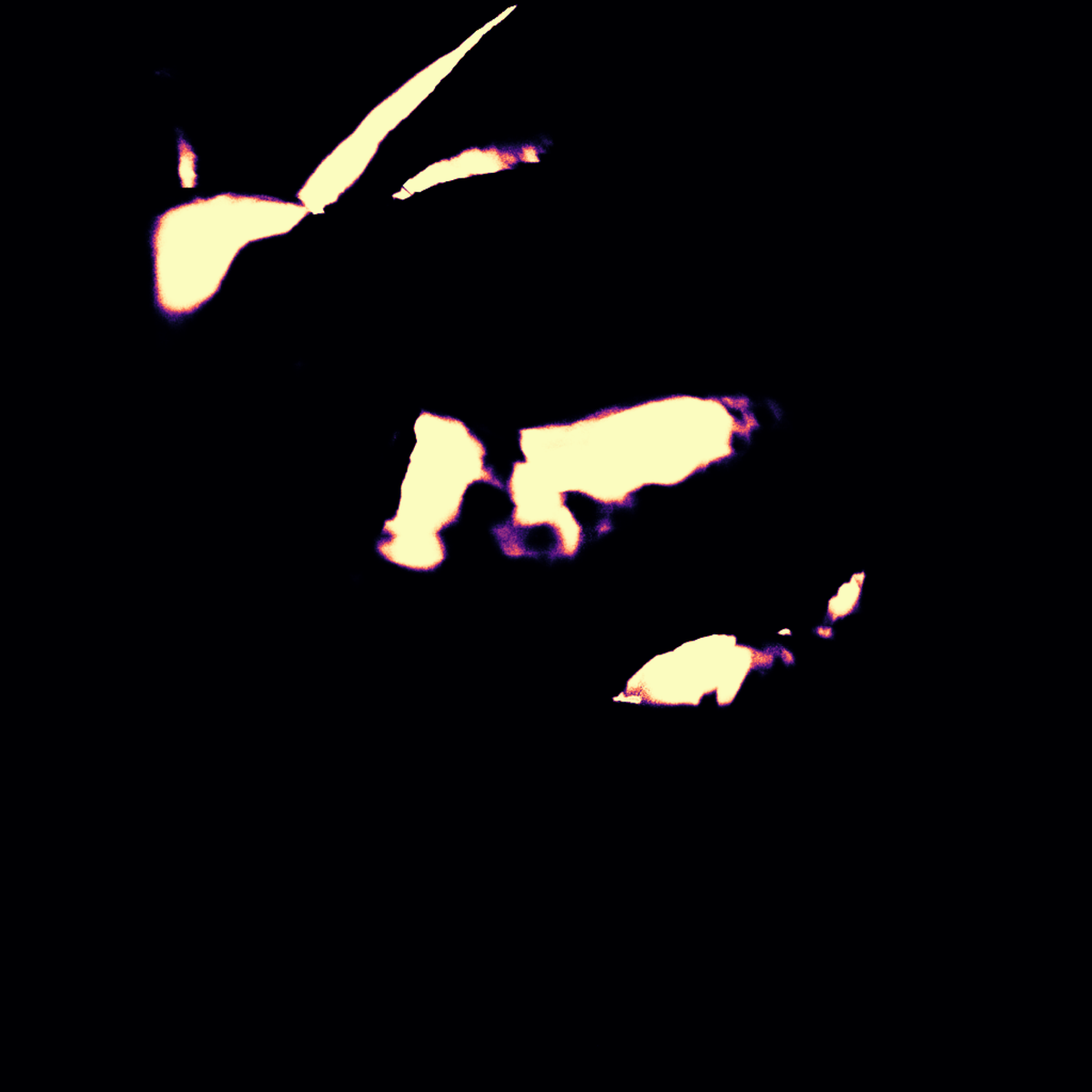}
        \caption{$k=15$}
    \end{subfigure}\hfill
    \begin{subfigure}[b]{0.23\linewidth}
        \centering
        \includegraphics[width=\textwidth]{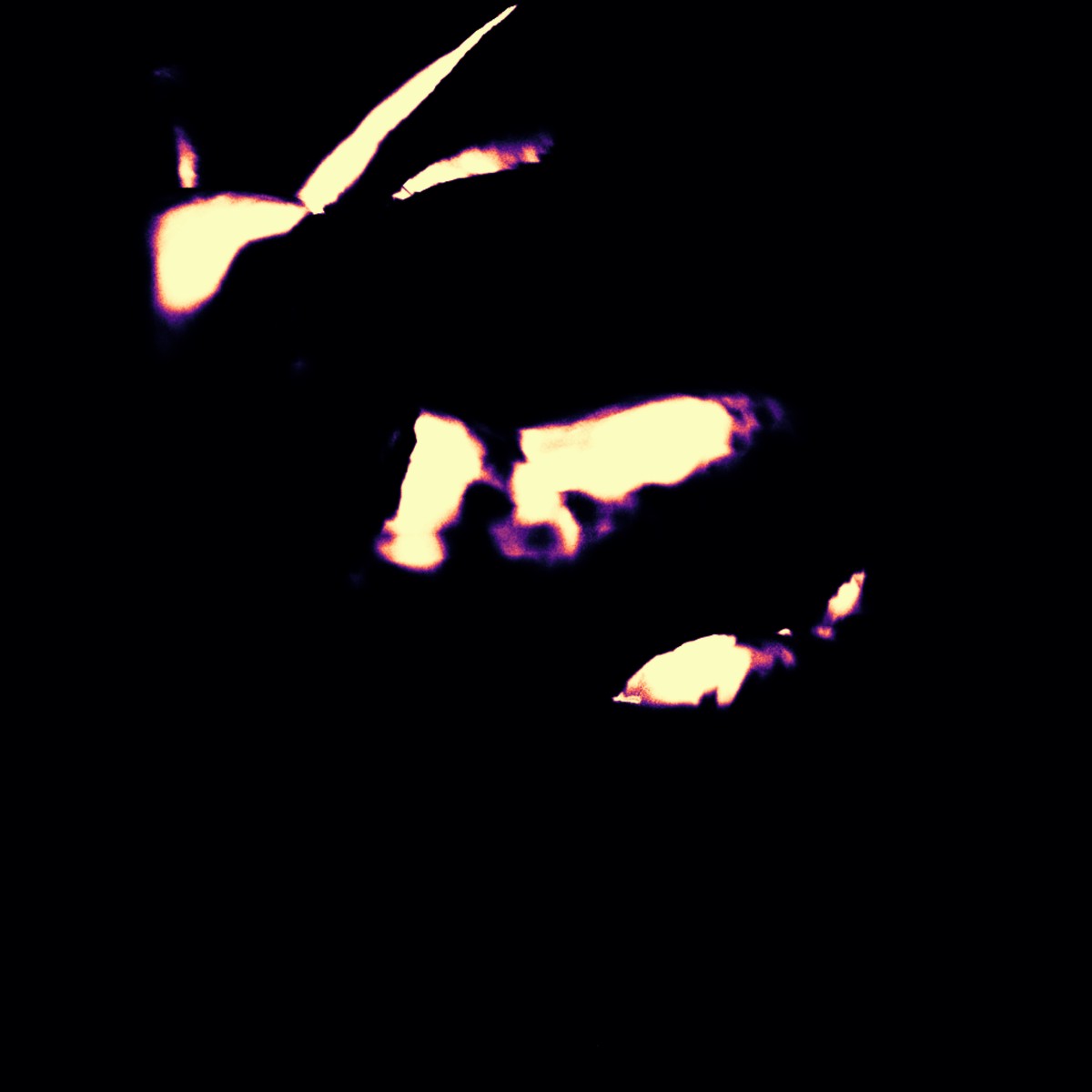}
        \caption{$k=10$}
    \end{subfigure}\hfill
    \begin{subfigure}[b]{0.23\linewidth}
        \centering
        \includegraphics[width=\textwidth]{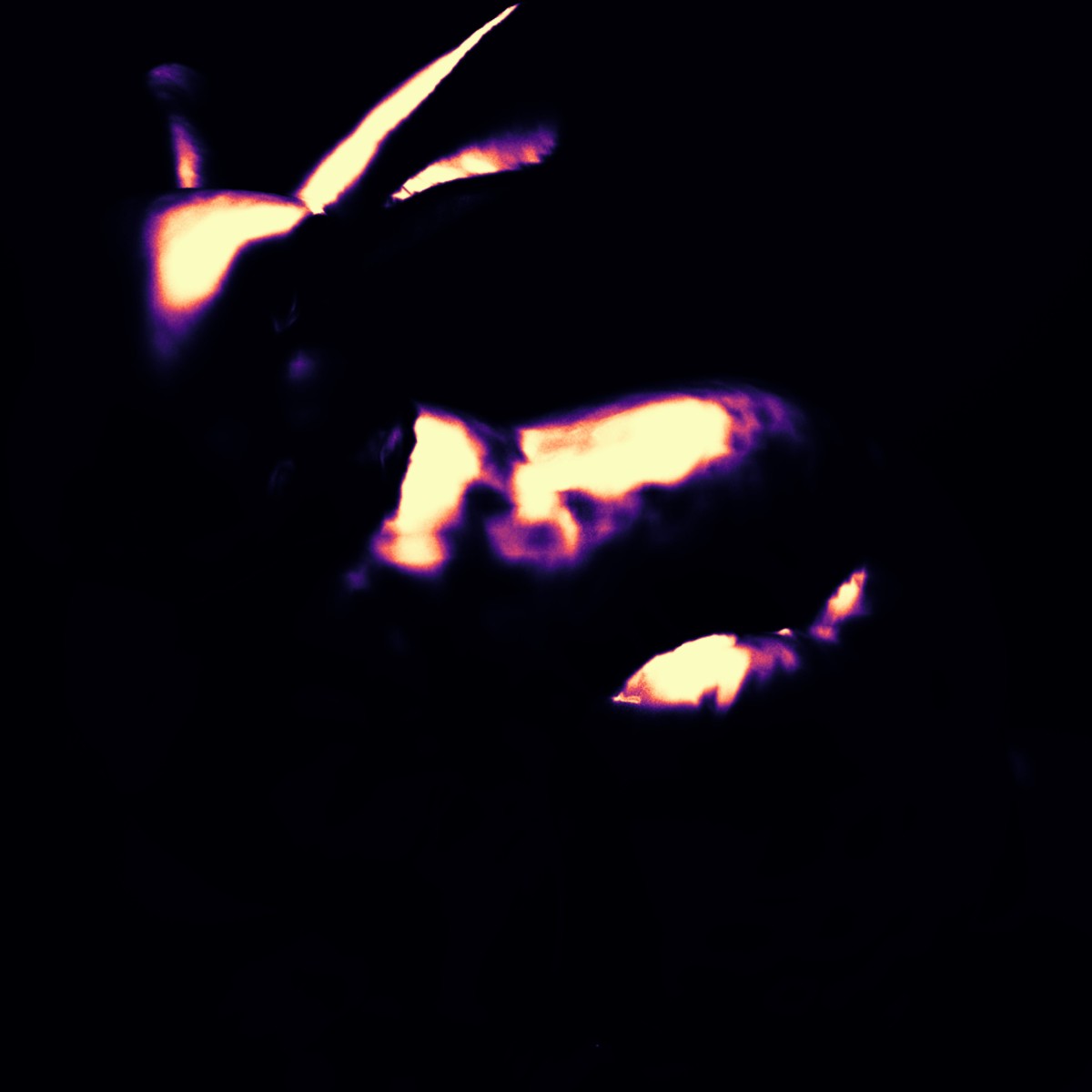}
        \caption{$k=5$}
    \end{subfigure}
    
    \caption{Differentiable boundary formulation via Sigmoid softening. (a, b) The baseline render and its perceptual luminance. (c) Even after resolving Monte Carlo variance, a standard binary threshold produces a Heaviside step function, preventing gradient propagation outside the immediate glare source. (d-f) By relaxing the sharpness parameter ($k$), we expand the functional penumbra into a continuous gradient. This ensures that scene parameters just outside the threshold receive meaningful optimization signals, smoothly guiding them toward compliance.}
    \label{fig:sigmoid_grid}
    \vspace{-2em}
\end{figure}

\subsection{Soft UGR Proxy}
\label{sec:soft_ugr}
To enable gradient-based optimization, we must replace the discrete $L > 5L_b$ classification with a continuous, differentiable proxy. We achieve this by formulating a soft masking function that assigns a fractional glare weight, $w_{glare} \in [0, 1]$, to every pixel in the image based on its relative luminance.

First, to establish a stable adaptation threshold without allowing the optimizer to trivially scale the global exposure to artificially minimize the loss, we compute the global mean luminance of the scene, $\bar{L}$, and detach it from the computational graph. The glare threshold is then defined as $L_{\mathrm{thresh}} = 5\bar{L}$.

We define the relative luminance difference for a given pixel $i$ as:
\begin{equation}
    \Delta_i = \frac{L_i - L_{\mathrm{thresh}}}{L_{\mathrm{thresh}}}.
\end{equation}

\noindent This formulation is mathematically equivalent to the \textit{Weber contrast} \cite{peli1990contrast}, a fundamental measure in psychophysics that describes the perceived difference between a stimulus and its background. By framing the glare source identification in terms of contrast rather than raw radiance, our framework aligns with established models of human visual perception.

To map this relative luminance difference into a continuous spatial mask, we apply a logistic sigmoid function\new{, following the sigmoid-based luminance remapping used in photographic tone reproduction~\cite{reinhard2002parameter}}:
\begin{equation}
    m_i = \frac{1}{1 + \exp(-k \Delta_i)},
\end{equation}
\noindent where $k$ is a hyperparameter dictating the steepness of the transition threshold. We set $k = 10.0$ to ensure non-zero derivatives during optimization while maintaining a strictly localized boundary around high-radiance sources. Consequently, a pixel whose luminance significantly exceeds the evaluation threshold approaches a mask value of $m_i \approx 1.0$, whereas pixels below the threshold approach $0.0$.

Using this continuous mask, we can smoothly separate the scene into glare and background components. The effective glare luminance for pixel $i$ becomes:
\begin{equation}
    L_{s, i} = L_i \cdot m_i.
\end{equation}
\noindent The background adaptation luminance $L_b$ must also be evaluated continuously to maintain energy conservation and prevent gradient discontinuities. We compute this global adaptation level by weighting the incident scene luminance against the complementary spatial mask, $(1 - m_i)$:
\begin{equation}
    L_b = \frac{\sum_{i=1}^{n} L_i (1 - m_i)}{\sum_{i=1}^{n} (1 - m_i) + \epsilon},
\end{equation}
\noindent where $\epsilon$ is a small numerical constant preventing division by zero.

By substituting these continuous formulations of $L_{s, i}$ and $L_b$ back into the original UGR equation, we obtain a differentiable UGR proxy. This formulation converges to the standard discrete measure at its boundary conditions, specifically, when a pixel's luminance is either decisively below or drastically above the evaluation threshold ($m_i \to 0$ and $m_i \to 1$, respectively). Between these boundaries, the proxy provides a continuous, differentiable loss landscape, allowing gradients to propagate directly from the psychophysical evaluation back to the physical scene parameters.

\section{Implementation and Optimization}
\label{sec:method}

\begin{figure}[t]
  \centering
  \includegraphics[width=\linewidth]{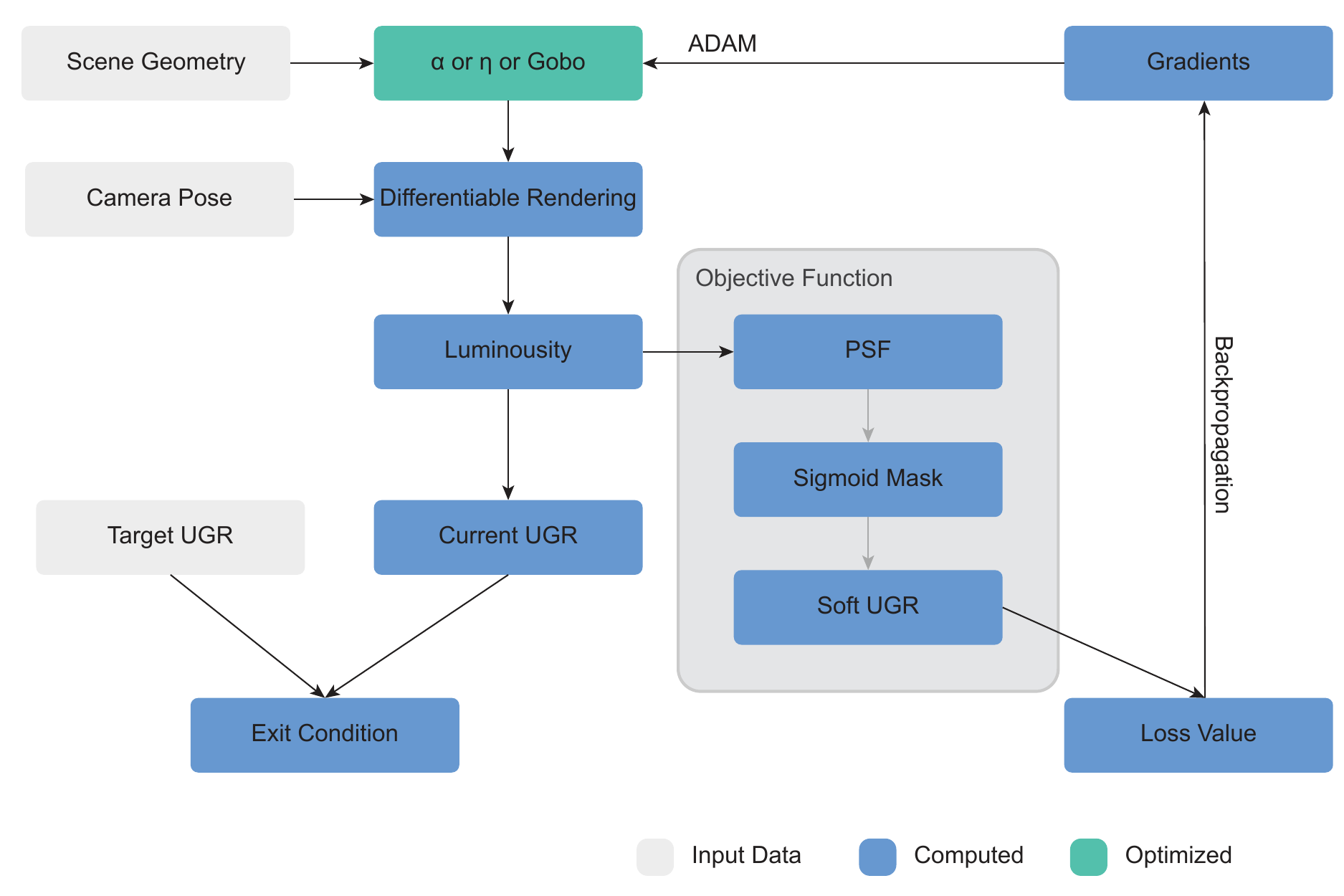}
  \caption{Our framework interfaces with the Mitsuba 3 renderer to compute forward-pass luminance and glare measures. Gradients are backpropagated through a soft UGR proxy to update scene parameters ($\alpha$, $\eta$, and emitter masks) using the Adam optimizer, subject to TV regularization and physical boundary constraints. The optimization exits once the target UGR has been achieved or up to a maximum iteration count.}
  \label{fig:pipeline_flow}
\end{figure}
With our continuous glare formulation established, this section details its integration into a fully differentiable rendering pipeline and its systematic application to physical scene parameters.

\subsection{Pipeline and Loss Formulation}
A high-level overview of our optimization pipeline is illustrated in Figure \ref{fig:pipeline_flow}. We implemented our automated glare mitigation framework in Mitsuba 3 \cite{mitsuba3}, leveraging the Dr.Jit compiler \cite{jakob2022drjit} to evaluate our continuous UGR proxy natively on the GPU. Global illumination gradients were computed using PRB \cite{vicini2021PathReplay}.

\begin{figure*}[t]
    \centering
    \setlength{\tabcolsep}{1.5pt}
    \renewcommand{\arraystretch}{0.5}
    
    \begin{tabular}{cccccc}
        & \small 0 Passes (No PSF) & \small 1 Pass & \small 2 Passes & \small 3 Passes & \small 4 Passes \\
        
        \vspace{0.5mm} \\

        \rotatebox{90}{\hspace{7mm} \small Final Render} &
        \includegraphics[width=0.185\linewidth]{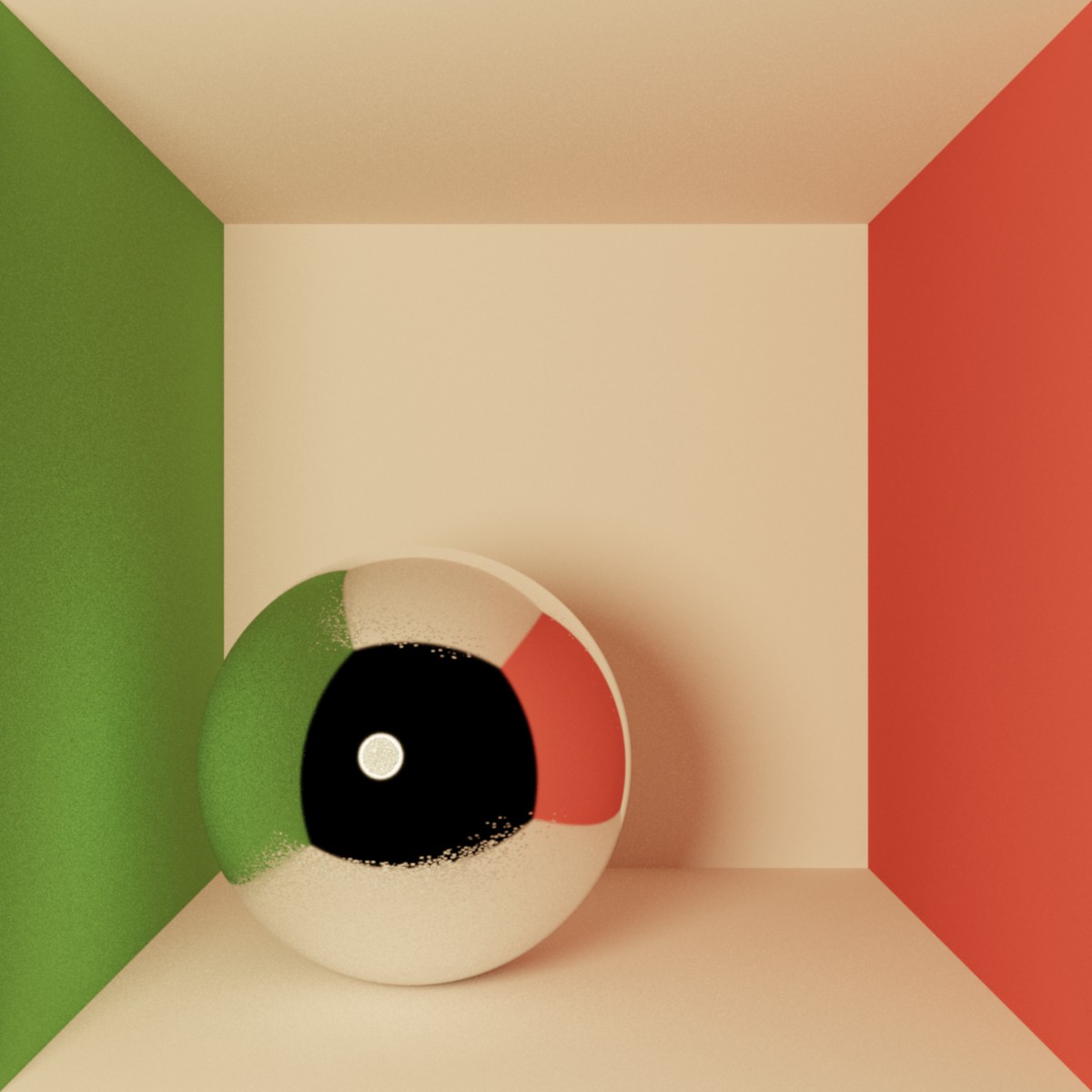} &
        \includegraphics[width=0.185\linewidth]{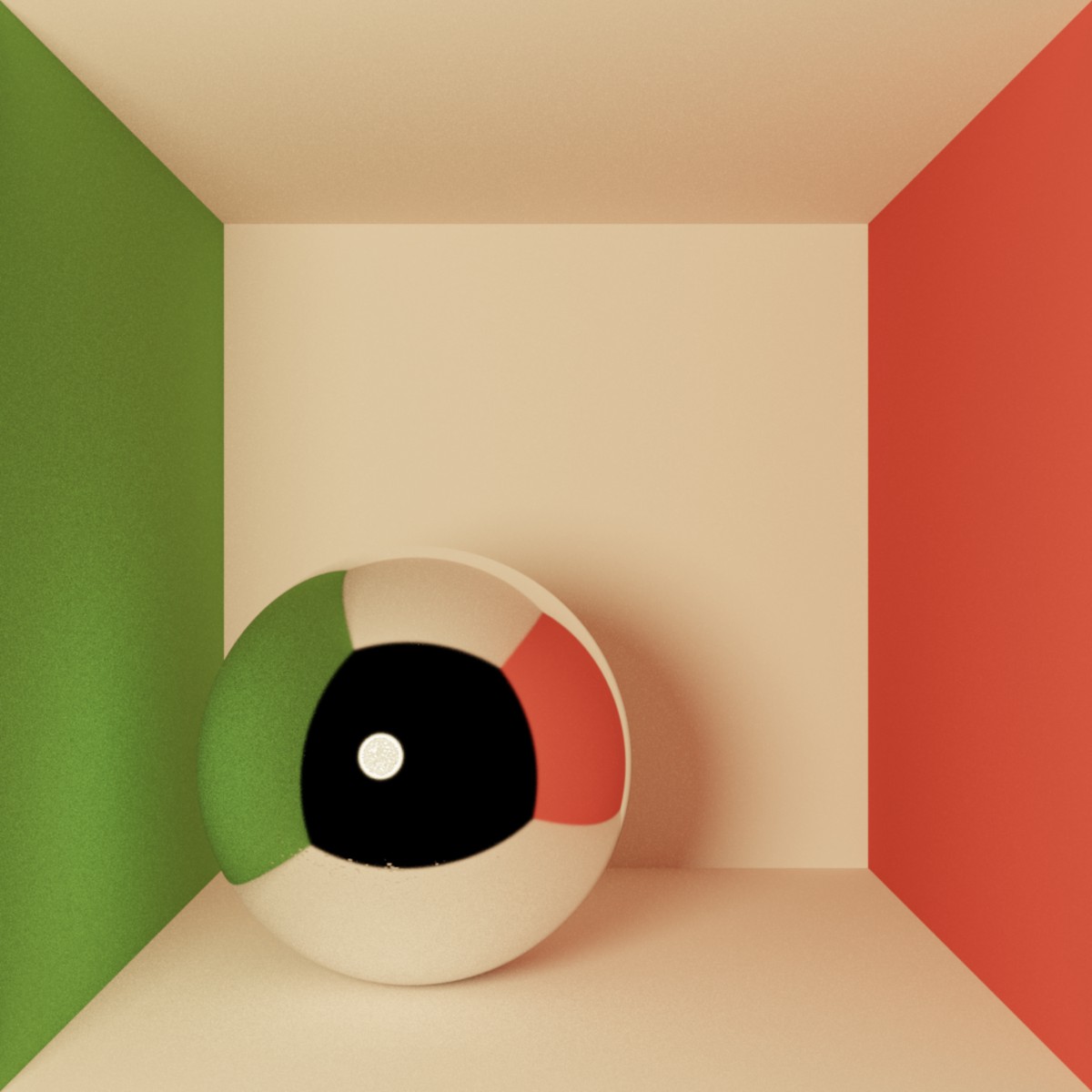} &
        \includegraphics[width=0.185\linewidth]{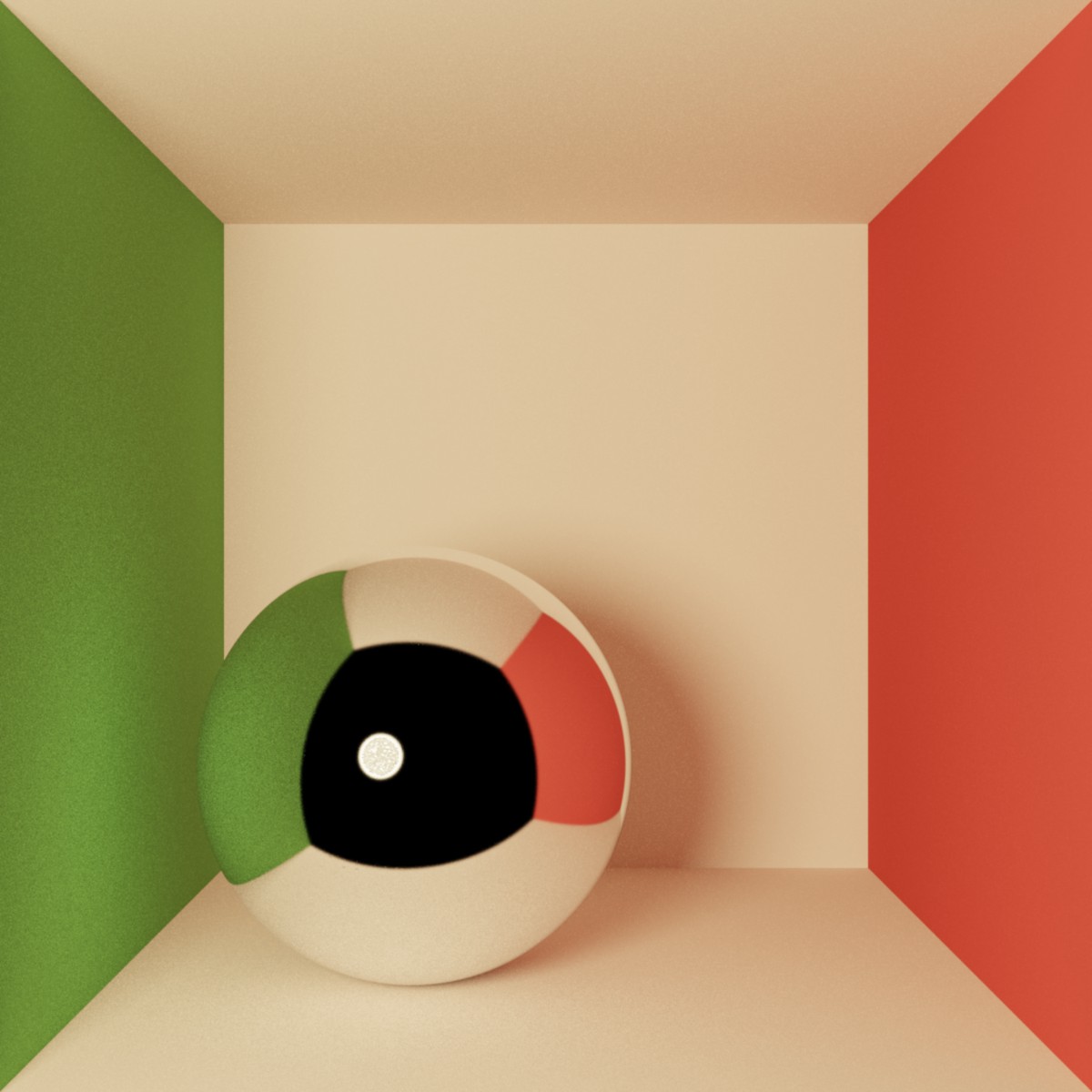} &
        \includegraphics[width=0.185\linewidth]{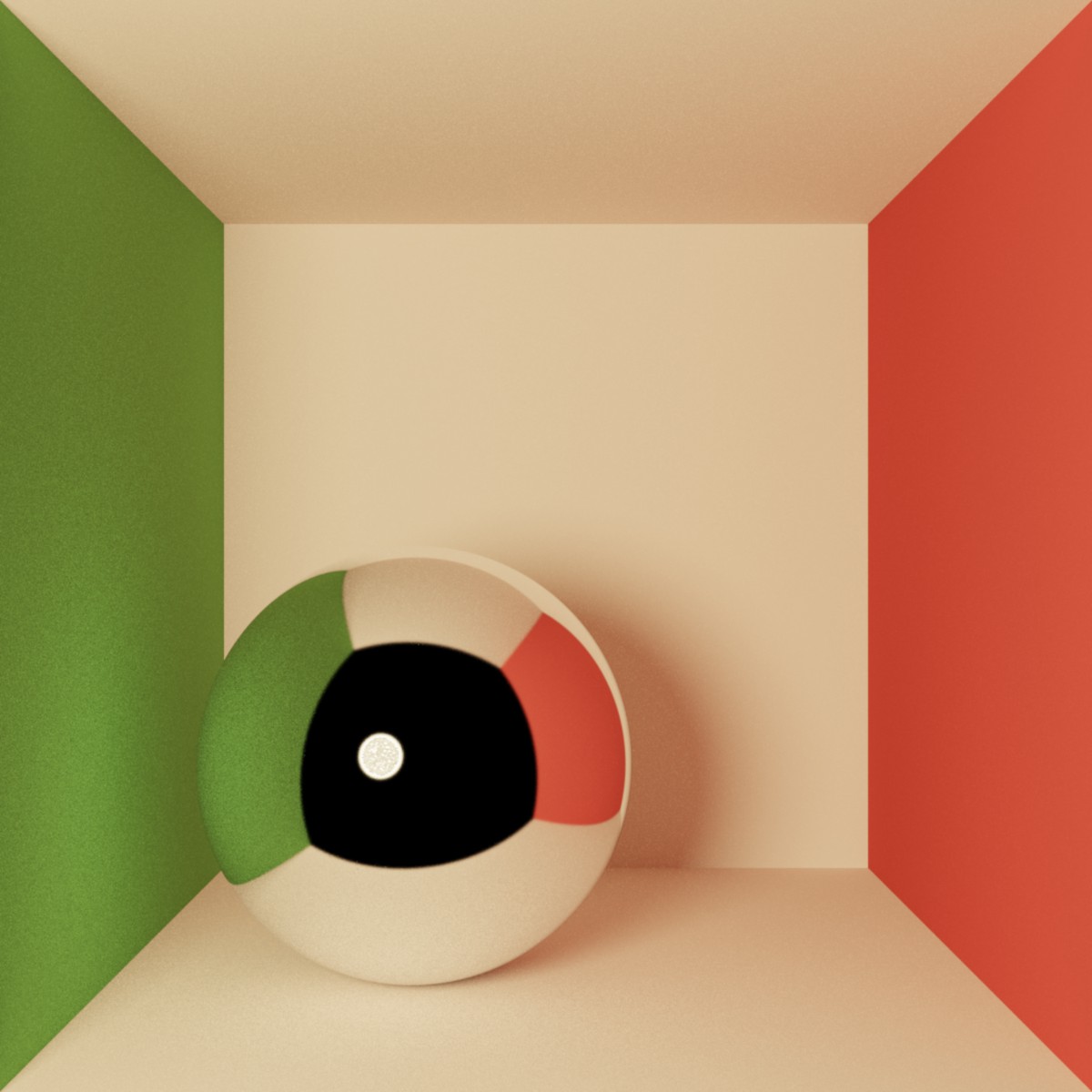} &
        \includegraphics[width=0.185\linewidth]{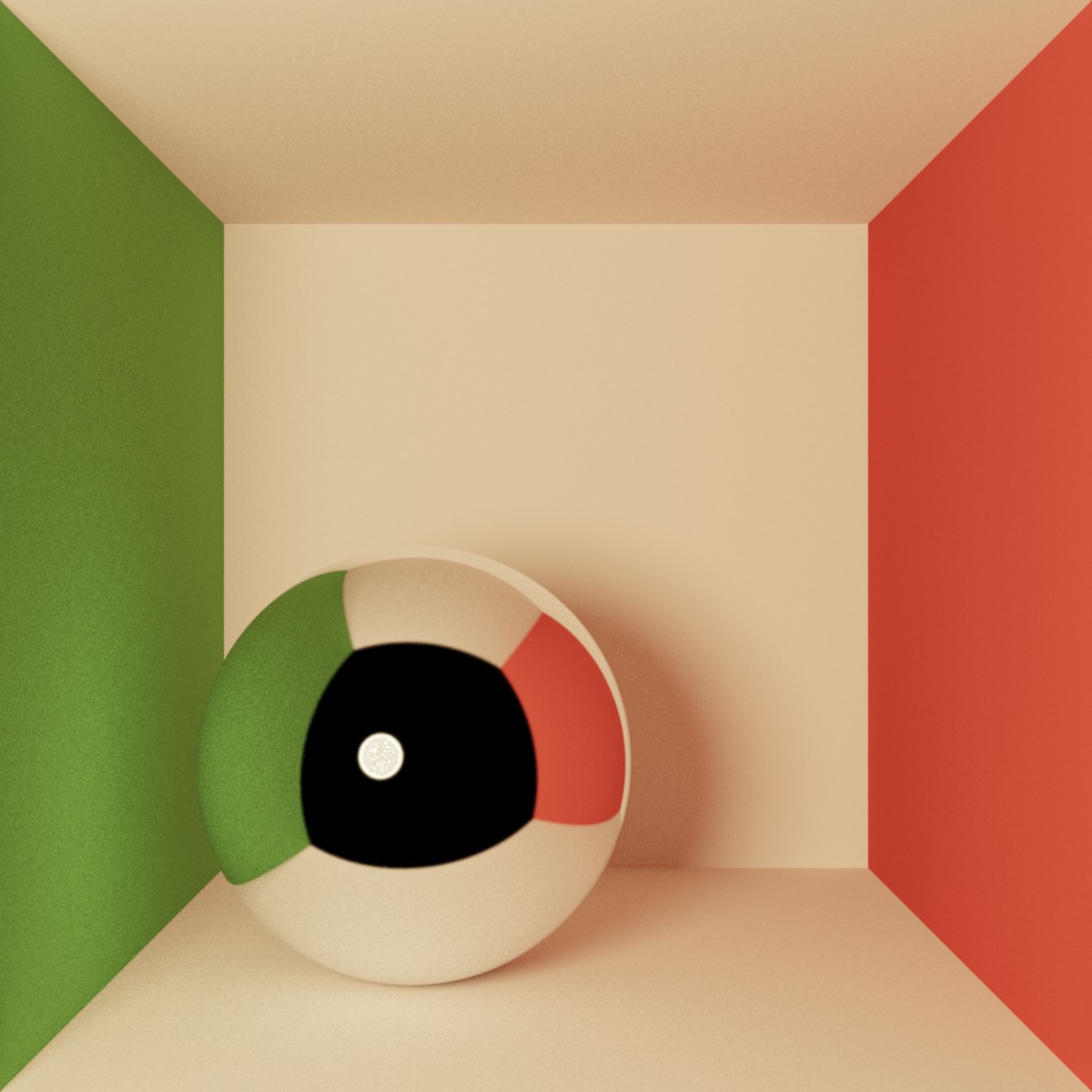} \\
        
        \rotatebox{90}{\hspace{7mm} \small Final Texture} &
        \includegraphics[width=0.185\linewidth]{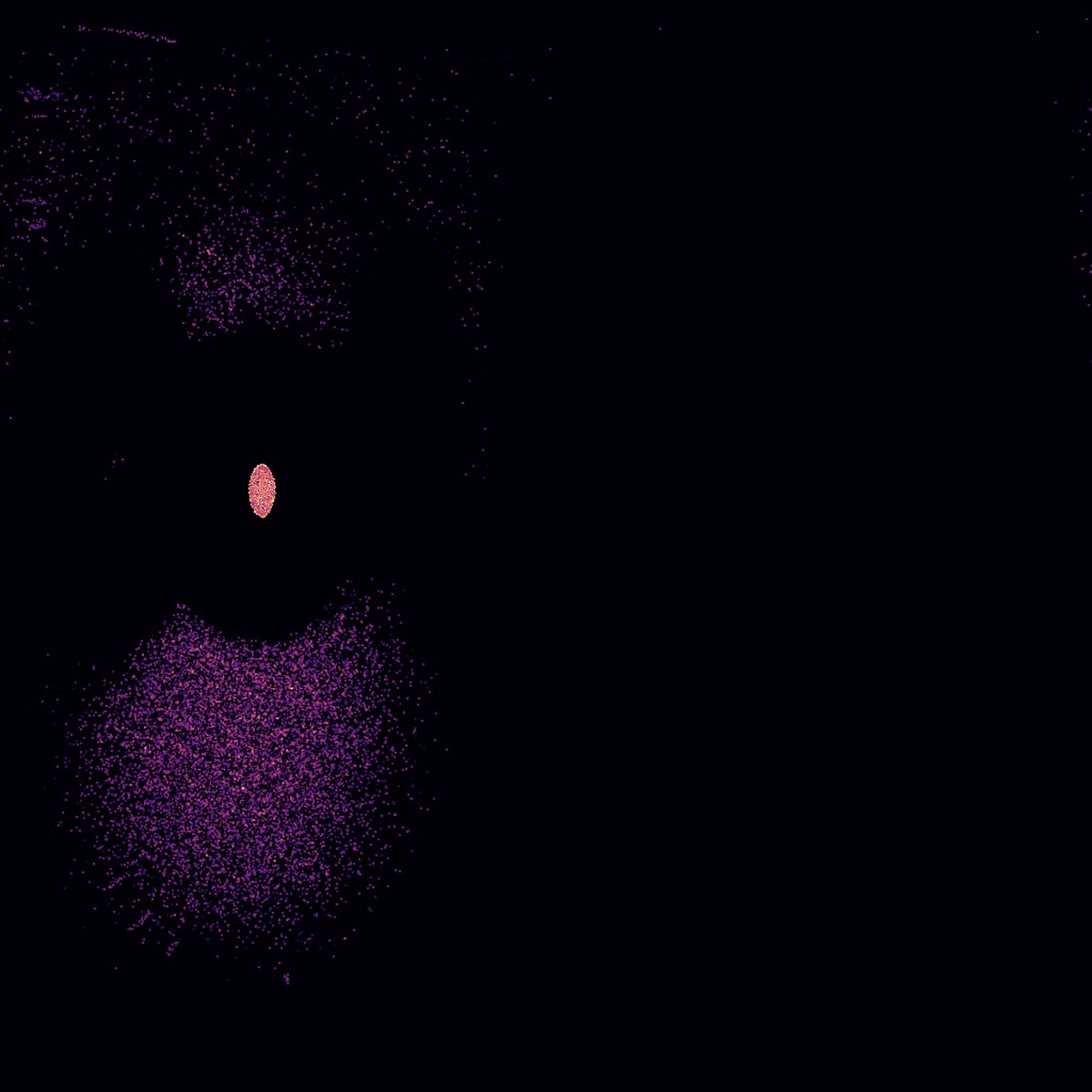} &
        \includegraphics[width=0.185\linewidth]{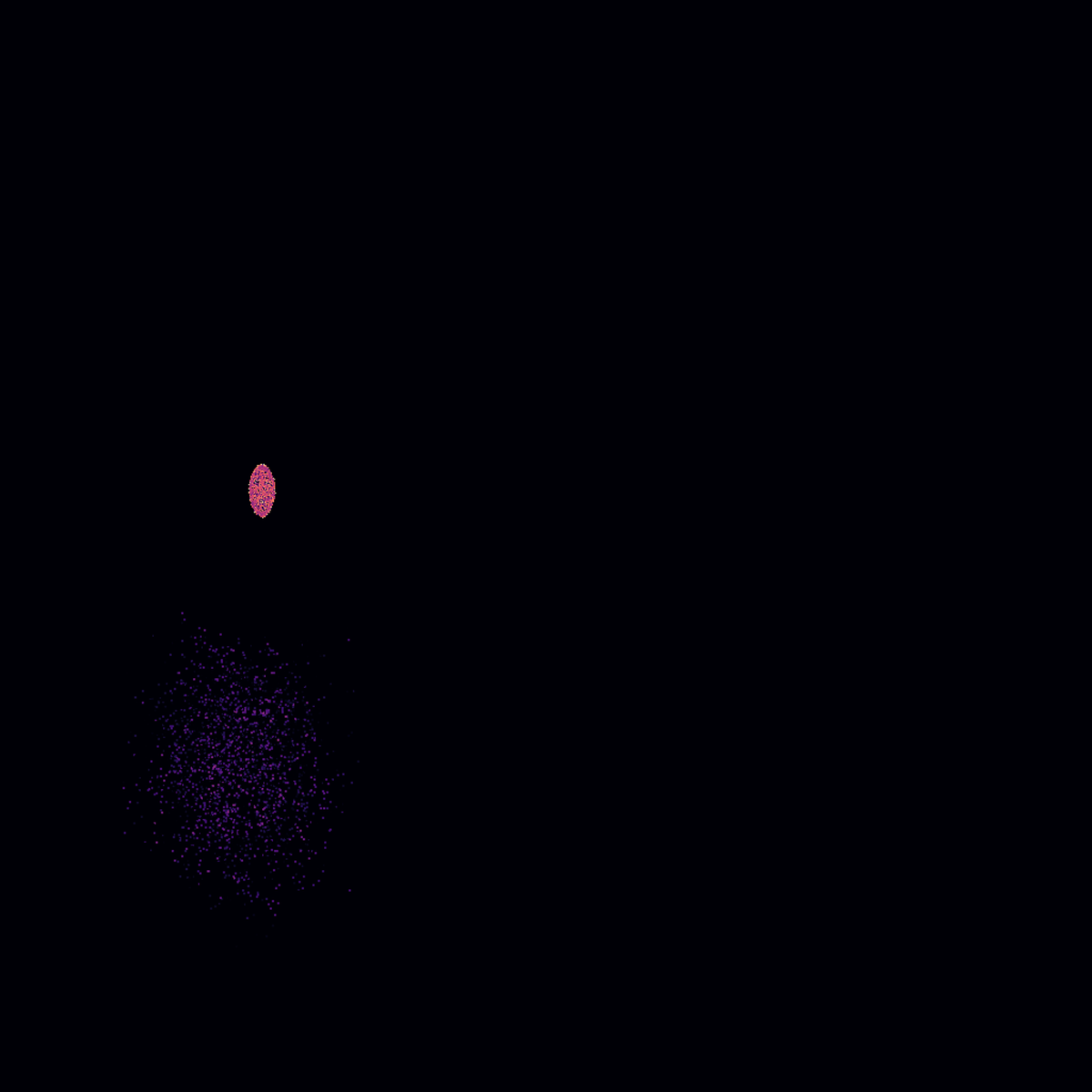} &
        \includegraphics[width=0.185\linewidth]{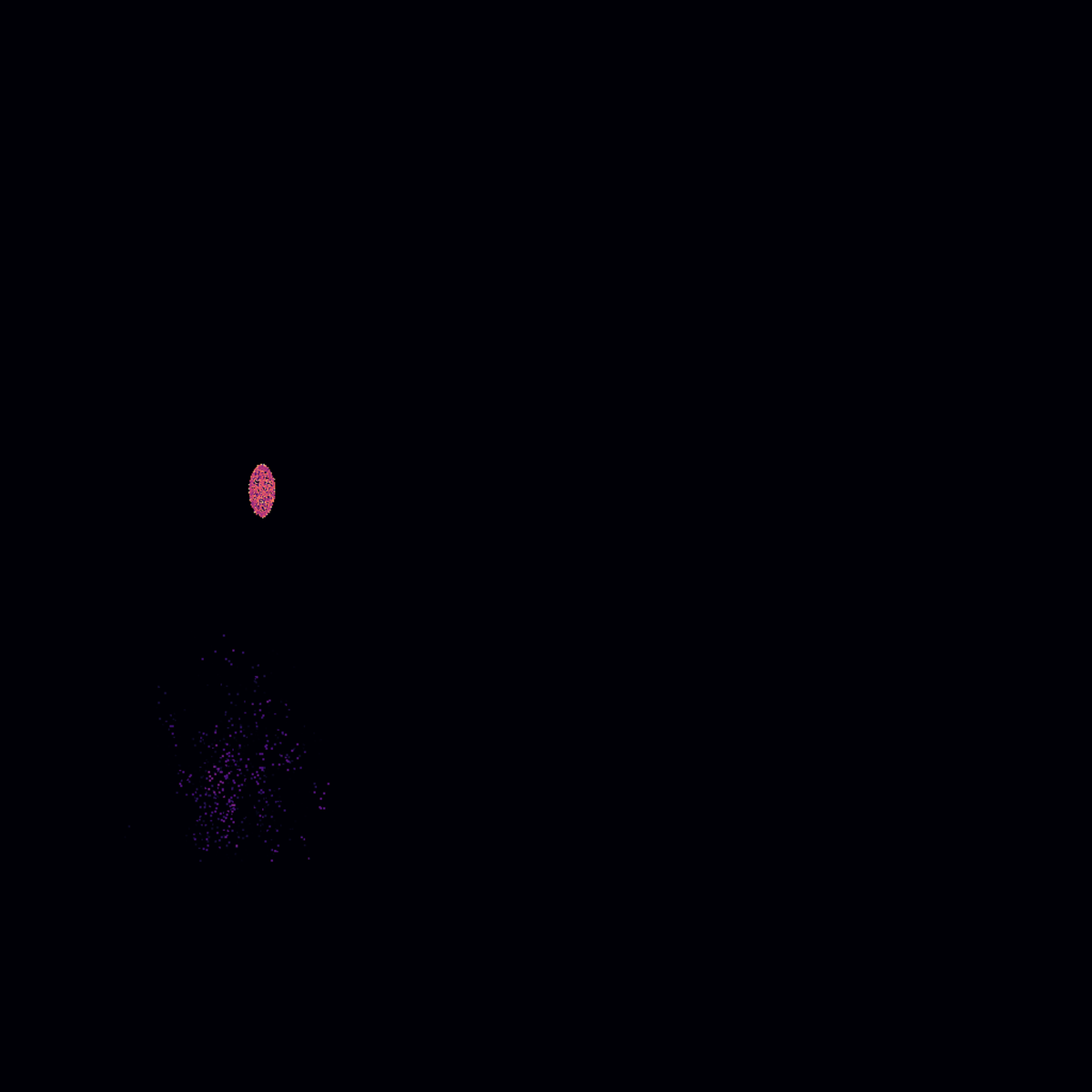} &
        \includegraphics[width=0.185\linewidth]{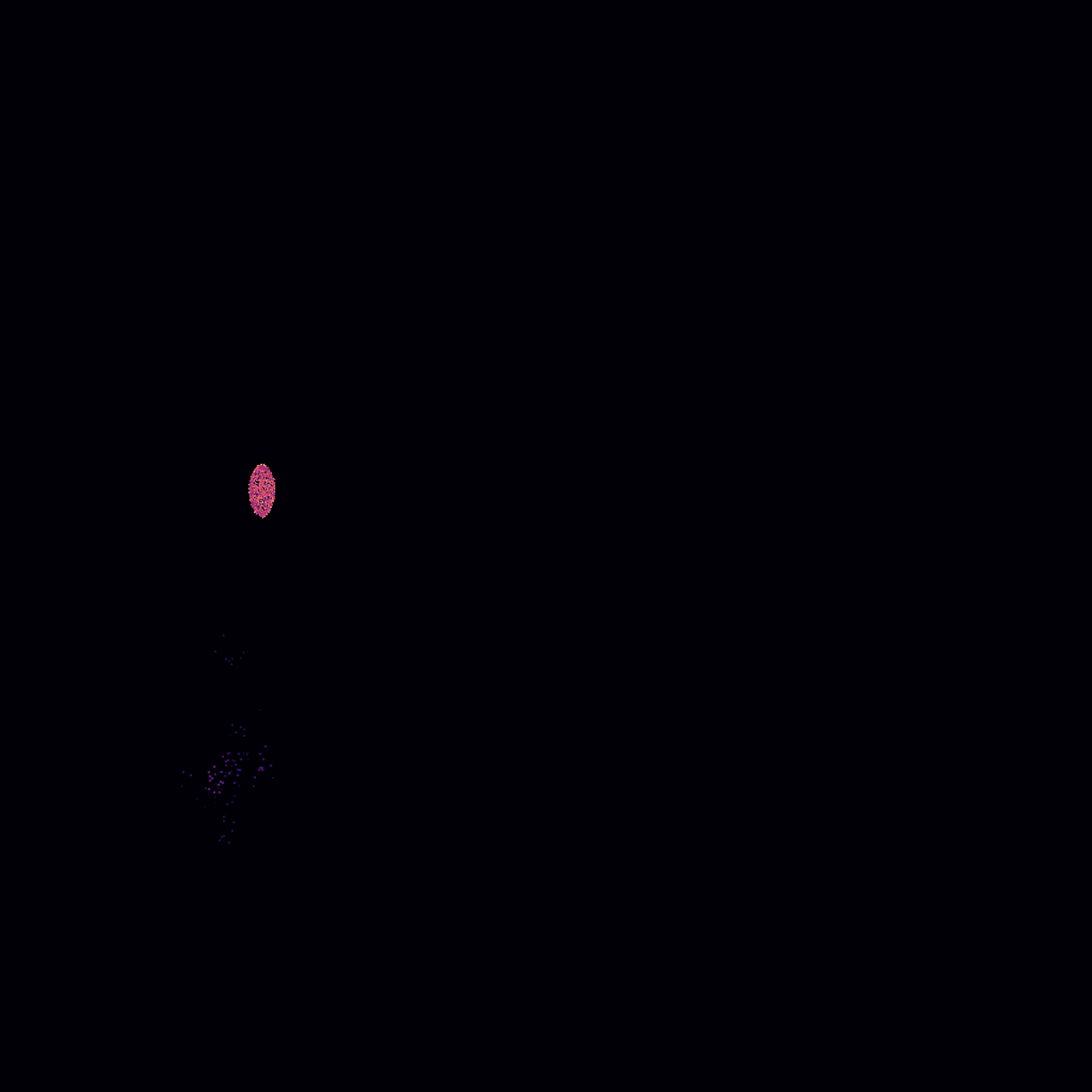} &
        \includegraphics[width=0.185\linewidth]{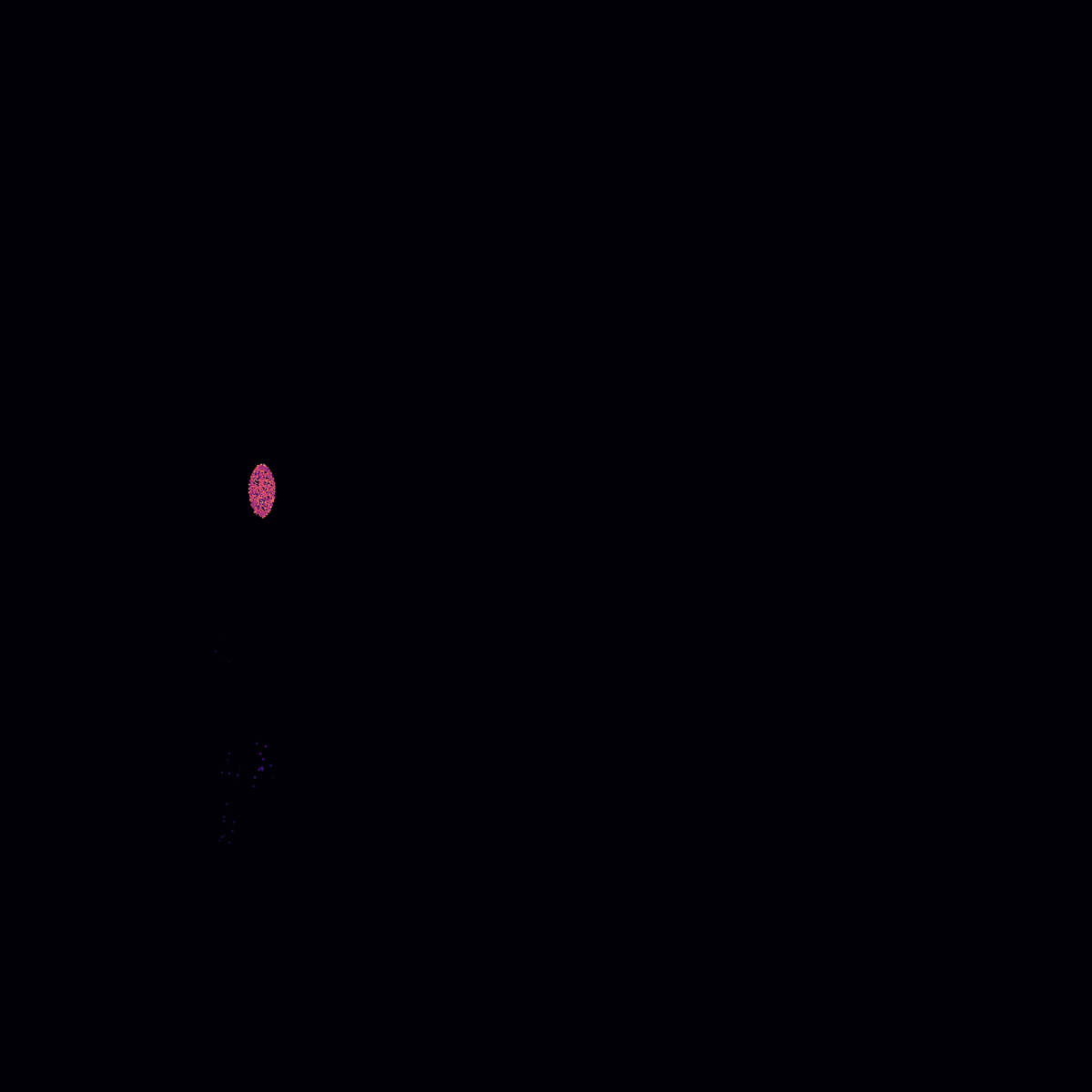} \\
        
    \end{tabular}

    \caption{Optimizing the roughness texture map of the sphere to reduce glare\newtwo{; the optimized roughness map is shown in the bottom row, with the resulting render above}. Optimizing without optical scattering (0 Passes) causes the optimizer to overfit to stochastic ``firefly'' pixels\newtwo{: isolated Monte Carlo noise spikes are misclassified as glare sources, so the optimizer roughens scattered, incoherent texels attempting to suppress glare that is not physically present,} destroying the spatial coherence of the resulting texture map. Applying sequential Point Spread Function passes acts as a low-pass spatial filter, distributing this high-intensity energy to heal the evaluation mask. This ensures stable gradient flow and yields a cleanly converged, physically plausible roughness map.}

    \label{fig:psf_passes}
\end{figure*}

The gradient descent process was driven by the Adam optimizer \cite{kingma2014adam} with the base learning rate dependent on the scene and the optimization parameter. Because inverse rendering via Monte Carlo integration is inherently noisy, optimizing high-resolution spatial textures is prone to severe overfitting. To prevent the optimizer from embedding high-frequency noise patterns into the materials, our total loss function incorporates Total Variation (TV) regularization for spatially varying parameters (emitter masks and AR blending distributions):
\begin{equation}
    \mathcal{L}_{\mathrm{total}} = \mathrm{UGR}_{\mathrm{soft}}(\mathbf{\theta}) + \lambda_{TV} \mathcal{R}_{TV}(\mathbf{\theta}_{\mathrm{spatial}}),
    \label{eq:total_loss}
\end{equation}
\noindent where $\mathbf{\theta}$ represents the full set of optimizable scene parameters, $\mathbf{\theta}_{\mathrm{spatial}} \subseteq \mathbf{\theta}$ denotes the subset of spatially-varying 2D textures subject to spatial regularization, $\mathcal{R}_{TV}$ is the TV penalty, and $\lambda_{TV}$ dictates the regularization strength. While we found TV regularization to be generally effective for enforcing manufacturable, piecewise-smooth physical boundaries, the framework is agnostic to the specific penalty; any appropriate spatial regularizer (e.g., Laplacian smoothing or sparsity constraints) can be chosen depending on the specific architectural context and material being optimized.

\newtwo{Because the UGR proxy applies non-linear functions of the rendered luminance $L$ to compute the rating, both the squared source term and the outer logarithm in Equation~\ref{eq:ugr}, care is needed when estimating the gradient of $\mathcal{L}_{\mathrm{total}}$ in Equation~\ref{eq:total_loss}. For a term $f(L)$, the gradient with respect to scene parameters $\theta$ is $df/dL \cdot dL/d\theta$; when $f$ is non-linear, evaluating $df/dL$ at a Monte Carlo estimate of $L$ yields a biased gradient, since only $L$ itself is estimated without bias~\cite{Nicolet2023Recursive}. We mitigate the effect by using a higher sample count in the forward pass to reduce the variance of $L$ (consistent with the deeper forward-pass path lengths discussed in Section~\ref{sec:results_gi_background}); debiasing constructions such as Russian-roulette estimators~\cite{misso2022unbiased} offer a complementary alternative.}

Finally, the gradient descent loop implements an early stopping criterion. The optimization halts immediately once the calculated UGR of the scene falls below a user-defined ergonomic threshold (e.g., UGR $\le$ 19 for standard office spaces). This prevents the optimizer from needlessly darkening the environment or over-altering materials once visual comfort has been achieved.

\new{Across all experiments we apply $p=4$ PSF scattering passes and a maximum of 100 optimization iterations, with early stopping once the target UGR (typically $\le 17$) is reached. The Adam learning rate was set in the range $0.05$--$0.1$ depending on the scene and parameter domain, and the TV regularization weight $\lambda_{TV}$ was set in the range $0.01$--$0.1$ depending on the spatial resolution of the optimized texture.}

\begin{figure}[t]
    \centering
    \begin{subfigure}[b]{0.32\linewidth}
        \centering
        \includegraphics[width=\linewidth]{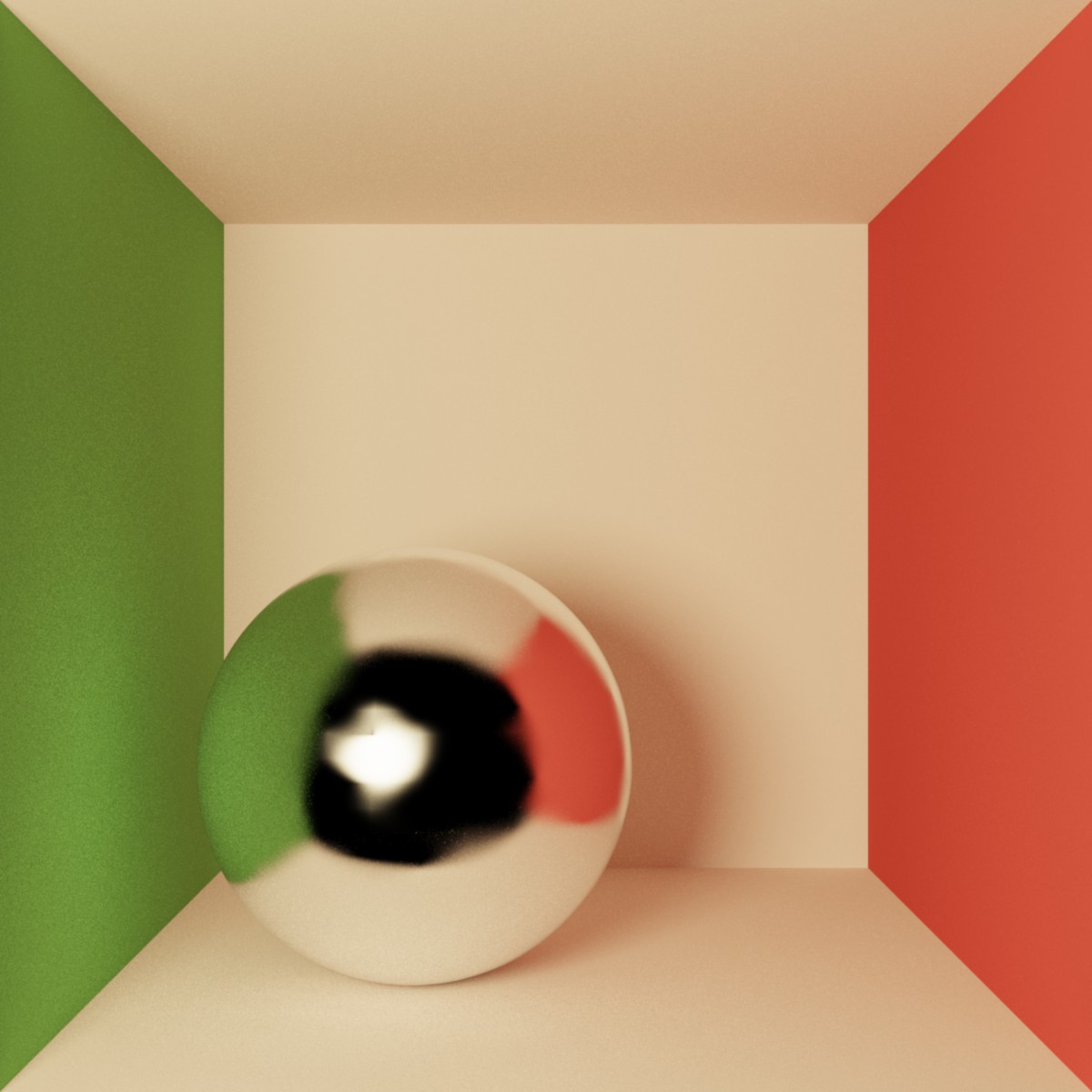}
        \caption{Initial Render}
    \end{subfigure}\hfill
    \begin{subfigure}[b]{0.32\linewidth}
        \centering
        \includegraphics[width=\linewidth]{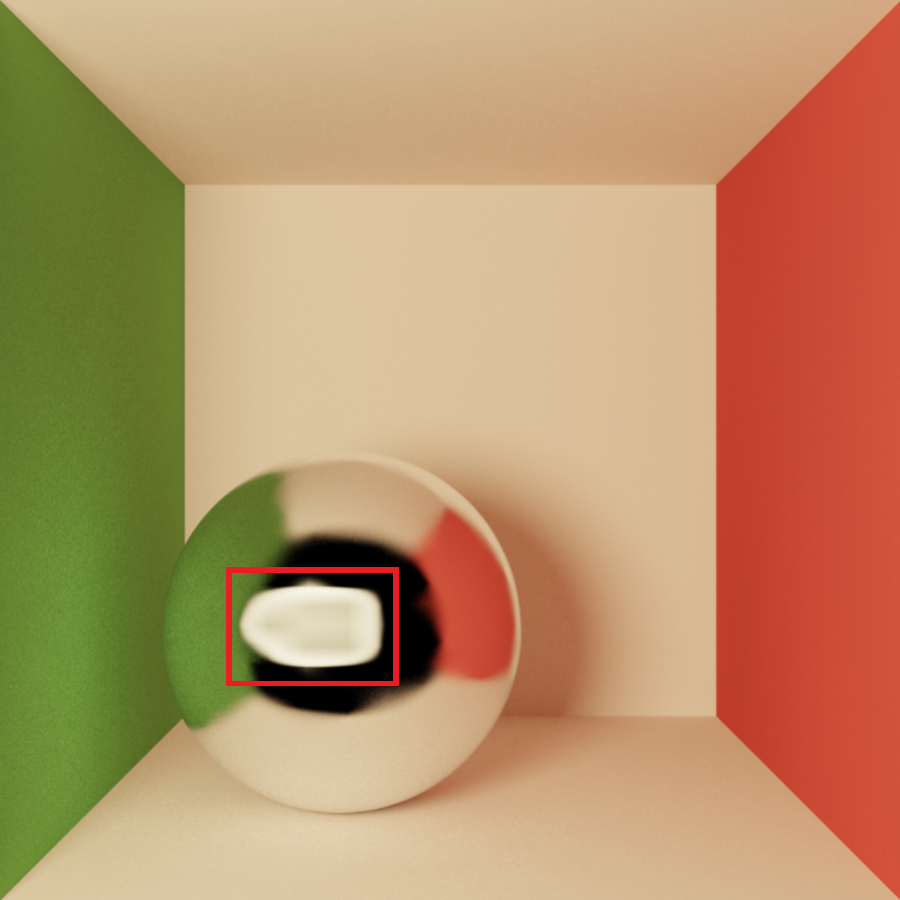}
        \caption{Hard Step}
    \end{subfigure}\hfill
    \begin{subfigure}[b]{0.32\linewidth}
        \centering
        \includegraphics[width=\linewidth]{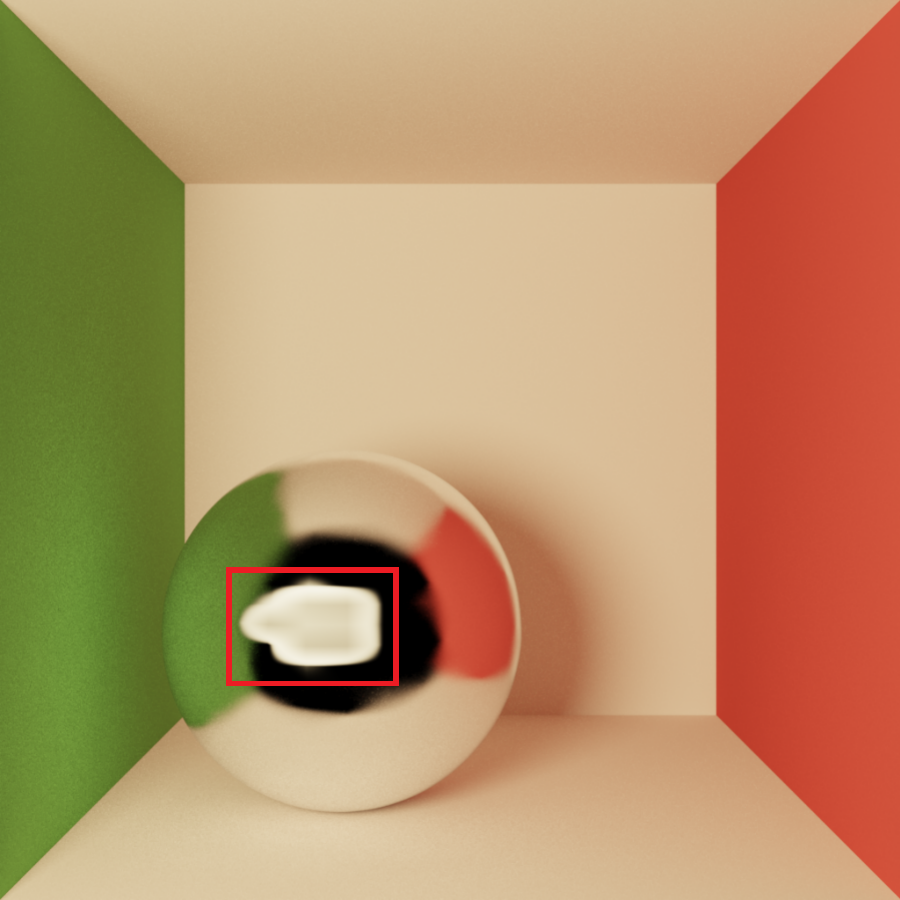}
        \caption{Soft Sigmoid}
    \end{subfigure}
    
    \caption{Ablation of the evaluation threshold. Compared to the glaring initial state (a), the discontinuous Hard Step function (b) forces the optimizer to over-compensate during gradient descent, resulting in an excessively blurred, unnatural surface finish. Our continuous Soft Sigmoid formulation (c) achieves a visually comfortable convergence without over-altering the material.}
    \label{fig:sigmoid_ablation_cornell}
\end{figure}

\subsection{Radiometric Mitigation Domains}
\label{sec:mitigation}
With a fully continuous loss landscape established and our pipeline configured, we can systematically apply gradient descent to the physical scene parameters responsible for visual discomfort. We categorize our automated mitigation strategies into three distinct radiometric domains.

\vspace{1em}
\noindent\textbf{Source Side: Emitter Radiance Masking} \par
\noindent The most direct intervention occurs at the light source. We parameterize the spatially varying emission profile of the luminaires as high-resolution textures. The optimizer selectively dims specific directional solid angles, effectively generating an automated, view-dependent \textit{computational gobo} that maximizes comfort without altering scene geometry.

\vspace{1em}
\noindent\textbf{Boundary Side: Anti-Reflective (AR) Optimization} \par
\noindent When light sources are fixed (e.g., daylighting), we target the physical boundary of the glaring object. 
\begin{itemize}
    \item \textit{Scalar Refraction:} We expose the relative index of refraction to the optimizer, guiding it toward minimizing Fresnel reflections.
    \item \textit{Spatial Blending:} For composite surfaces, we optimize a high-resolution mask interpolating between specular and AR-coated materials. To ensure physically plausible, contiguous patterns and prevent high-frequency noise, we apply the aforementioned TV regularization loss to the spatial gradients of the mask.
\end{itemize}

\vspace{1em}
\noindent\textbf{Surface Side: Anti-Glare (AG) Scattering} \par
\noindent We simulate industry standard scattering (e.g., frosted glass) by exposing the microfacet roughness parameter ($\alpha$). 


\section{Results and Discussion}
\label{sec:results}
We evaluate our automated glare mitigation framework through a series of controlled ablations and complex architectural scenarios. Because inverse rendering performance is fundamentally bound by Monte Carlo sample counts and scene complexity, our empirical analysis prioritizes the robustness of the optimization landscape over raw wall-clock execution time.

\subsection{Differentiable Formulation and Stability}
\label{sec:results_stability}

Our initial experiments validate the necessity of modifying the standard UGR formulation for gradient-based optimization, specifically regarding Monte Carlo noise and discontinuous thresholding.

\textit{Resolving Variance via Optical Scattering.}
To evaluate the impact of Monte Carlo noise on spatial material optimization, we optimized a roughness texture to a target UGR using identical setups, isolating the presence of our PSF scattering passes as the sole variable. As demonstrated in Figure \ref{fig:psf_passes}, optimizing without PSF scattering causes the objective function to misidentify stochastic ``fireflies'' as true glare sources. Consequently, the optimizer updates incorrect regions of the texture map, attempting to mitigate non-existent glare and destroying spatial coherence. Simulating intraocular scattering acts as a critical low-pass filter, dilating these noise spikes and yielding cleanly converged texture maps.

\textit{Continuous Gradients via Soft Sigmoid.}
We also performed a controlled ablation to compare our continuous Soft Sigmoid formulation against a standard Hard Step function. In Figure \ref{fig:sigmoid_ablation_cornell} both optimizations were initialized from the identical high-glare Cornell Box state and halted upon reaching the same target UGR. While the Hard Step function technically succeeds in reducing the UGR in this extreme scenario, it produces noticeable visual artifacts. Because each iteration causes new pixels to abruptly cross the binary glare threshold, those pixels are suddenly hit with the full force of the Adam optimizer's gradient update. This discontinuity causes the optimizer to over-compensate, resulting in excessive and unnecessary surface blurring. Conversely, our Soft Sigmoid formulation provides a smooth gradient penumbra, reaching the target UGR while preserving the intended physical specularity of the material.

\begin{figure}[t]
    \centering
    \includegraphics[width=1.0\columnwidth]{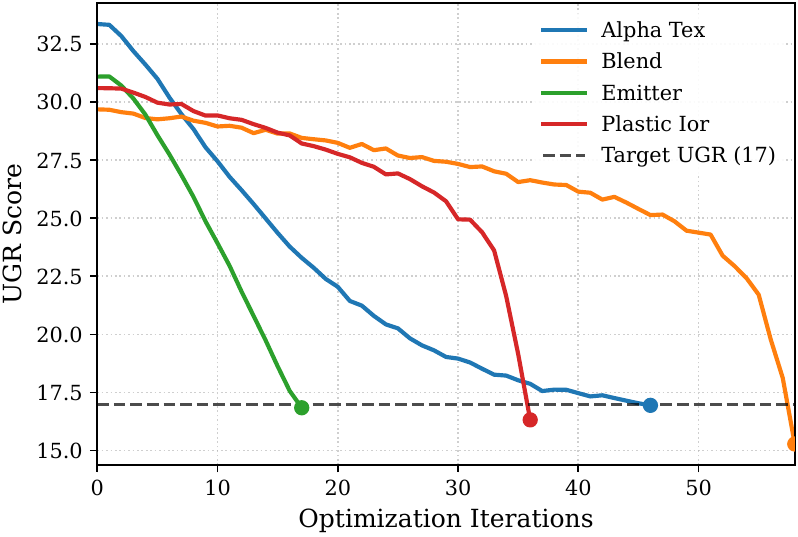}
    \caption{We plot the UGR reduction over time for each parameter isolated in Figure \ref{fig:ablation_matrix}. Our framework demonstrates consistent, smooth convergence regardless of whether the target is a surface property (Roughness, IOR) or a light control parameter (Radiance, Blend), reaching the target UGR }
    \label{fig:convergence_graph}
\end{figure}

\begin{figure*}[t]
    \centering
    \setlength{\tabcolsep}{2pt}
    \renewcommand{\arraystretch}{0.5}
    
    \begin{tabular}{ccccc}
        & \small Initial Render & \small Optimized Render & \small Initial Luminance & \small Optimized Luminance \\
        
        \vspace{0.5mm} \\
        \rotatebox{90}{\hspace{10mm} \small Conductor $\alpha$} &
        \includegraphics[width=0.22\linewidth]{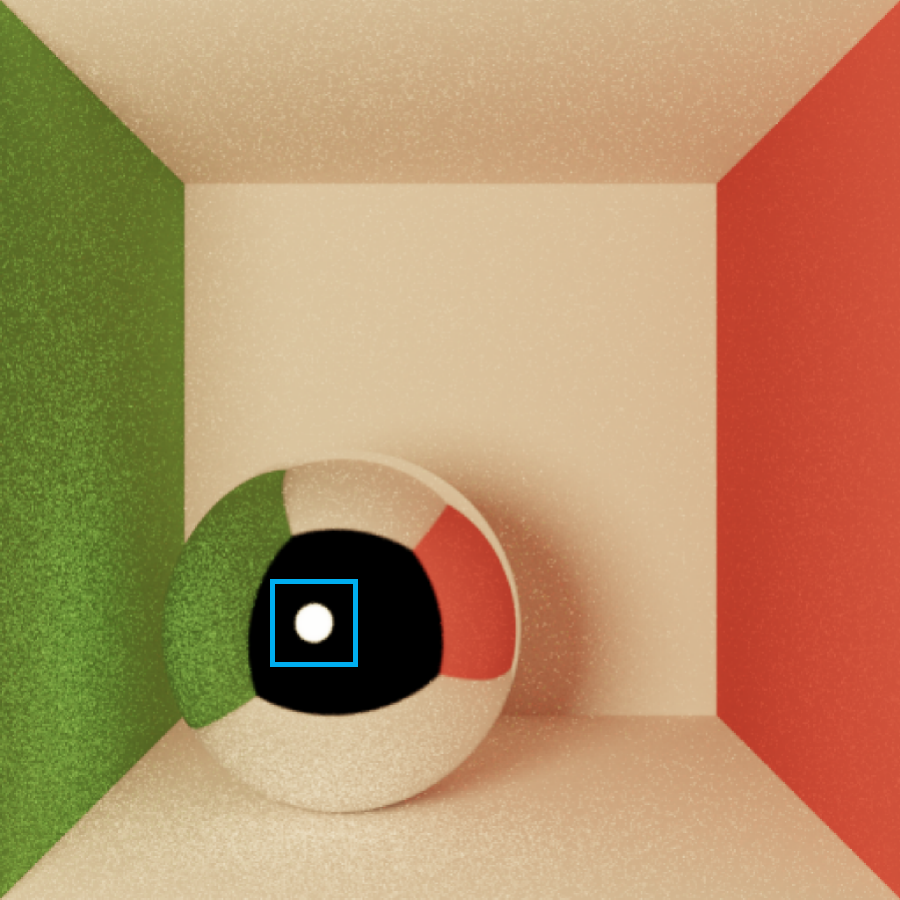} &
        \includegraphics[width=0.22\linewidth]{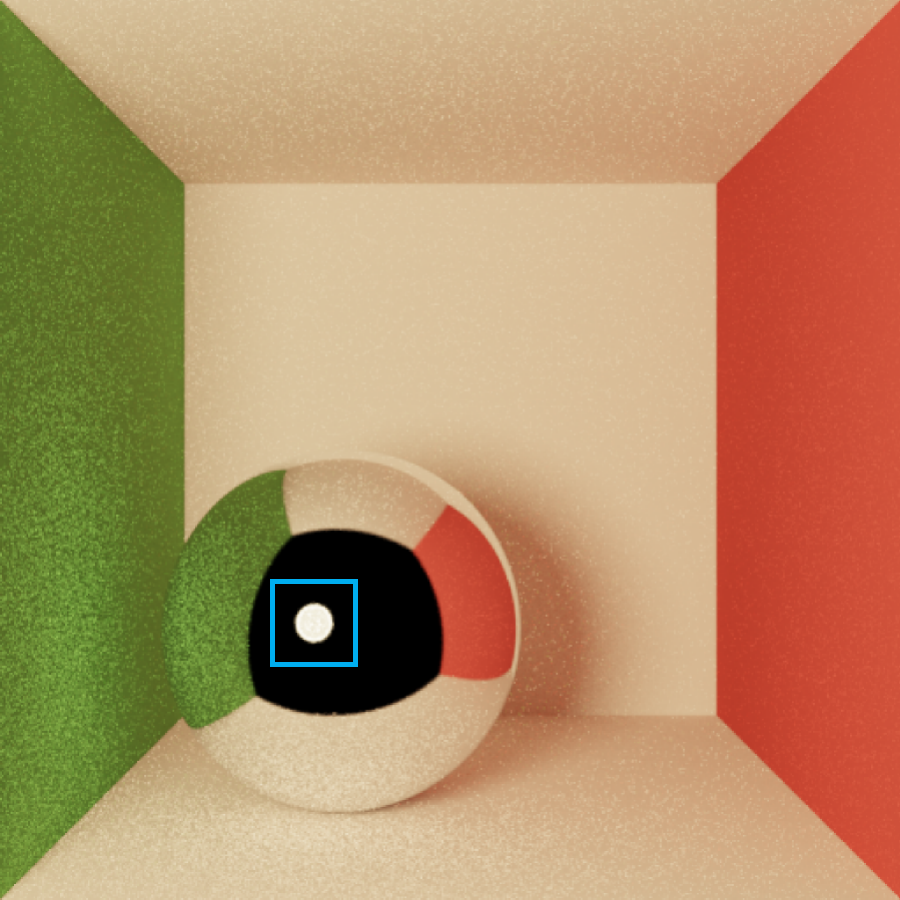} &
        \includegraphics[width=0.22\linewidth]{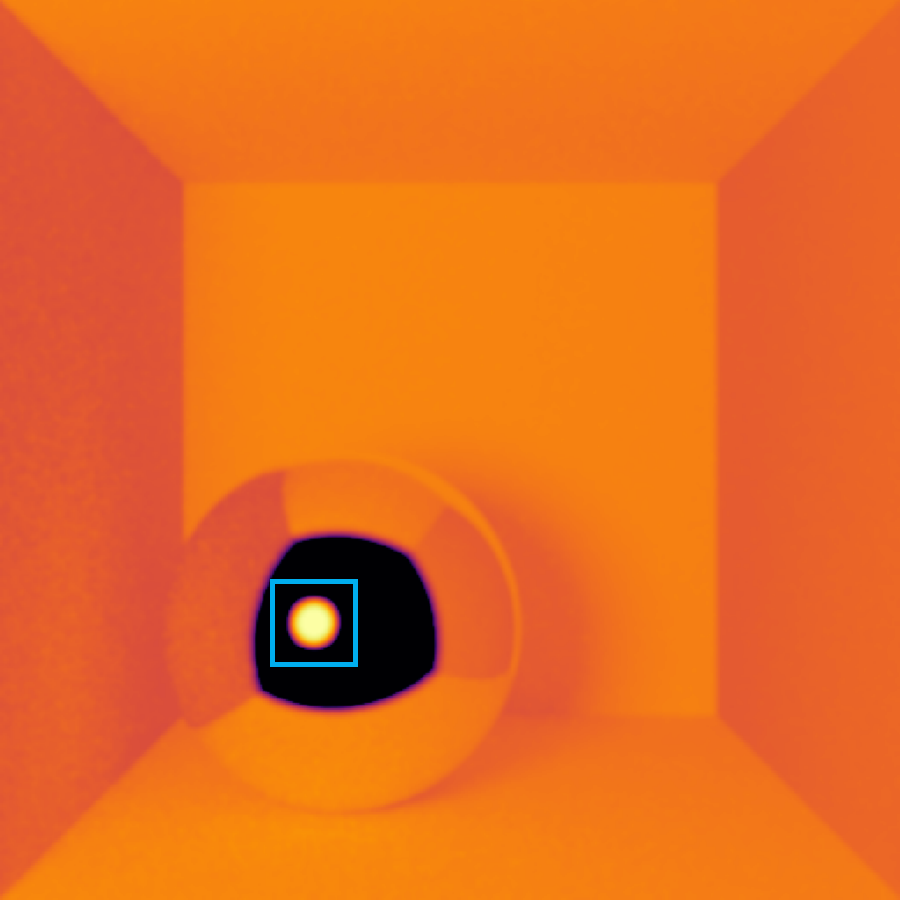} &
        \includegraphics[width=0.22\linewidth]{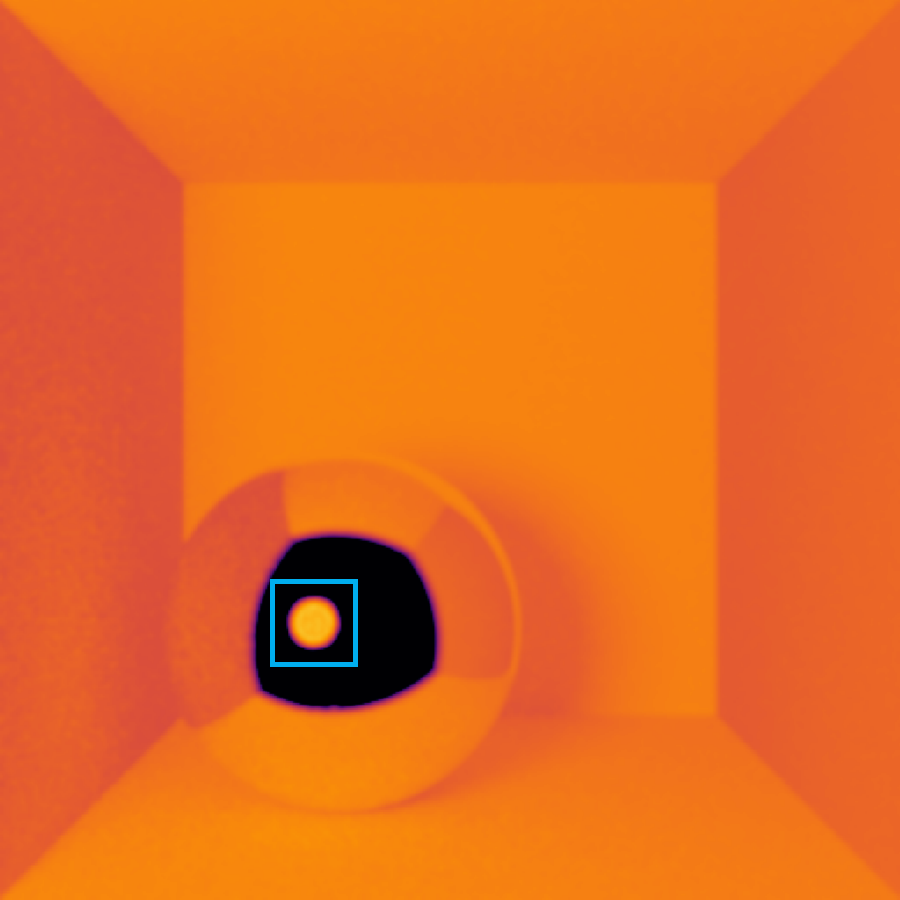} \\
        \rotatebox{90}{\hspace{12mm} \small Plastic $\eta$} &
        \includegraphics[width=0.22\linewidth]{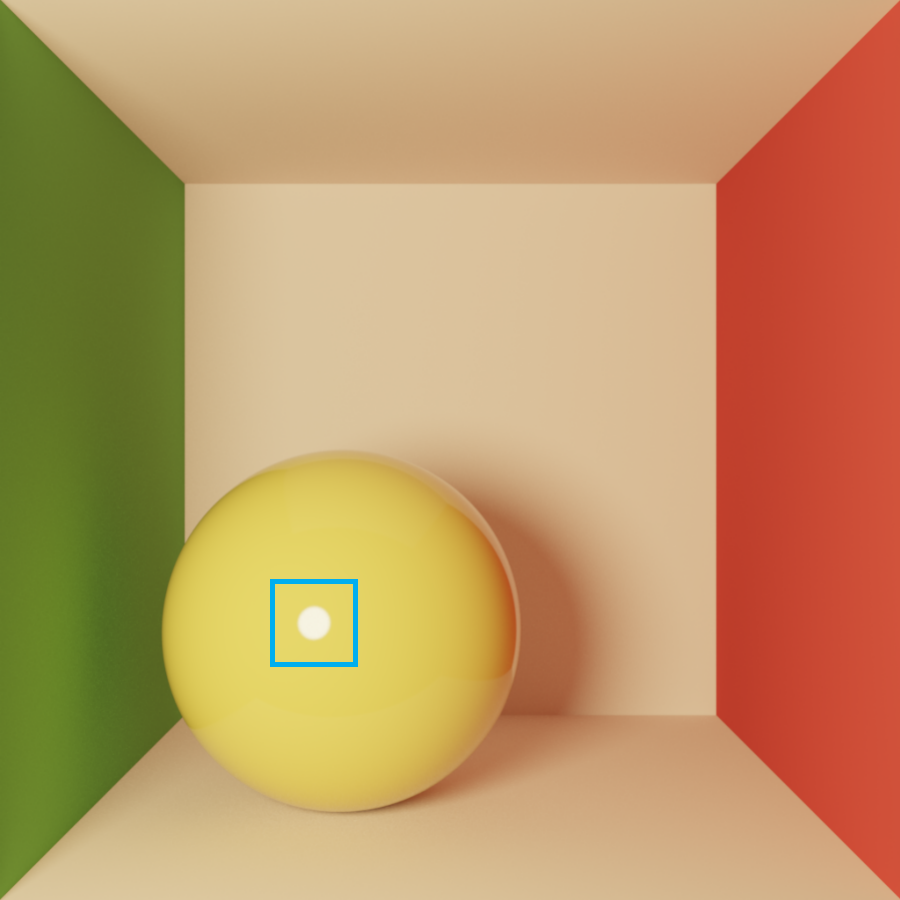} &
        \includegraphics[width=0.22\linewidth]{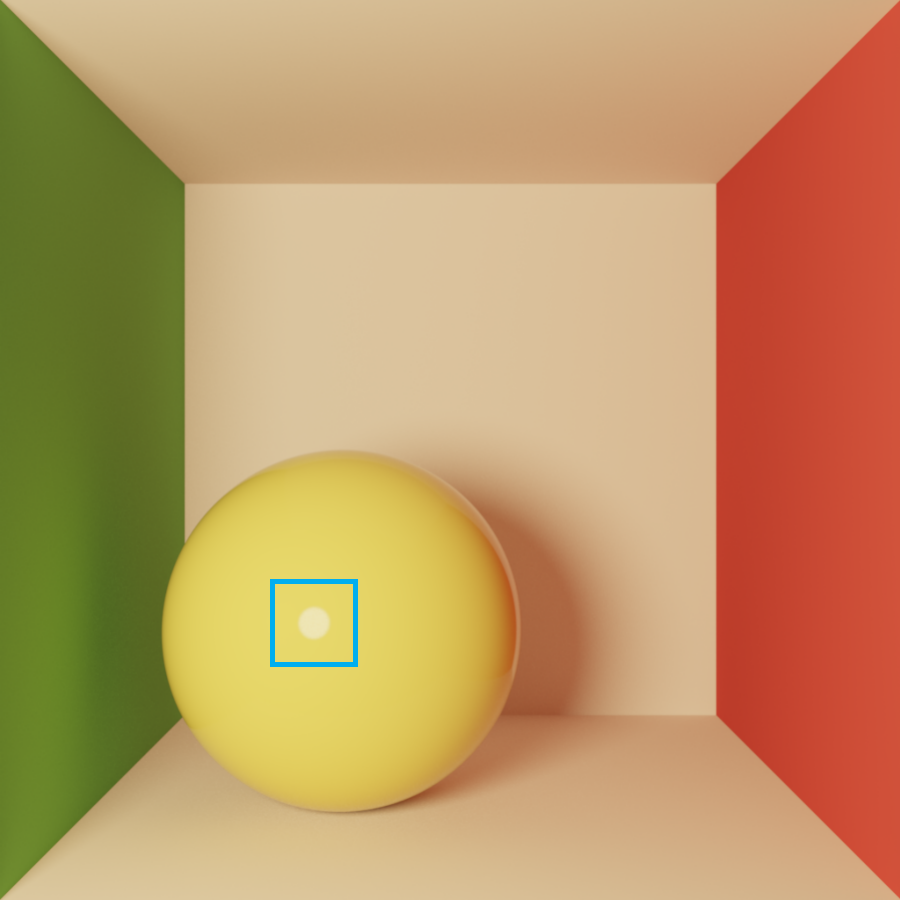} &
        \includegraphics[width=0.22\linewidth]{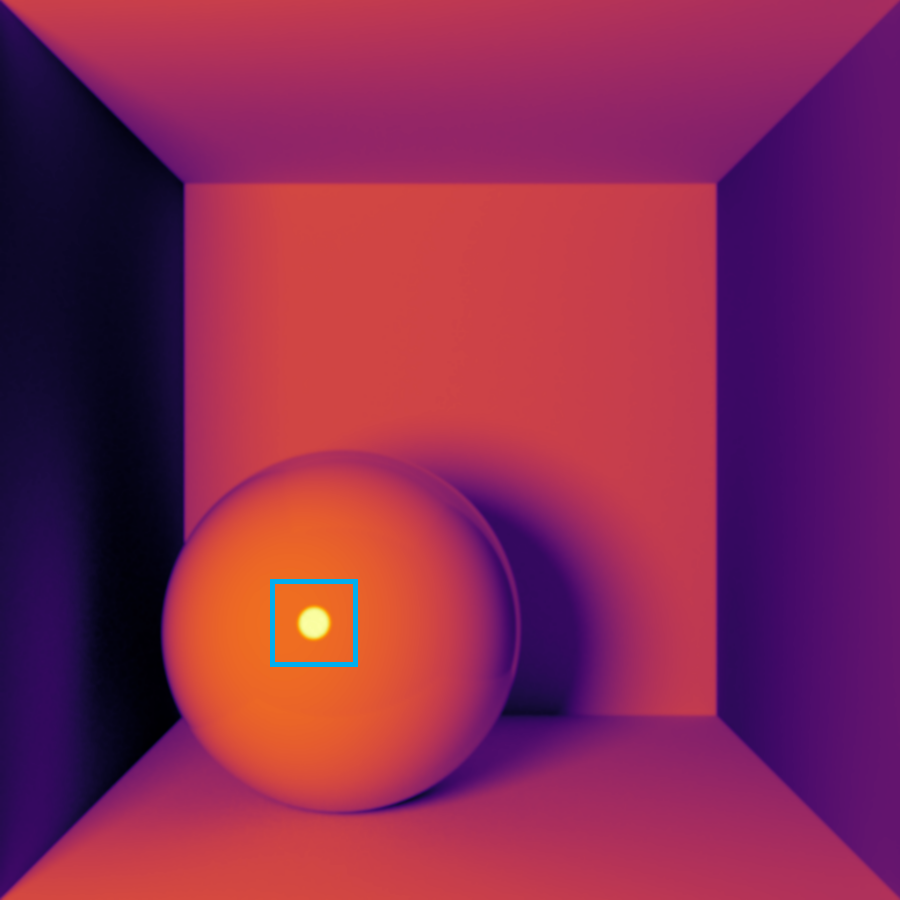} &
        \includegraphics[width=0.22\linewidth]{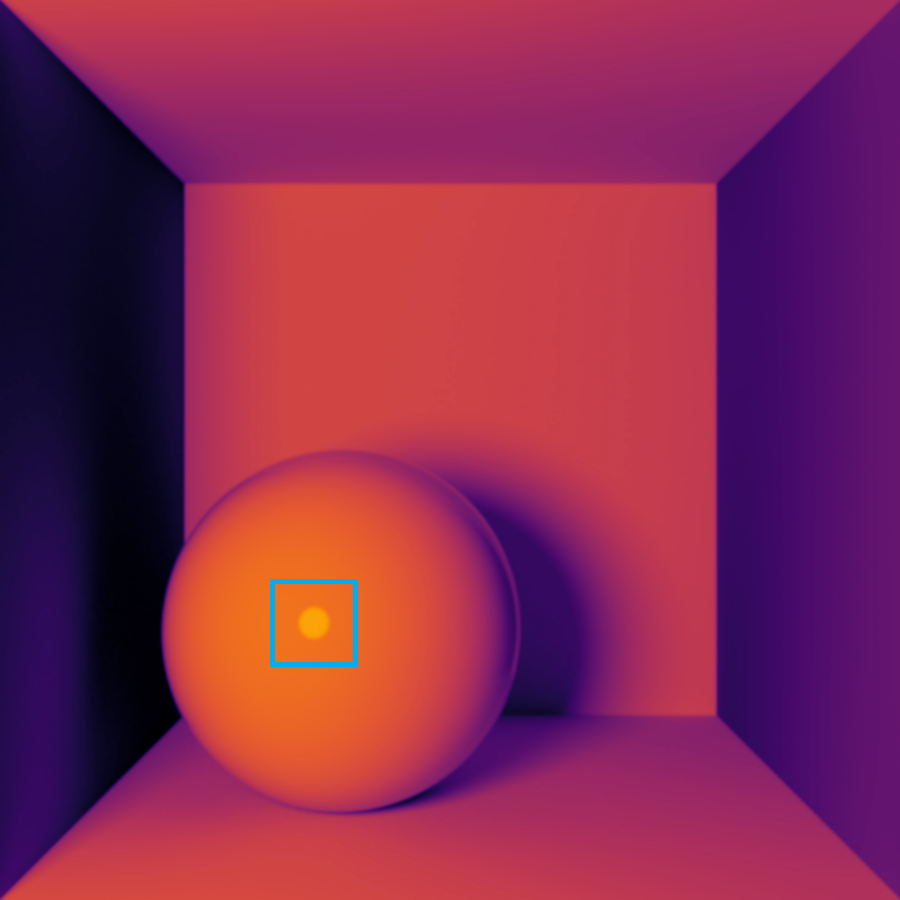} \\
        \rotatebox{90}{\hspace{12mm} \small Radiance} &
        \includegraphics[width=0.22\linewidth]{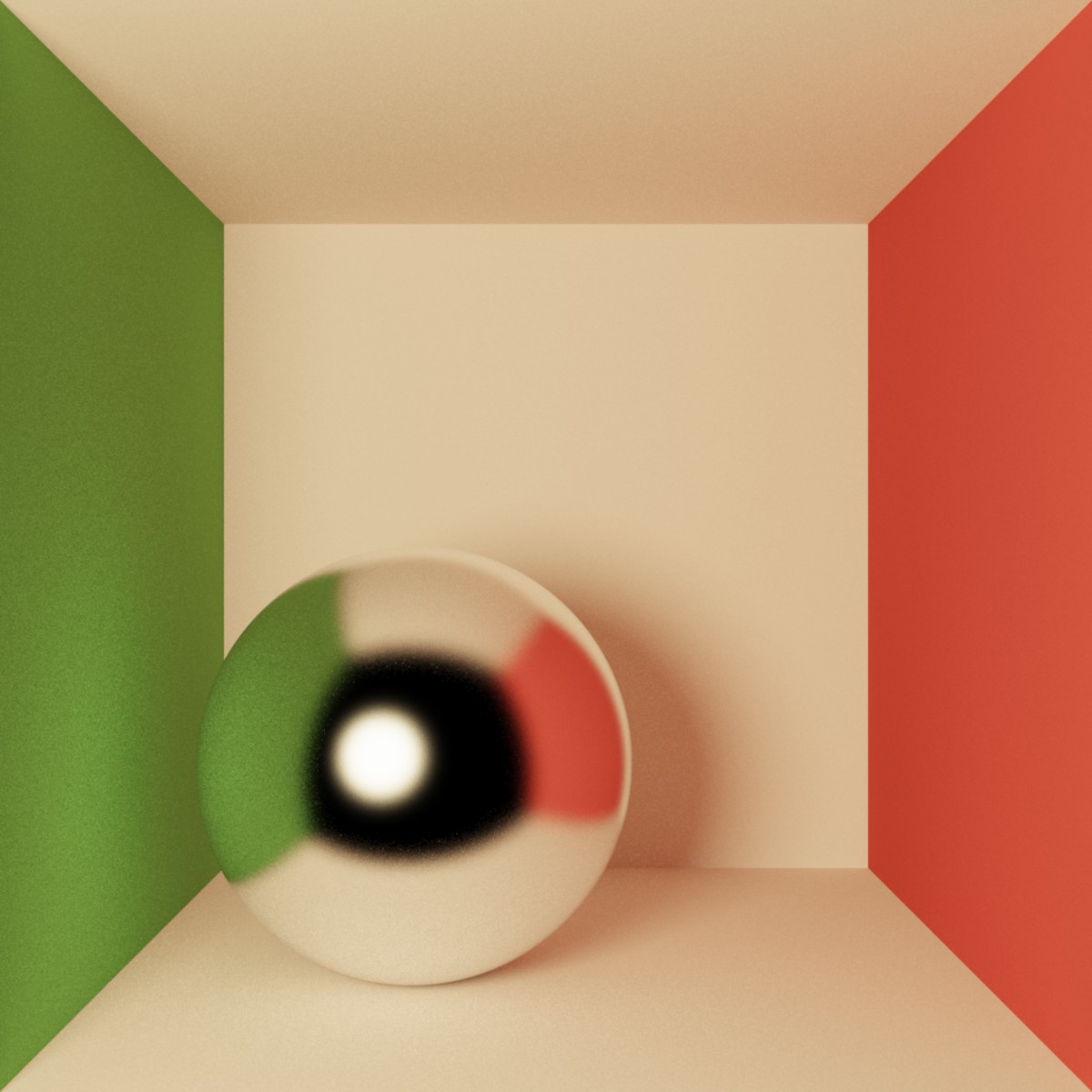} &
        \includegraphics[width=0.22\linewidth]{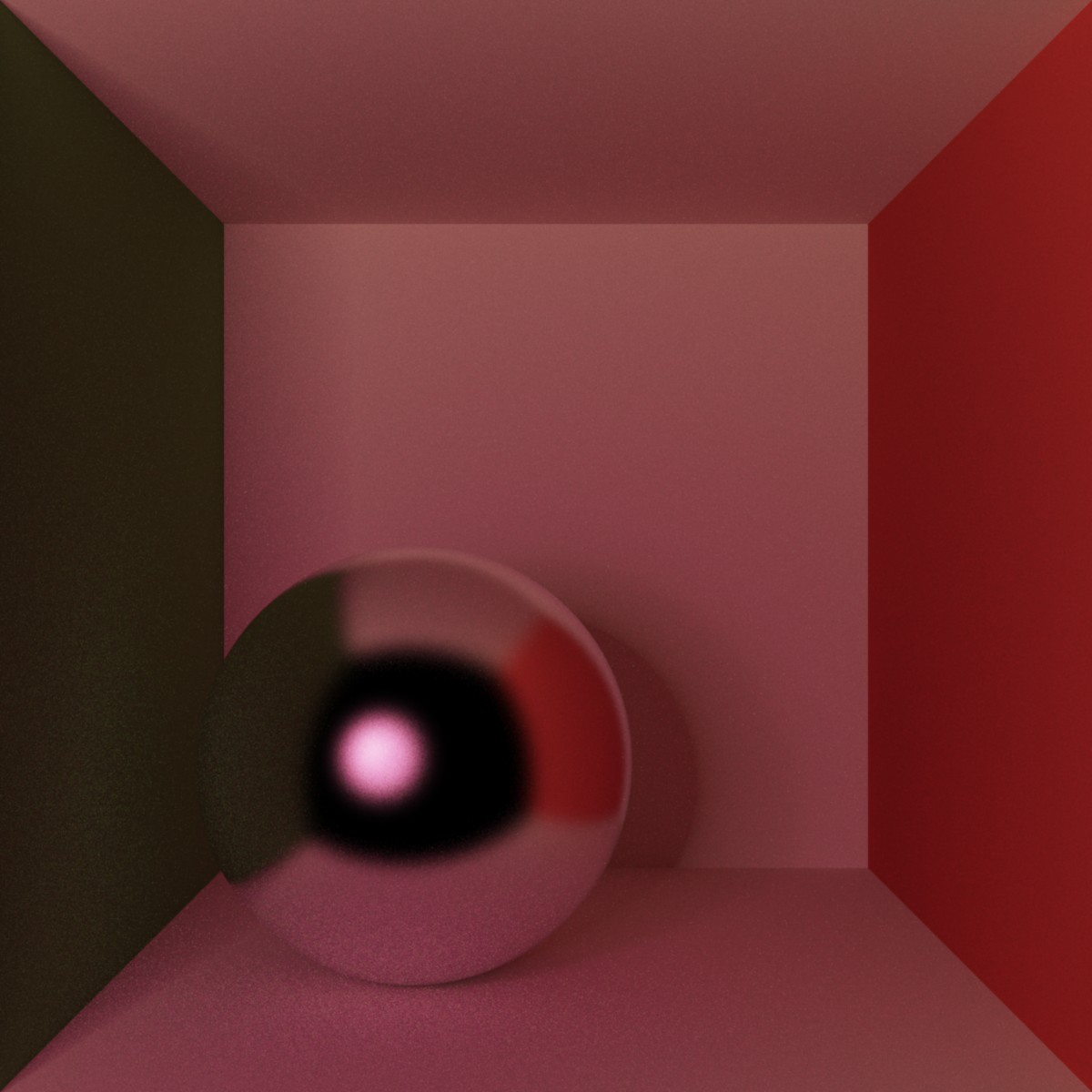} &
        \includegraphics[width=0.22\linewidth]{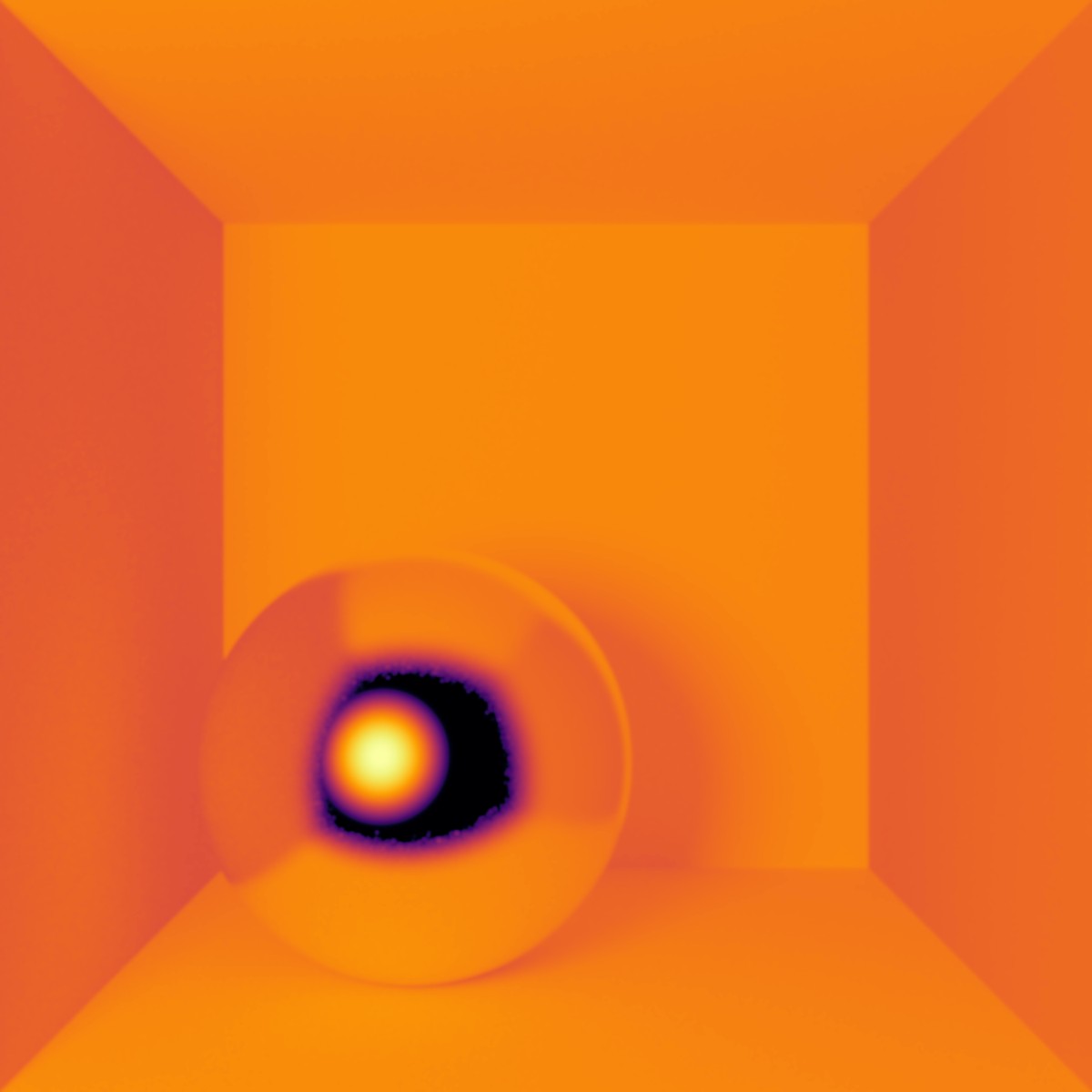} &
        \includegraphics[width=0.22\linewidth]{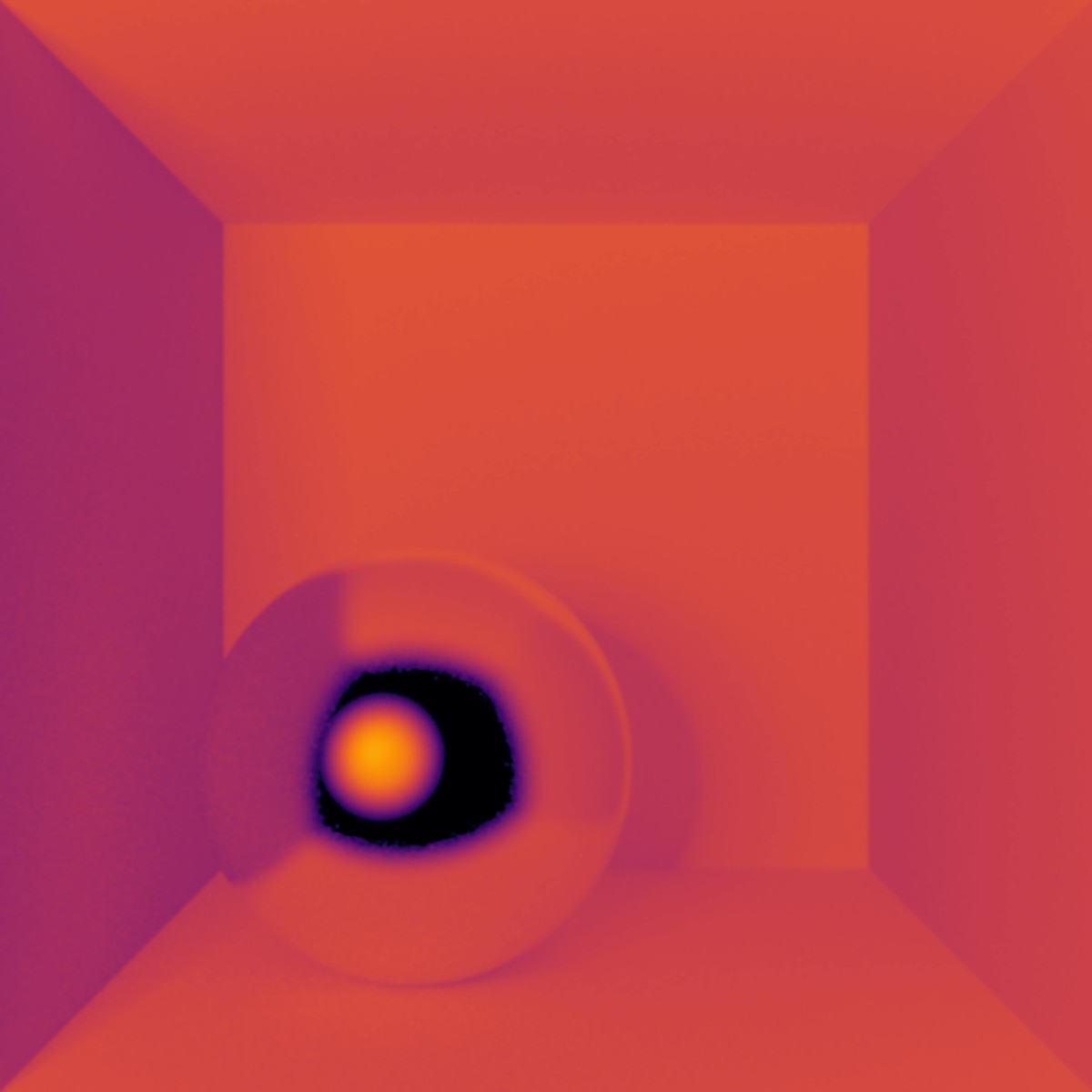} \\
        \rotatebox{90}{\hspace{14mm} \small Blend} &
        \includegraphics[width=0.22\linewidth]{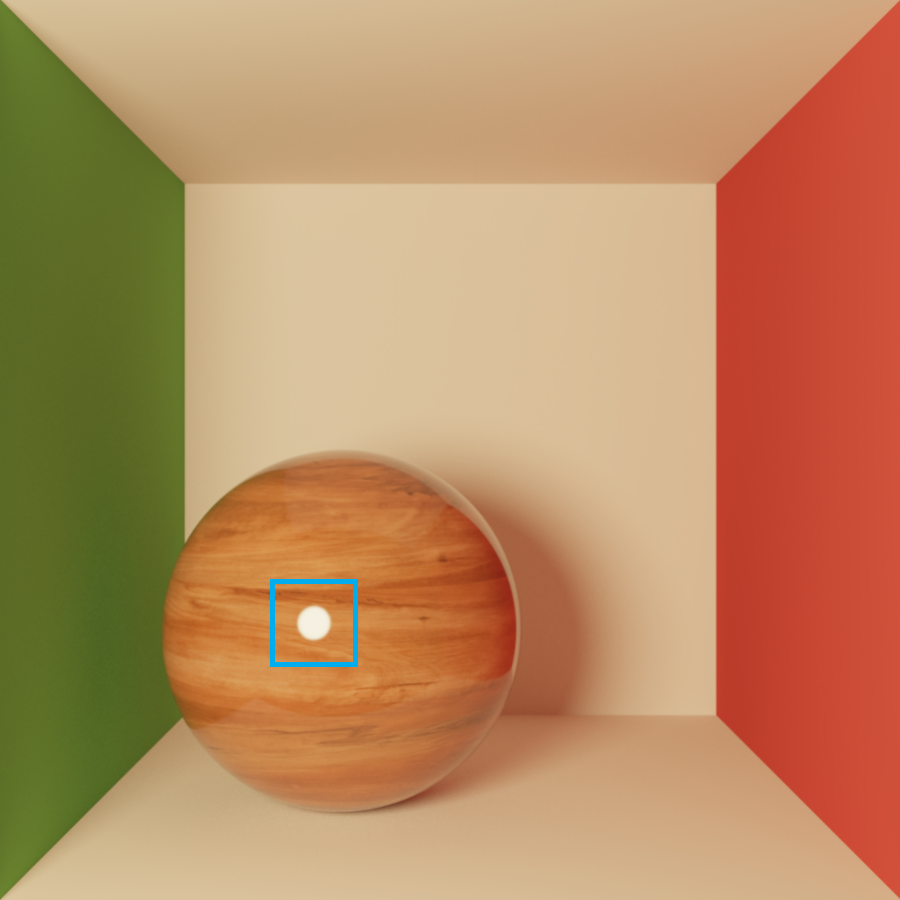} &
        \includegraphics[width=0.22\linewidth]{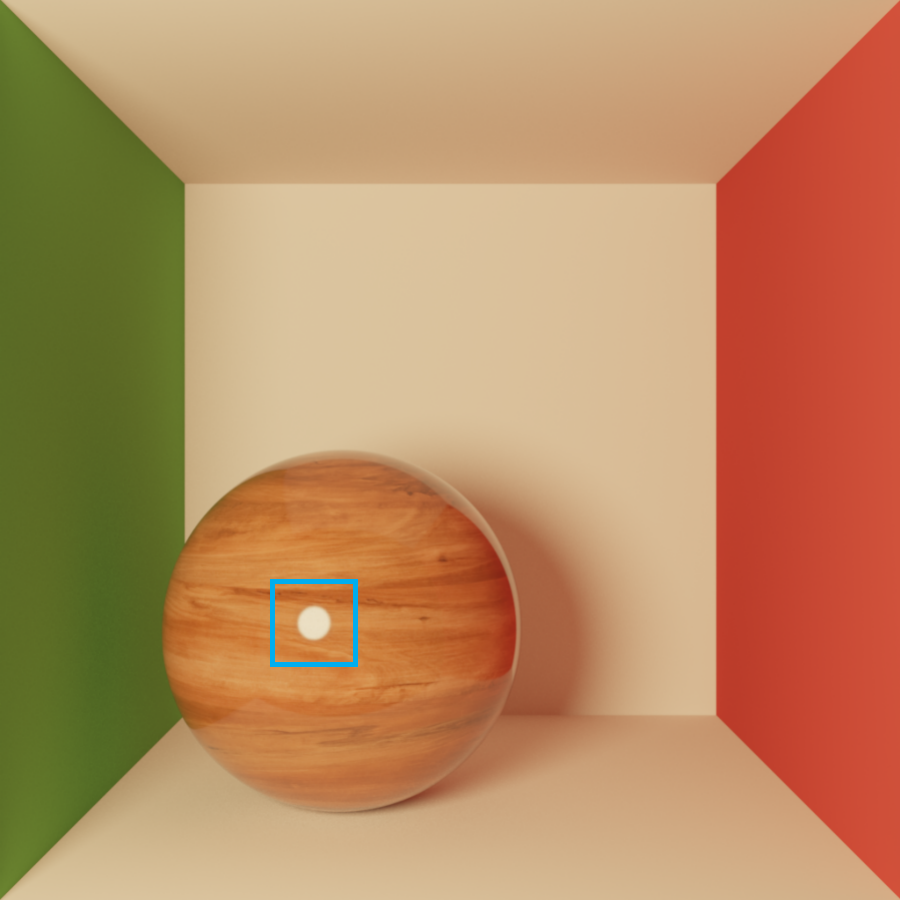} &
        \includegraphics[width=0.22\linewidth]{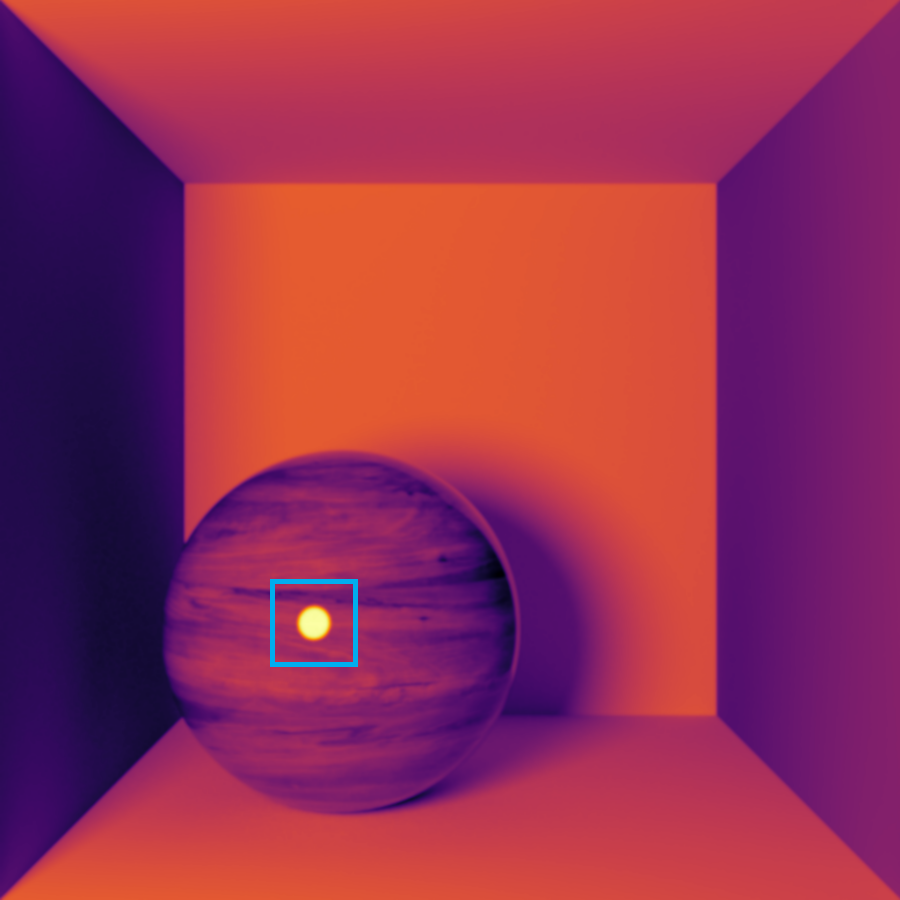} &
        \includegraphics[width=0.22\linewidth]{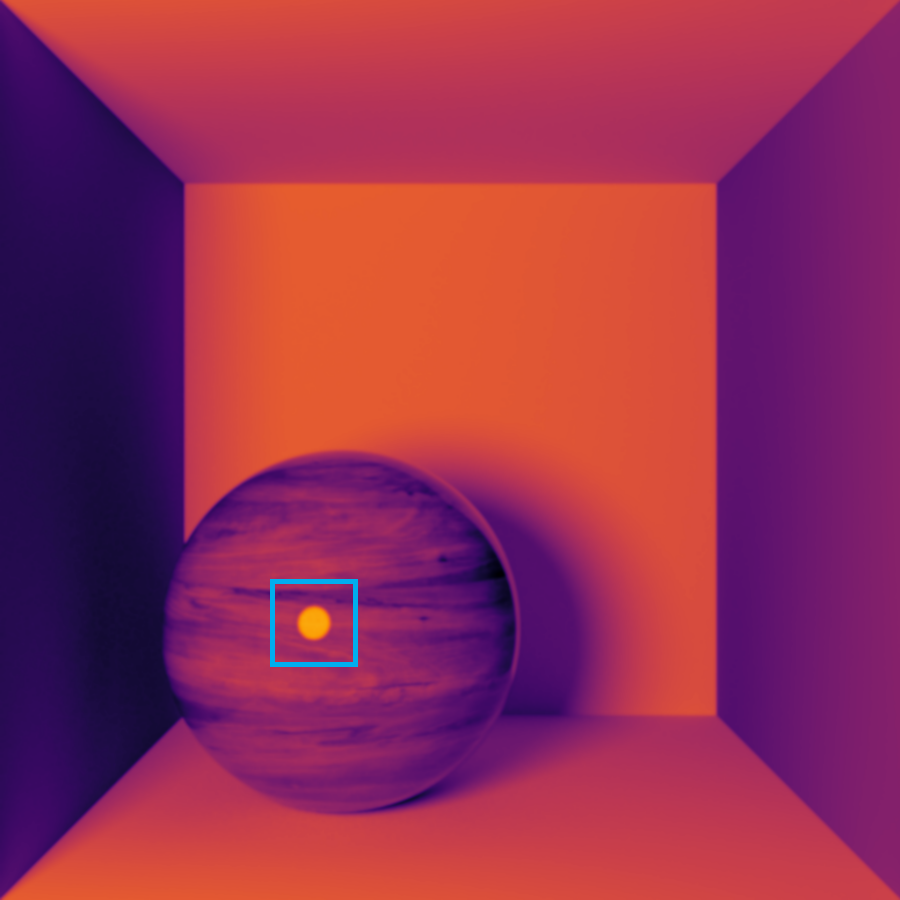} \\
    \end{tabular}
	\caption{We evaluate our differentiable UGR framework across four distinct physical domains. Each row demonstrates the transition from a high-glare state to an optimized state. Pairing initial and optimized renders side-by-side highlights the subtle material and radiance adjustments, while the corresponding luminance maps demonstrate the resulting shift in high-intensity energy distributions used by the UGR metric. Notably, direct emitter optimization (row 3, optimized render) inherently darkens the overall scene and induces a hue shift, a trade-off discussed further in Section \ref{sec:color-preservation}.}
    \label{fig:ablation_matrix}
\end{figure*}

\subsection{Validation Across Radiometric Domains}

To demonstrate the versatility of our differentiable UGR framework, we validated its performance across four distinct physical domains within the controlled environment of the Cornell Box. Our objective was to prove that visual discomfort can be systematically mitigated regardless of the specific architectural constraint; whether by altering surface microgeometry, boundary refraction, or source emission. 

As illustrated in Figure \ref{fig:ablation_matrix}, we successfully optimized the following parameters from an initially high-glare state down to an ergonomic target (UGR $\le$ 17):
\begin{itemize}
    \item \textbf{Surface Scattering:} Increasing the microfacet roughness ($\alpha$) of a metallic conductor to diffuse the specular highlight.
    \item \textbf{Boundary Refraction:} Adjusting the Index of Refraction ($\eta$) of a dielectric plastic to minimize Fresnel boundary reflections.
    \item \textbf{Source Emission:} Modulating the radiance of the primary area emitter to selectively dim the problematic solid angles.
    \item \textbf{Spatial Blending:} Optimizing a continuous 2D mask to blend between a highly specular base material and a diffuse fallback (simulating localized anti-reflective coating). 
\end{itemize}

Figure \ref{fig:convergence_graph} tracks the UGR reduction over time for each of these four domains. The results demonstrate that our formulation provides a stable, continuous loss landscape regardless of the underlying physical parameter being optimized. Whether the target is a scalar material property (Roughness, Index of Refraction) or a high-resolution spatial texture (Radiance, Blend), the framework converges to the target ergonomic threshold.

Beyond geometric and physical versatility, we evaluated the computational scalability of our adjoint framework against standard derivative-free optimizers, specifically CMA-ES \cite{hansen2016cma} and Powell's method \cite{powell1964efficient}. While these gradient-free baselines can successfully optimize global scalar values, they suffer from exponential scaling constraints. As parameter dimensionality increases, particularly for spatial textures at resolutions of $64 \times 64$ ($N=4096$) and beyond, these solvers exhaust their evaluation budgets and are prone to timing out. In contrast, our adjoint formulation computes gradients for all parameters simultaneously, allowing execution time to remain near-constant regardless of the spatial resolution. We provide a comprehensive quantitative and qualitative breakdown of this dimensionality scaling and baseline comparison in Appendix A.

While the isolated Cornell Box experiments validate the theoretical stability of our framework, practical architectural design potentially requires mitigating glare across multiple surfaces simultaneously. To demonstrate the scalability of our method, we applied it to a highly non-linear, multi-bounce kitchen environment, as shown in Figure \ref{fig:teaser}. Starting from an initial high-glare state, the framework performed a joint optimization across four distinct parameters: the scalar roughness ($\alpha$) of the metallic extractor hood, the scalar index of refraction ($\eta$) of the wooden table top (plastic BSDF), the high-resolution radiance texture of the primary emitter, and the spatially-varying blend BSDF mask of the secondary worktop.

\begin{figure}[t]
    \centering
    \includegraphics[width=0.9\columnwidth]{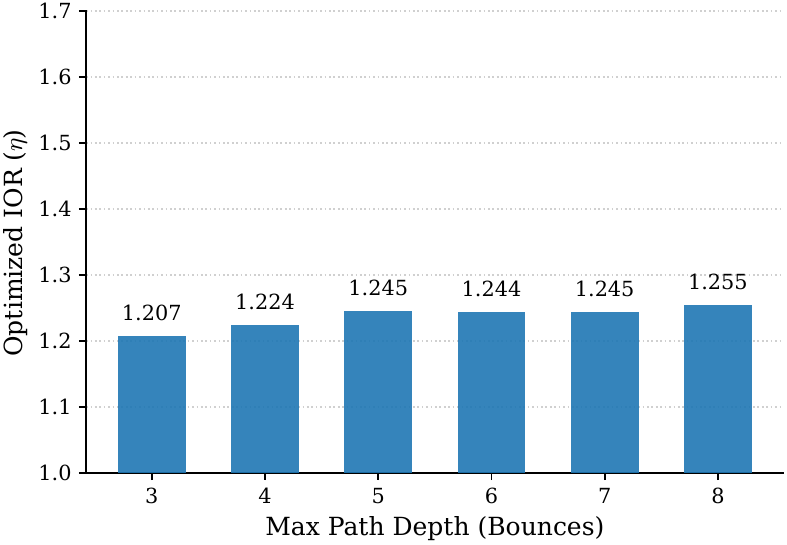}
	\caption{We optimize the Index of Refraction ($\eta$) of a dielectric sphere under varying maximum path lengths. At low path lengths, the integrator underestimates the ambient background luminance ($L_b$), causing the optimizer to overcompensate by aggressively reducing the IOR. Deeper path lengths provide sufficient $L_b$ to naturally mitigate the glare ratio, allowing the optimizer to reach the target UGR while preserving the material's physical specularity.}
    \label{fig:gi_bounces}
    \vspace{-1em}
\end{figure}

\subsection{Global Illumination and Background Regularization}
\label{sec:results_gi_background}
Because the UGR formulation is fundamentally a ratio between localized source luminance ($L_s$) and global ambient adaptation ($L_b$), accurately simulating multiple bounce light transport is paramount. Interestingly, our experiments reveal that whether the background luminance is attached to the computational graph dictates how well the optimization can succeed and profoundly influences the physical plausibility of the resulting materials.

\begin{figure}[t]
    \centering
    
    \includegraphics[width=1.0\columnwidth]{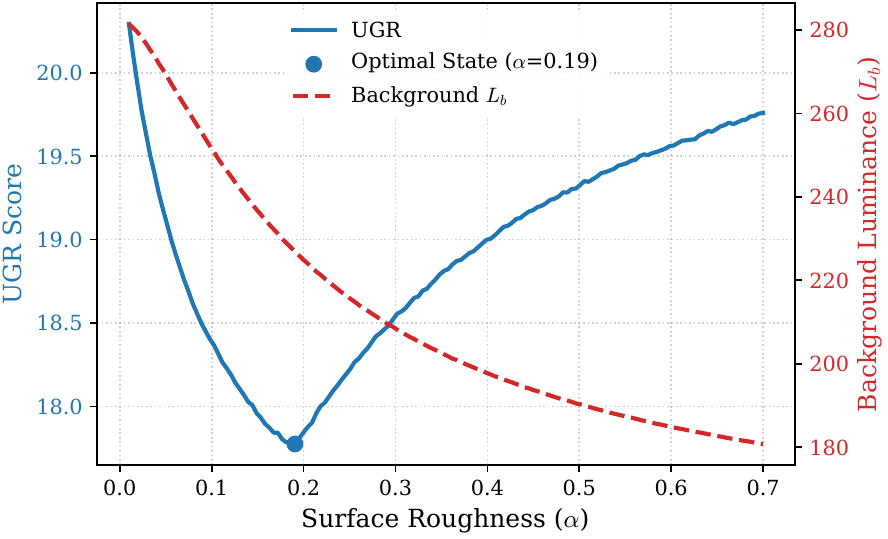}
    
    \vspace{3mm}
    
    \setlength{\tabcolsep}{1pt}
    \begin{tabular}{cccc}
        \includegraphics[width=0.24\columnwidth]{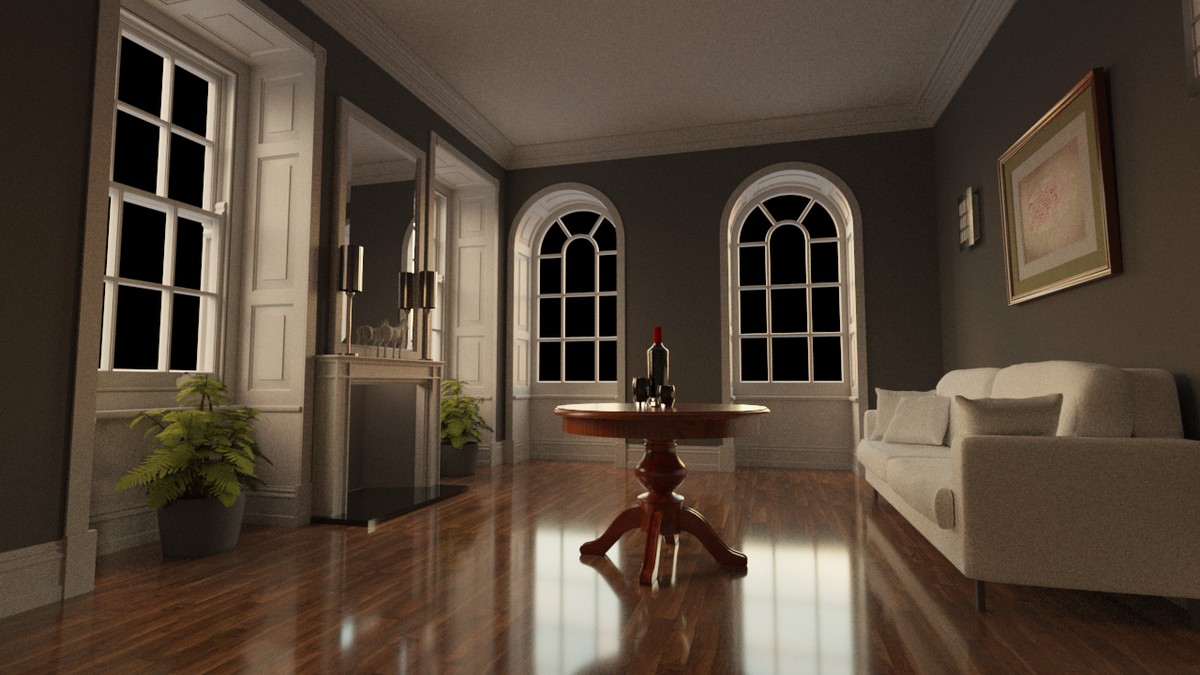} &
        \includegraphics[width=0.24\columnwidth]{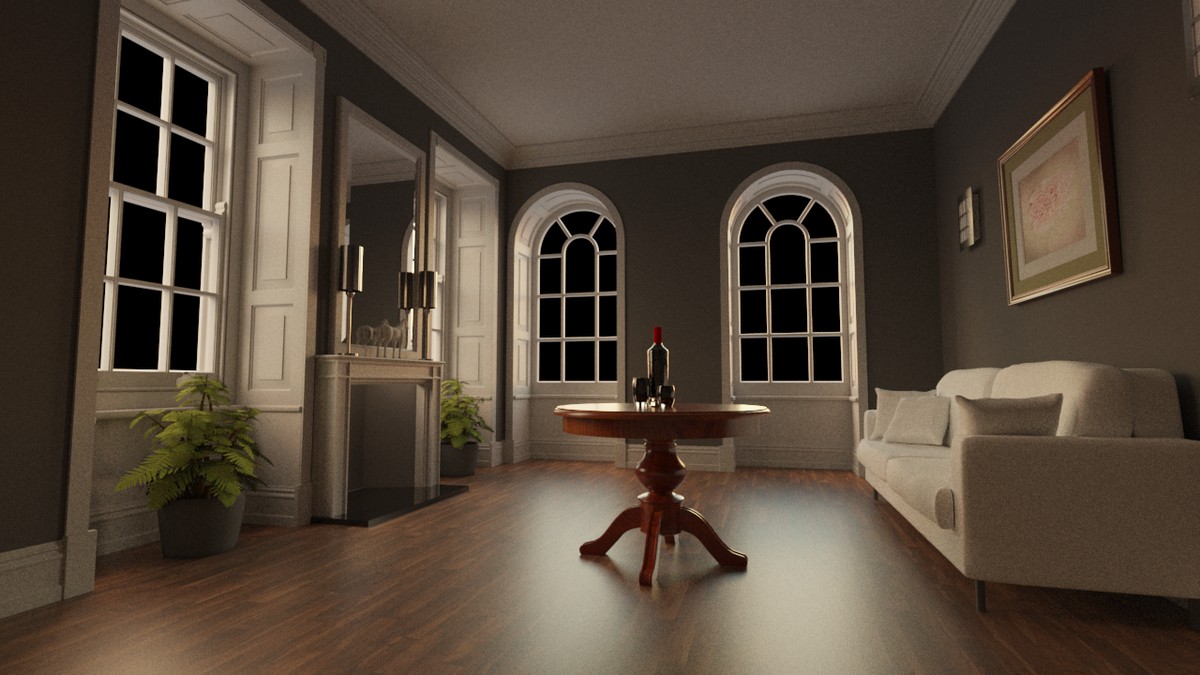} &
        \includegraphics[width=0.24\columnwidth]{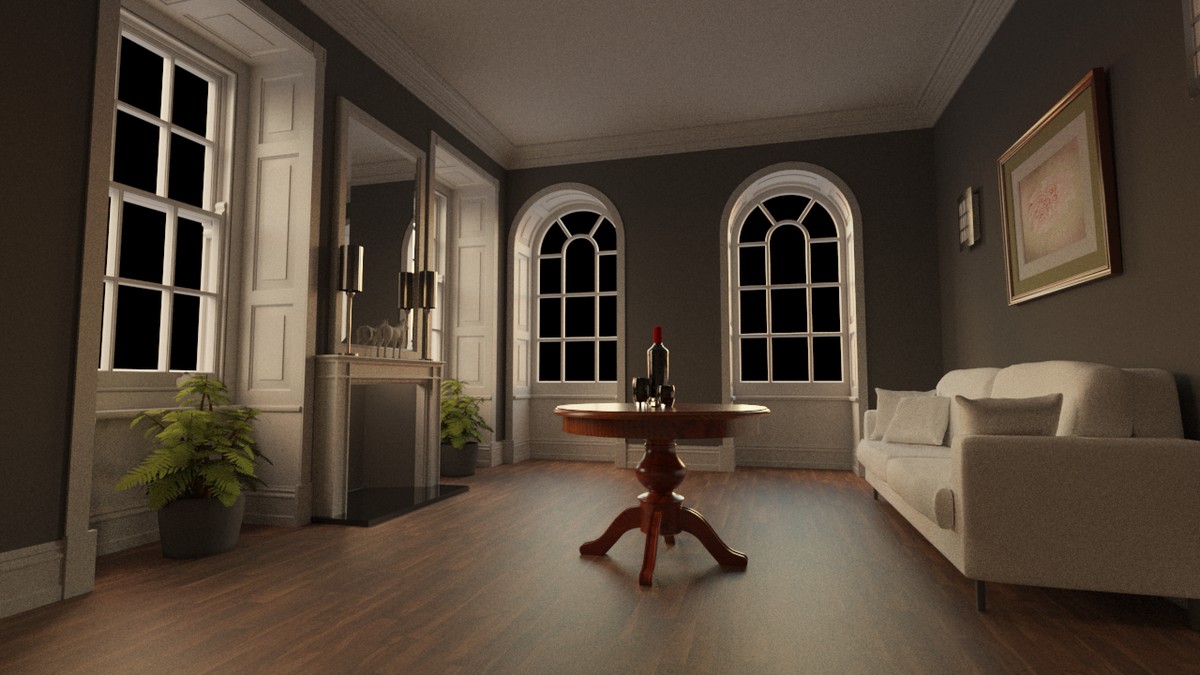} &
        \includegraphics[width=0.24\columnwidth]{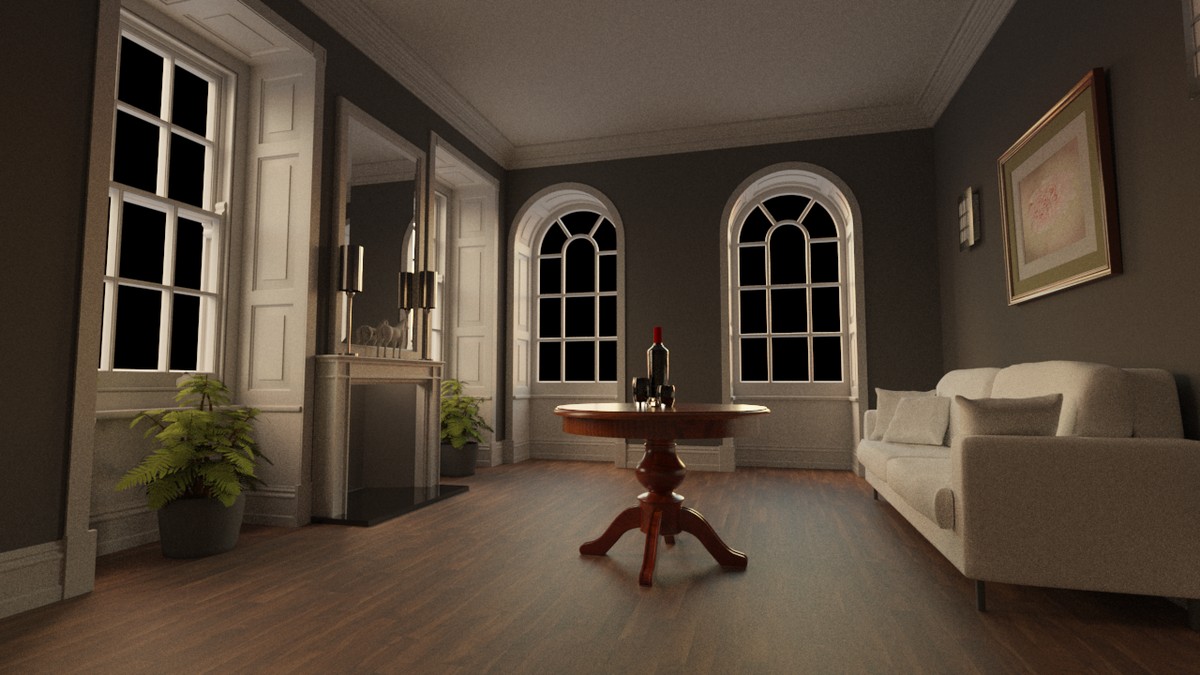} \\
        {\tiny $\alpha=0.01$} & {\tiny $\alpha=0.15$} & {\tiny $\alpha=0.30$} & {\tiny $\alpha=0.40$}
    \end{tabular}

    \vspace{1mm}

    \begin{tabular}{cccc}
        \includegraphics[width=0.24\columnwidth]{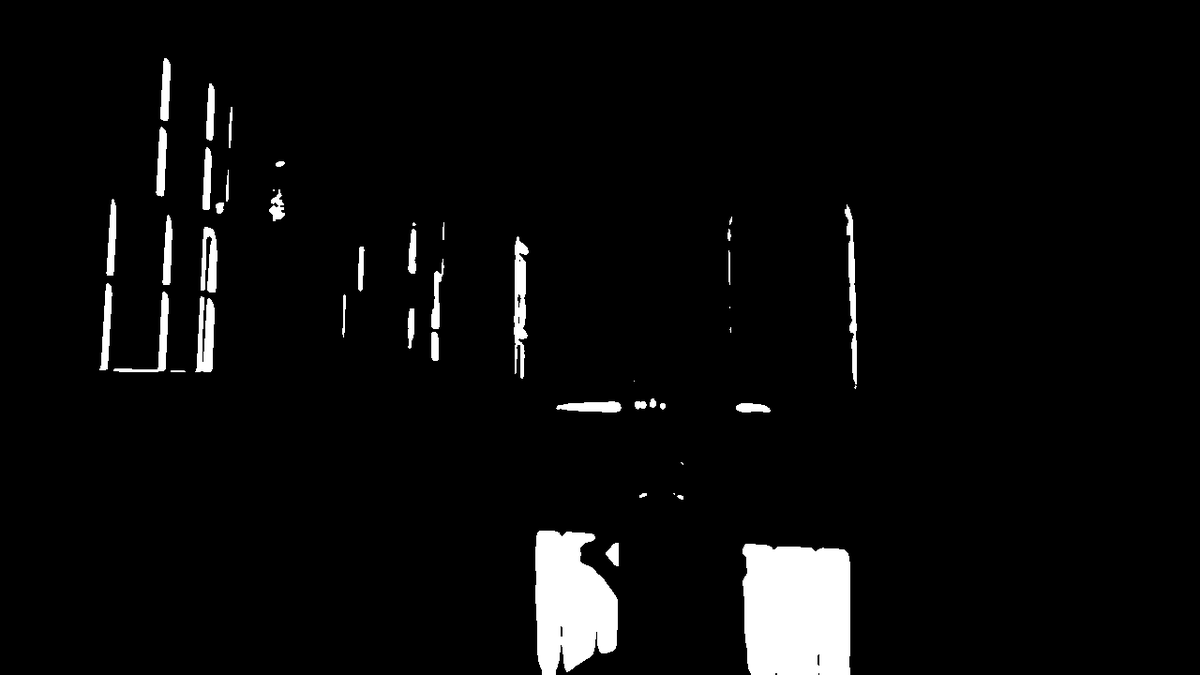} &
        \includegraphics[width=0.24\columnwidth]{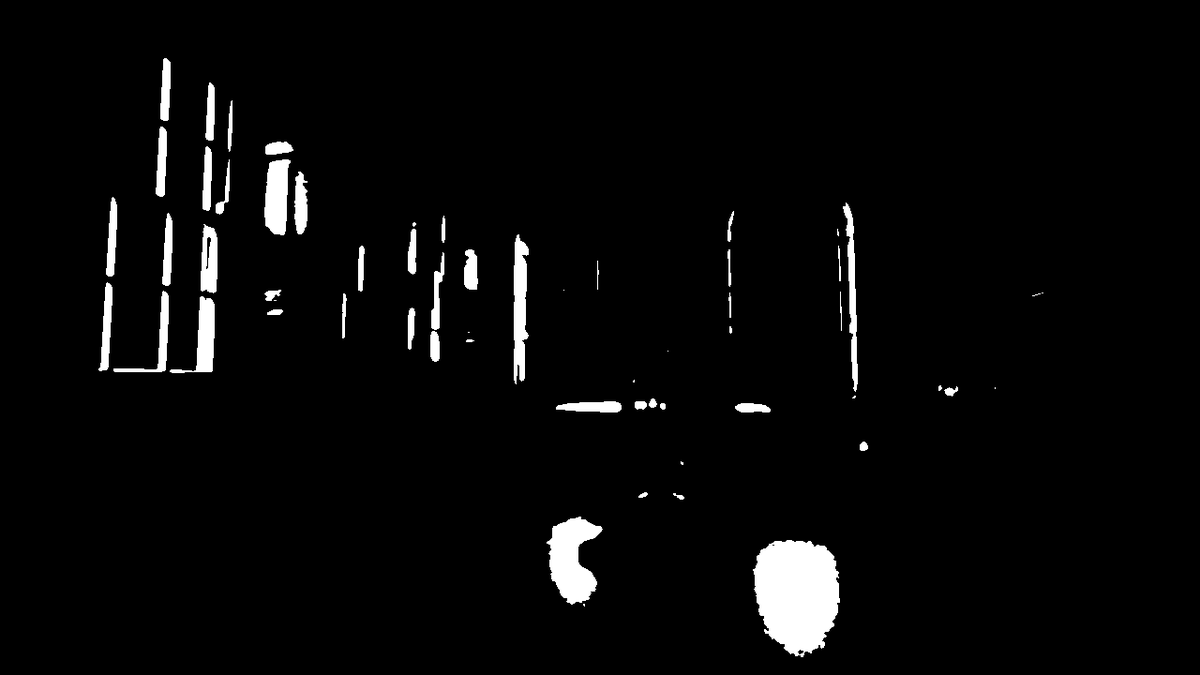} &
        \includegraphics[width=0.24\columnwidth]{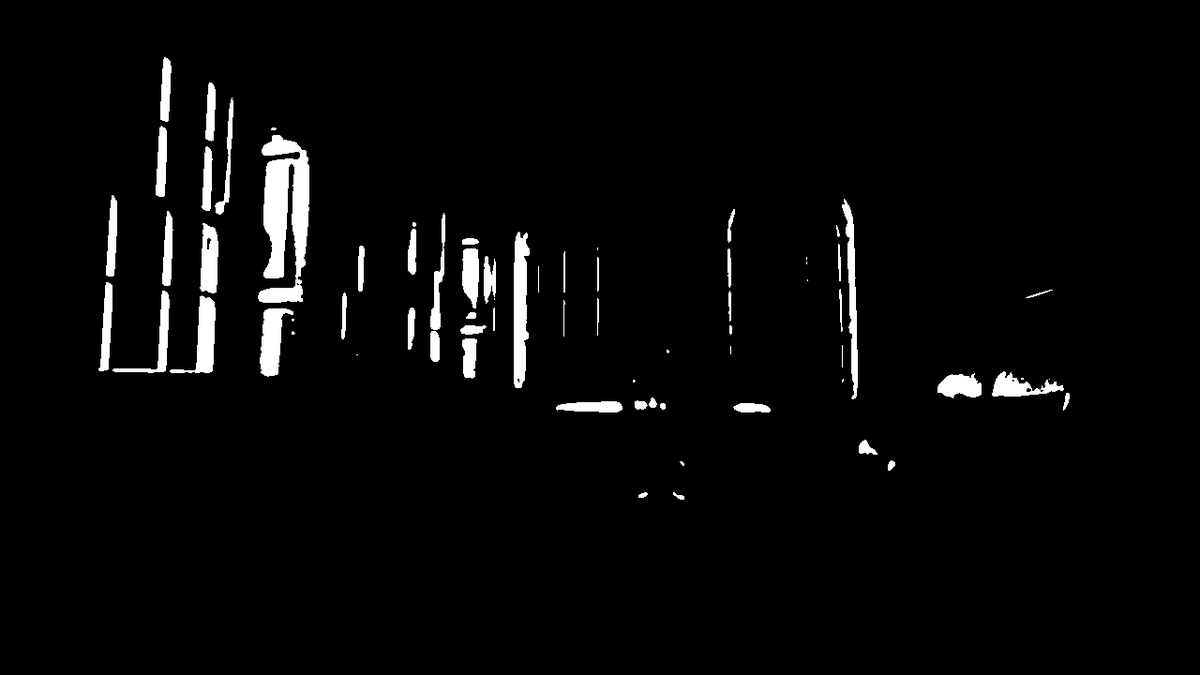} &
        \includegraphics[width=0.24\columnwidth]{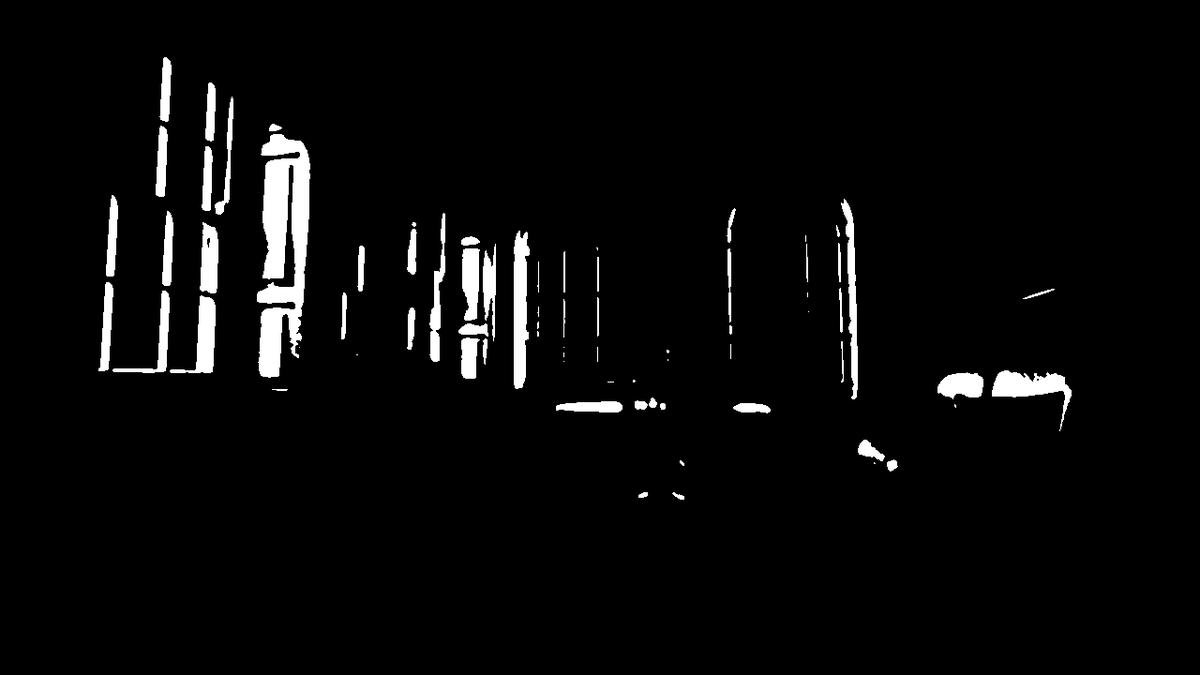} \\
        \multicolumn{4}{c}{\small Corresponding Glare Masks}
    \end{tabular}

    \caption{(Top) Sweeping surface roughness $\alpha$ reveals a highly non-convex UGR loss landscape. While initially increasing $\alpha$ diffuses the primary specular reflection and reduces visual discomfort, excessive roughening starves the room of global ambient light ($L_b$). (Middle/Bottom) As the room progressively darkens ($\alpha = 0.30 \rightarrow 0.40$), the background adaptation luminance $L_b$ drops so severely that secondary, previously comfortable light sources now exceed the relative evaluation boundary. This causes new regions to actively trigger the spatial glare mask, paradoxically spiking the final UGR score.}
    \label{fig:alpha_trap}
\end{figure}

\textit{The Influence of Path Length.}
To investigate the relationship between global illumination accuracy and optimization behavior, we conducted an ablation study varying the maximum path length during the inverse rendering process. As demonstrated in Figure \ref{fig:gi_bounces}, optimizing the Index of Refraction ($\eta$) of a dielectric surface at short path lengths leads the optimizer to heavily reduce the IOR to remove specular reflections. This occurs because the integrator underestimates the ambient background luminance ($L_b$).

In order to obtain a better background luminance estimation, we must include longer path lengths. However, simulating higher path lengths increases the computational budget and exacerbates the variance for the gradients which can have adverse effects on the optimization. To balance accuracy and stability, we simulated deeper path lengths during the forward pass to provide accurate ambient lighting, but only record gradients for a small number of path lengths.

\textit{Background Gradient Routing and the Alpha Trap.}
An interesting discovery from our empirical analysis involves the routing of gradients through the $L_b$ term. As established in Figure \ref{fig:alpha_trap}, the UGR objective landscape is nonconvex with respect to spatial roughening ($\alpha$). Sweeping the roughness parameter reveals that initially increasing $\alpha$ diffuses the glare and lowers the UGR. However, if roughening continues, it eventually starves the room of ambient light, causing $L_b$ to drop and the UGR to paradoxically spike, creating a trap for the optimizer.

\begin{figure}[t]
    \centering  
    \includegraphics[width=1.0\columnwidth]{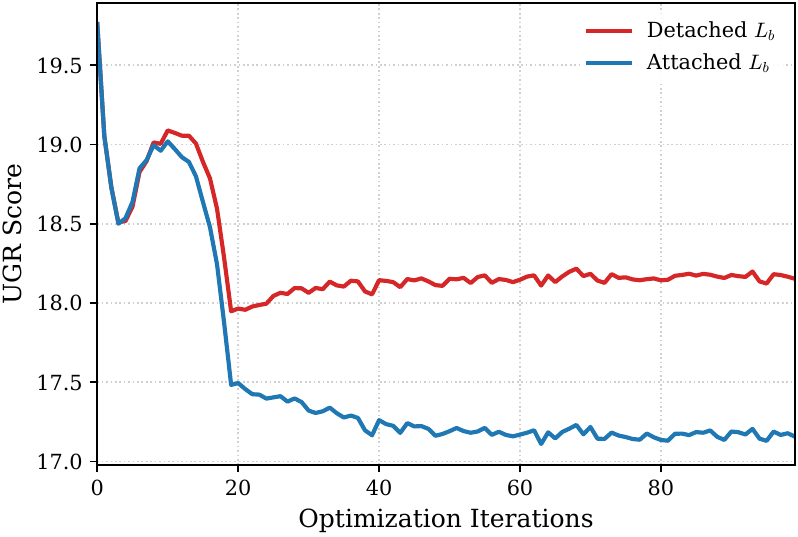}  
    \vspace{3mm}   
    \setlength{\tabcolsep}{1.5pt}
    \renewcommand{\arraystretch}{0.5}   
    \begin{tabular}{cccc}
        & \small Optimized Render & \small Luminance Map & \small Glare Mask \\       
        \vspace{0.5mm} \\
        \rotatebox{90}{\hspace{2mm} \scriptsize Detached} &
        \includegraphics[width=0.3\columnwidth]{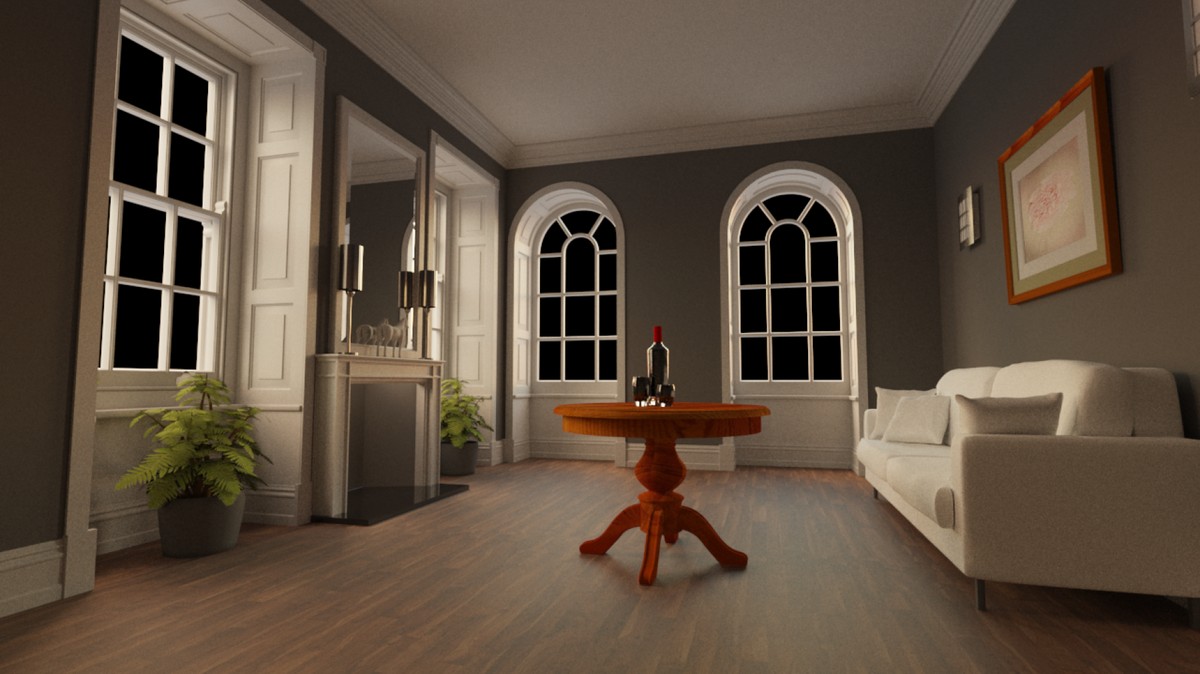} &
        \includegraphics[width=0.3\columnwidth]{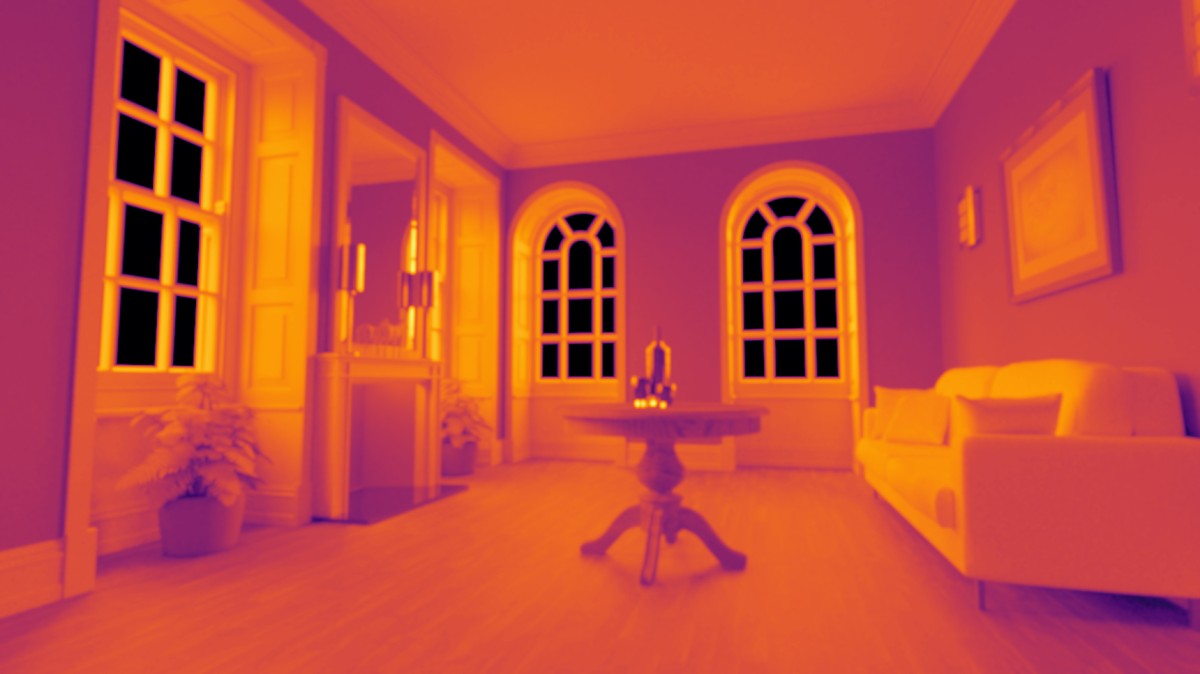} &
        \includegraphics[width=0.3\columnwidth]{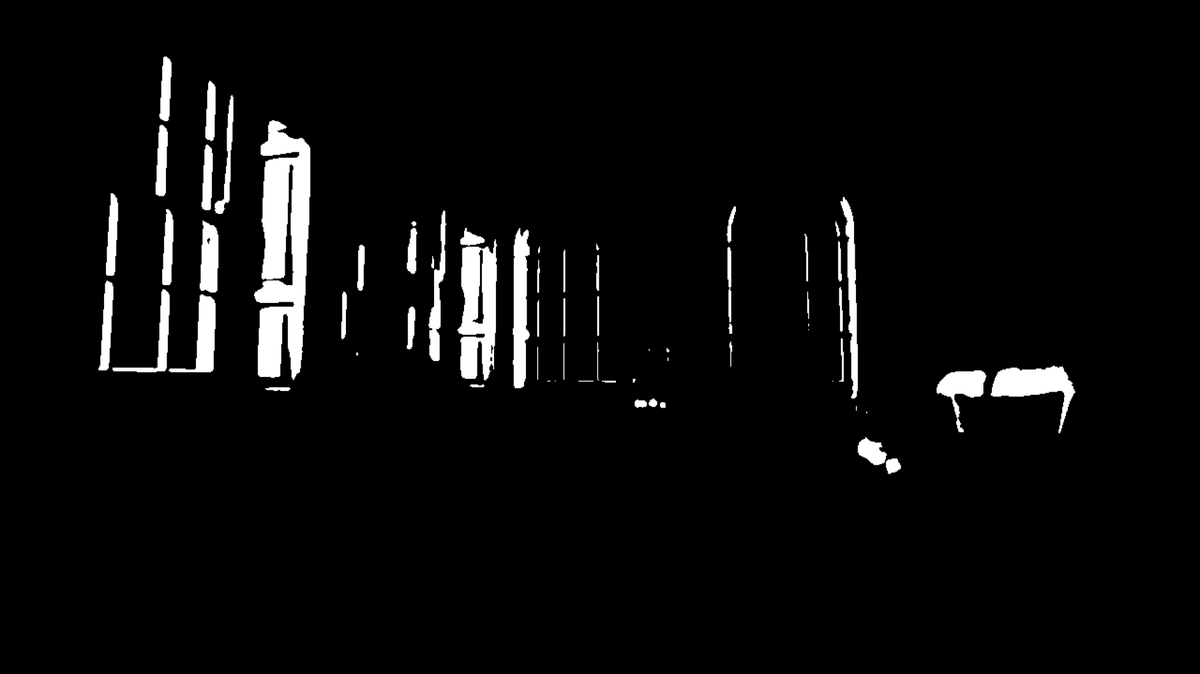} \\       
        \vspace{0.5mm} \\
        \rotatebox{90}{\hspace{2mm} \scriptsize Attached} &
        \includegraphics[width=0.3\columnwidth]{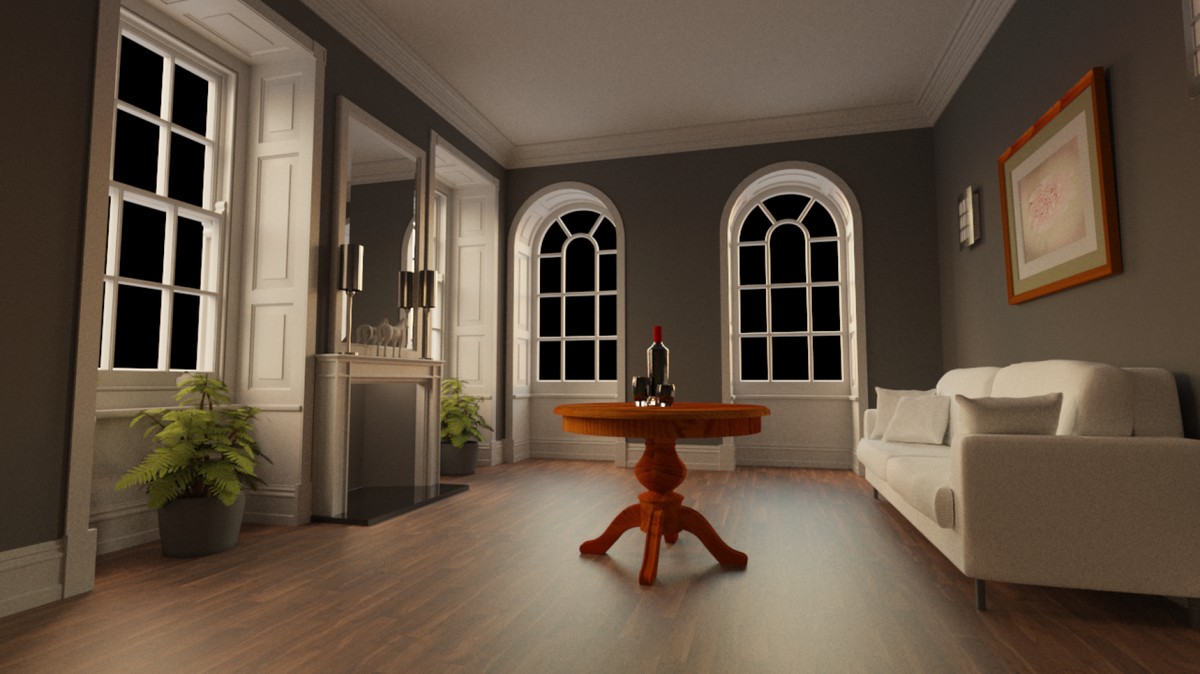} &
        \includegraphics[width=0.3\columnwidth]{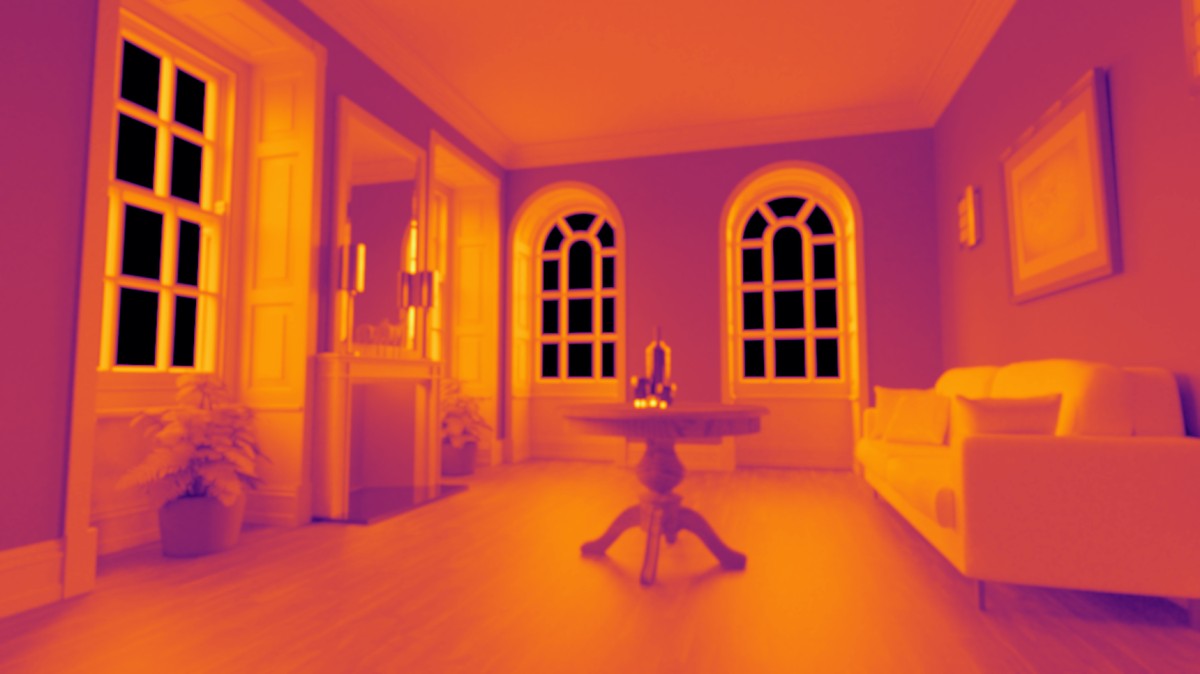} &
        \includegraphics[width=0.3\columnwidth]{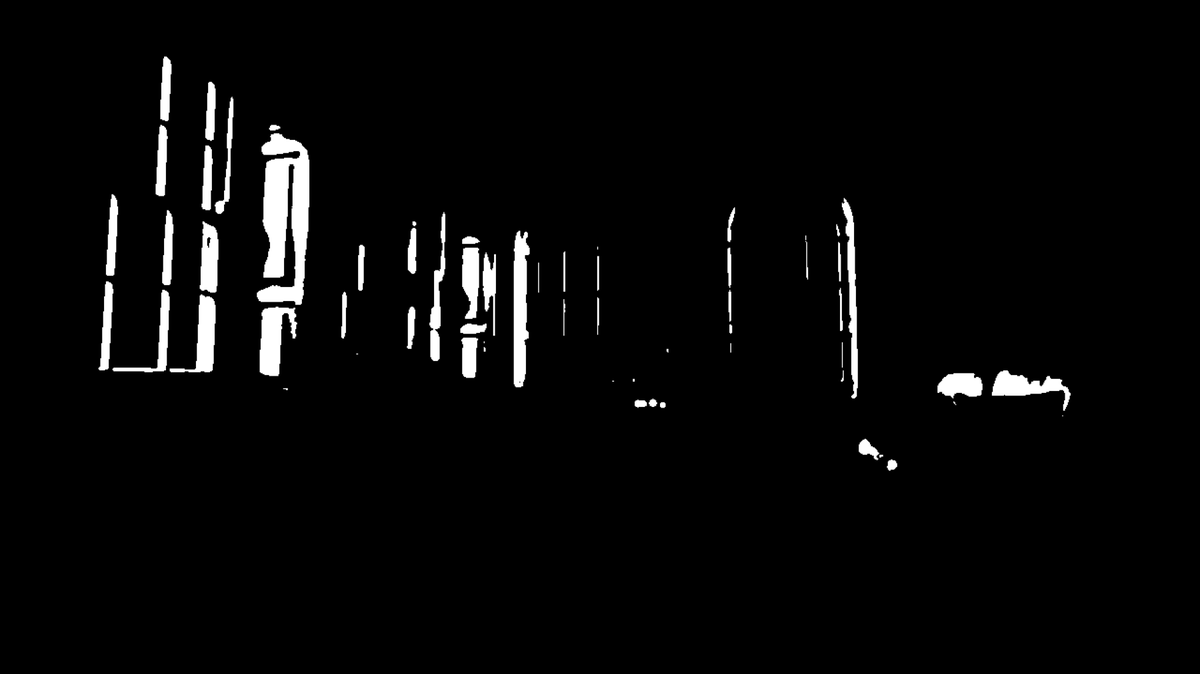} \\ 
    \end{tabular}
    \caption{(Top) Convergence plot demonstrating the "over-frosting" trap. When $L_b$ is detached (red), the optimizer greedily minimizes the UGR by darkening the room. When $L_b$ is attached (blue), the loss of ambient light penalizes the UGR score, forcing the optimizer to plateau. (Bottom) Visual results of the optimization. The detached approach results in an unnaturally dark room and an over-blurred highlight. Retaining $L_b$ in the autodiff graph allows the system to autonomously find a ``leveled'' roughness that diffuses the glare mask while preserving physically plausible global illumination.}
    \label{fig:lb_attachment}
\end{figure}

How the framework handles this nonconvexity depends entirely on whether $L_b$ is attached or detached from the autodiff computational graph. As shown in Figure \ref{fig:lb_attachment}, detaching $L_b$, treating it as a constant during the backward pass, results in an over frosting trap. The optimizer minimizes the UGR by continually increasing roughness, unnaturally darkening the room and heavily blurring the highlight. 

Conversely, keeping $L_b$ attached to the graph introduces a helpful self regulating effect. Attaching the background luminance allows the optimizer to find a leveled area of the roughness where it makes the material rough enough to reduce the highlighted glare but not too rough to starve the room of ambient light. This attached state allows the system to autonomously diffuse the glare mask while preserving physically plausible global illumination.

\textit{Contextual Optimization Constraints.}
While attaching $L_b$ is highly effective for global scene parameters like the living room floor, we observed that care must be taken when applying this regularization to localized texture maps. If $L_b$ is attached while optimizing a spatial roughness texture on a specific object like the sphere in the Cornell Box, the optimizer may exploit the metric by roughening nonglare regions of the texture simply to increase ambient scattering and artificially inflate $L_b$. Therefore, the decision to attach or detach the background luminance is contextually driven. We detach the background for localized object textures to prevent spatial artifacts, and attach it for broad architectural surfaces to prevent global over darkening.

\new{\textit{Relationship to Luminance-Only Optimization.}
When $L_b$ is detached from the computational graph, the optimizer's effective 
loss reduces to minimizing $\sum_i L_i m_i$ subject to a fixed threshold, 
functionally equivalent to a luminance-only objective applied to the glare 
hotspots. The over-frosting failure demonstrated in Figure~\ref{fig:lb_attachment} 
is precisely the consequence: without the ambient-preservation signal carried by 
$L_b$, the optimizer darkens the room to suppress hotspot luminance with no regard 
for global illumination. When $L_b$ is attached, the optimization becomes 
contrast-aware; the optimizer must reduce the source-to-background ratio rather 
than absolute luminance, which has no analogue in a luminance-only objective.}

\subsection{Source-Side Constraints and Color Preservation}
\label{sec:color-preservation}
\begin{figure}[t]
    \centering
    \setlength{\tabcolsep}{2pt}
    \renewcommand{\arraystretch}{0.5}
    
    \begin{tabular}{ccc}
        & \small Optimized Render & \small Optimized Texture \\
        
        \vspace{0.5mm} \\

        \rotatebox{90}{\hspace{10mm} \small Unconstrained} &
        \includegraphics[width=0.43\columnwidth]{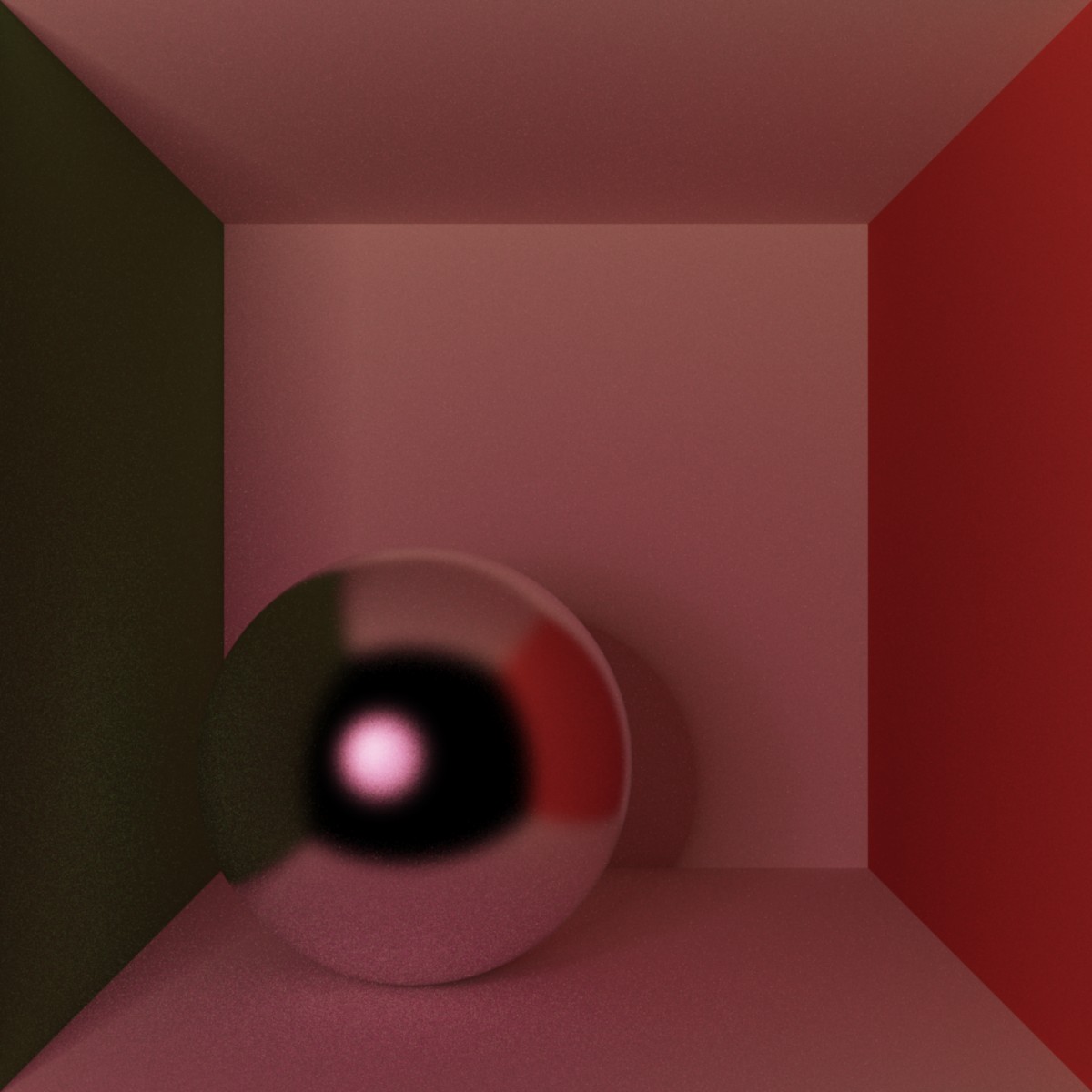} &
        \includegraphics[width=0.43\columnwidth]{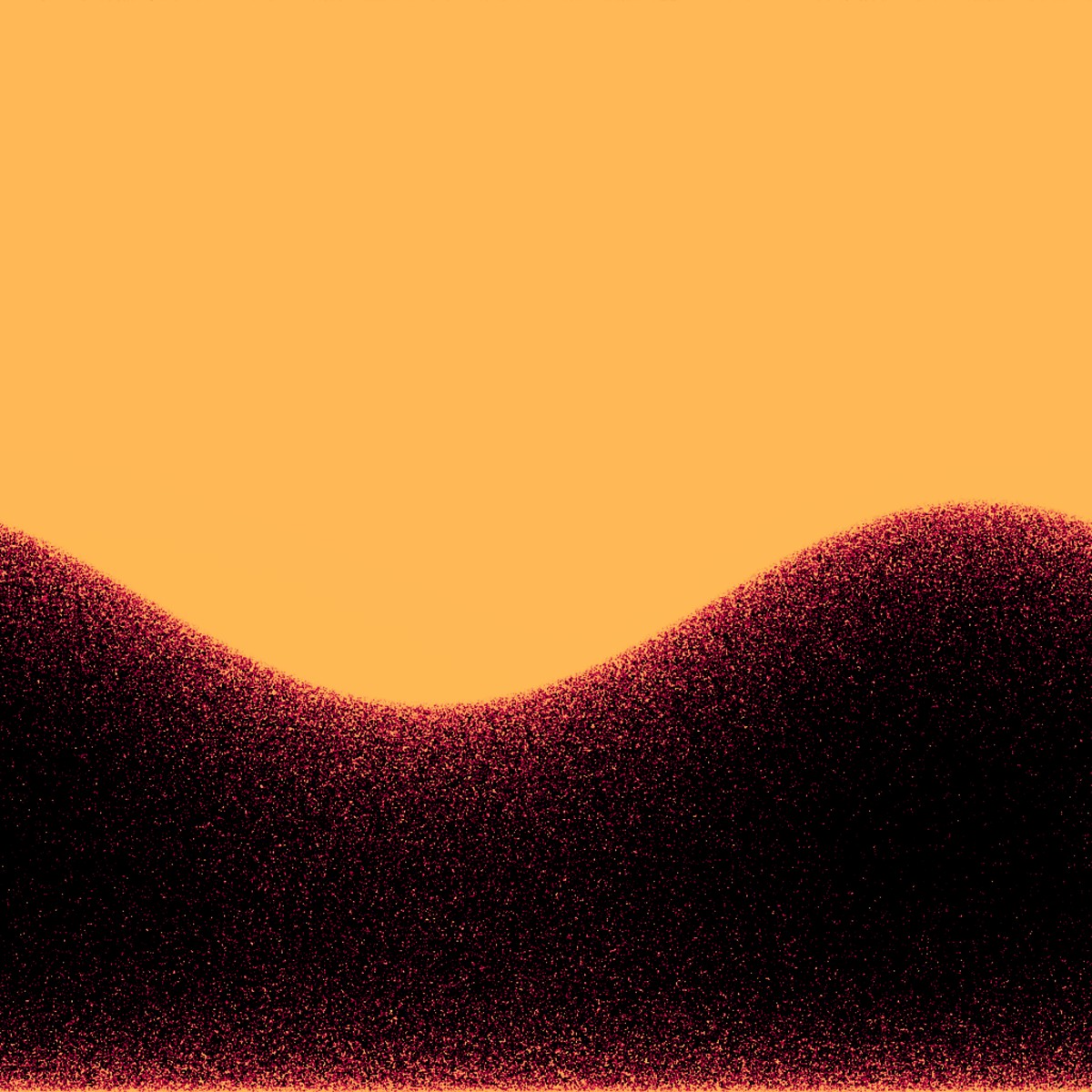} \\
        \rotatebox{90}{\hspace{11 mm} \small Constrained} &
        \includegraphics[width=0.43\columnwidth]{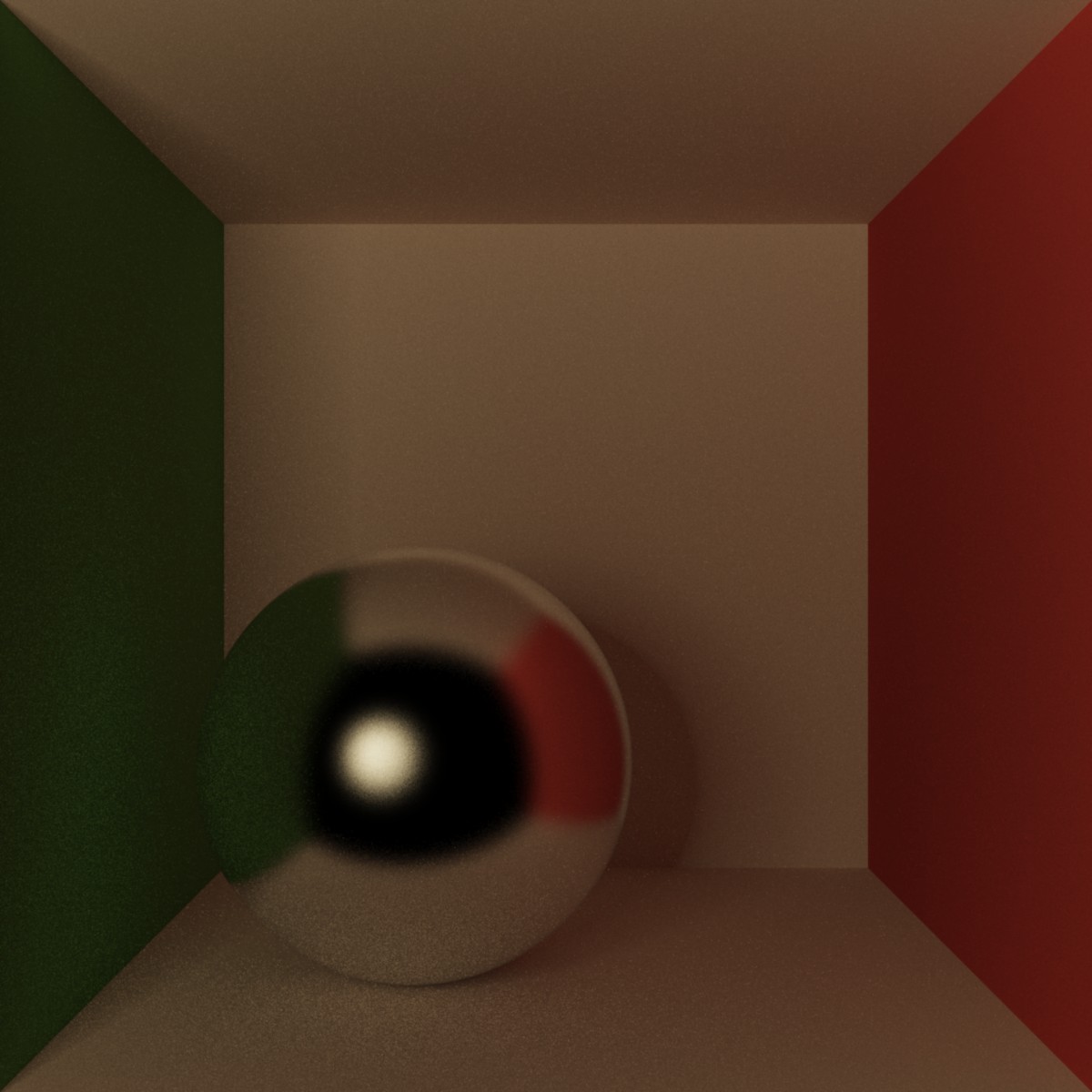} &
        \includegraphics[width=0.43\columnwidth]{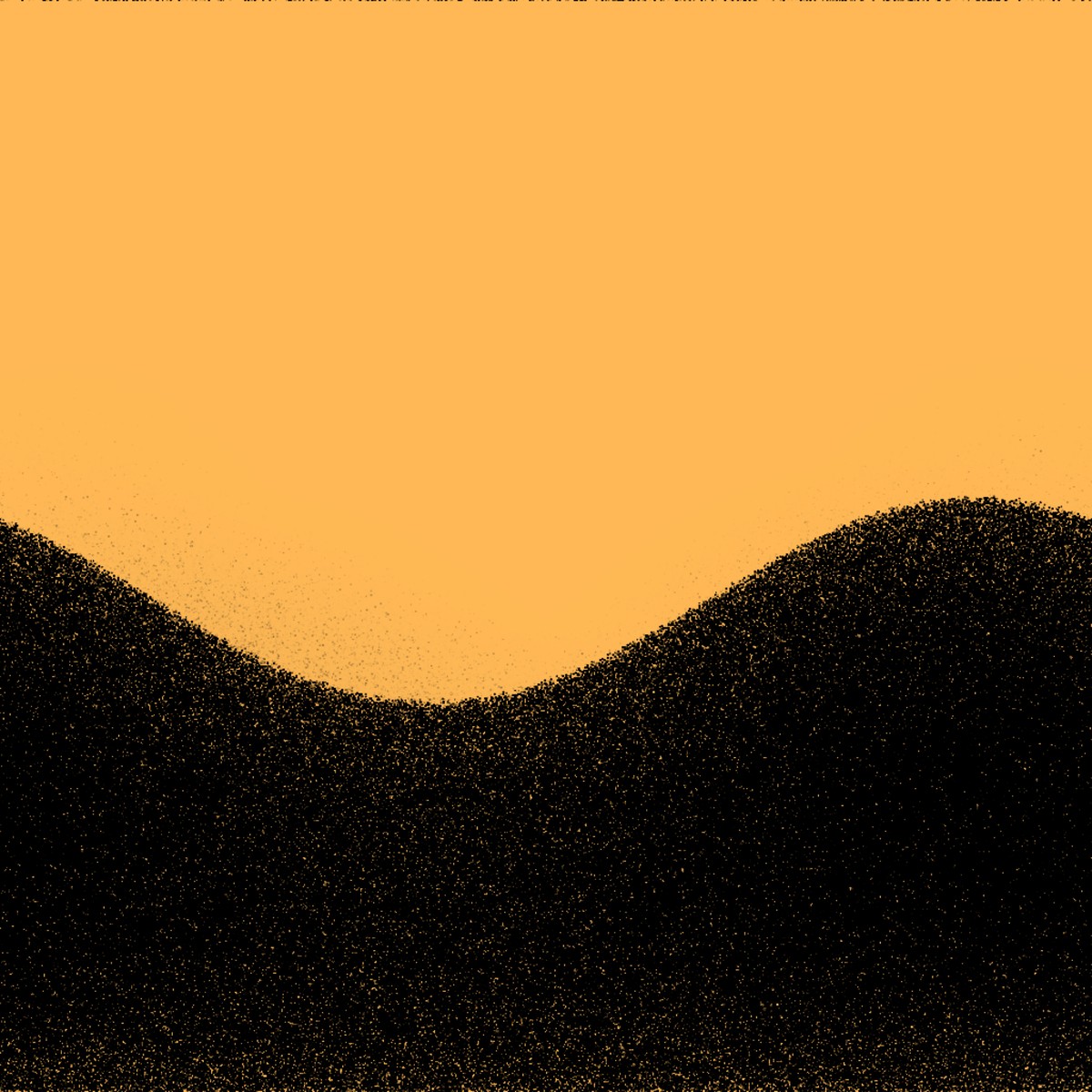} \\
        
    \end{tabular}

    \caption{(Top) Unconstrained RGB optimization exploits the psychophysical green-bias of the UGR metric, resulting in unnatural chromaticity shifts. (Bottom) By constraining the latent space to a 1D grayscale mask, the optimizer acts as a neutral density filter, mitigating glare while preserving the original light color.}
    \label{fig:color_preservation}
    \vspace{-1em}
\end{figure}

When optimizing source side emission, we discovered a unique failure mode stemming from the definition of perceptual luminance. Because photometric luminance ($L$) is derived from radiometric radiance by integrating against the human eye's spectral sensitivity curve, specifically the CIE photopic luminous efficiency function $V(\lambda)$ \cite{wyszecki2000color}, which heavily favors green wavelengths, an unconstrained optimization will exploit this biological bias. 

As shown in Figure \ref{fig:color_preservation} (Top), if the optimizer is granted free control over the spectral distribution of an emitter, it will aggressively reduce the wavelengths corresponding to peak human sensitivity to rapidly lower the perceived UGR score, while leaving other wavelengths relatively untouched. This mathematical exploit successfully mitigates the calculated glare, but results in severe unnatural chromaticity shifts in the rendered environment.

To ensure physical and aesthetic plausibility, we found it necessary to restrict the optimizer's latent space when dealing with emission. By mathematically constraining the optimizable parameter to a 1D scalar multiplier applied uniformly across all wavelengths, we force the optimizer to act as a programmable neutral density filter.

\section{Limitations and Future Work}
\label{sec:limitations}

As detailed in Section \ref{sec:results_gi_background}, the decision to attach or detach the background luminance ($L_b$) from the computational graph is currently a manual, context-driven choice. While attaching $L_b$ acts as a powerful self-regularizer for global parameters, it can introduce spatial artifacts in localized textures. Future work should investigate automated mechanisms for balancing this trade-off. This could involve developing localized background estimators or adaptive gradient weighting schemes that automatically determine the optimal $L_b$ routing based on the specific parameter being optimized, eliminating the need for manual intervention.

Currently, our framework requires the user to manually select which scene parameters (e.g., a specific floor's roughness or a specific window's IOR) are exposed to the optimizer. While our adjoint method can inherently compute gradients for all scene parameters simultaneously, natively optimizing the entire scene would lead to unacceptable, widespread alterations of the original architectural design. A truly autonomous lighting design tool should instead leverage these global gradients to perform an automated sensitivity analysis, identifying and isolating a sparse set of materials or emitters that contribute most significantly to visual discomfort. \new{More broadly, our framework is an inverse design tool rather than a manufacturing pipeline: it identifies which physical parameters to alter and by how much, but does not guarantee that an optimized $\eta$ or roughness map corresponds to an off-the-shelf material.} While we impose basic physical realizability limits (e.g., $\alpha \ge \alpha_{init}$, scalar emitter multipliers), we do not constrain the optimization to pre-manufactured material catalogues. Future research should focus on integrating these discrete constraints, ensuring the optimizer selects a physically manufacturable intervention rather than simply finding the mathematical path of least resistance.

Our current implementation optimizes static scenes from a single, fixed observer viewpoint under static lighting conditions. Discomfort glare, however, is an inherently dynamic phenomenon; occupants move through spaces, and daylighting conditions change continuously throughout the day. Extending our framework to perform joint optimization across multiple viewpoints or temporal daylighting sequences represents a critical next step for holistic architectural design. While our method is theoretically compatible with such extensions, evaluating complex global illumination across multiple temporal states simultaneously would require significant advances in differentiable rendering frameworks to remain within practical computational and memory budgets.

\section{Conclusion}
\label{sec:conclusion}
We introduced a framework that bridges psychophysical lighting metrics and inverse rendering, replacing the discrete binary thresholds of the Unified Glare Rating with a continuous, differentiable proxy that turns UGR from a passive post-process evaluation into an active loss landscape for optimizing physical scene parameters. Evaluating mitigation across source-side emission, boundary-side refraction, and surface-side scattering, we show that naive roughening is bottlenecked by ambient light starvation, and that retaining the background adaptation luminance in the computational graph acts as a critical self-regularizer that navigates this non-convex trap without over-frosting; in contrast, refractive and emitter-masking strategies yield inherently stable landscapes. Together, these provide lighting designers, architectural engineers, and material manufacturers with a physically based automated pipeline for mitigating human visual discomfort in complex rendering environments.

\section{Acknowledgements}
The authors are grateful to the anonymous reviewers for their many constructive suggestions. We particularly commend the insightful comments and suggestions of the primary reviewer. We acknowledge with gratitude the funding provided for this research by Simon Fraser University as well as a Discovery Grant provided by the Natural Sciences and Engineering Research Council of Canada. 

Additionally, we would like to thank Benedikt Bitterli for providing the rendering resources~\cite{resources16} (including the Cornell Box, Grey and White Room, Country Kitchen, and Contemporary Bathroom models), and the Stanford Computer Graphics Laboratory for the Stanford Bunny model from the Stanford 3D Scanning Repository~\cite{stanford3d}.

\bibliographystyle{eg-alpha-doi}
\bibliography{egsrbib}



\clearpage
\appendix

\section{Baseline Evaluation and Scalability}
\label{sec:appendix_baselines}

In this section, we provide an empirical evaluation of our adjoint gradient descent framework against standard derivative-free black-box optimizers (CMA-ES and SciPy's Powell method). The objective is to demonstrate the necessity of gradient-based optimization for high-dimensional spatial mitigation tasks.

\subsection{Dimensionality Scaling}
To evaluate scalability, we optimize the spatial roughness map ($\alpha$) across exponentially increasing grid resolutions, from a uniform scalar ($N=1$) to a $64 \times 64$ texture ($N=4096$). To ensure a rigorous comparison, the optimization objective was formulated as a target-seeking task (a symmetric $L_2$ loss targeting an exact UGR of 17.0) rather than a simple minimization threshold. This prevents derivative-free solvers from trivially satisfying the objective by immediately clamping the parameter to its maximum value. 

\begin{figure}[h]
    \centering
    \includegraphics[width=\linewidth]{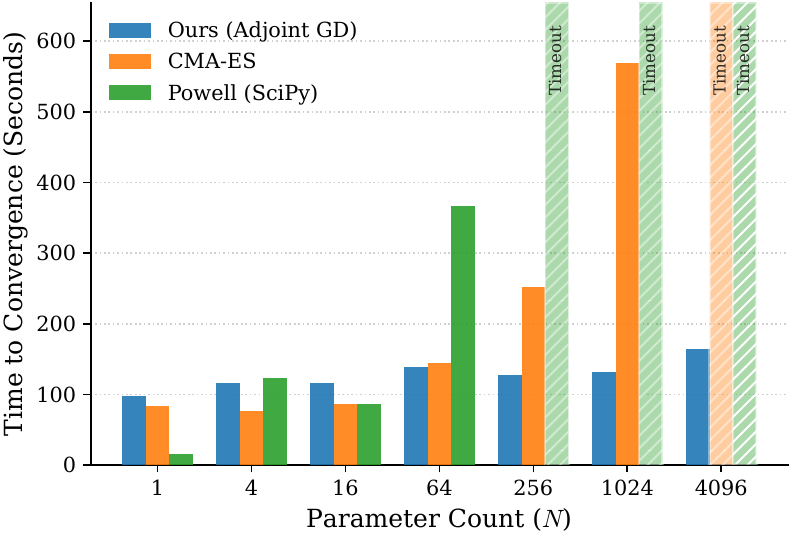}
    \caption{Optimization time vs. parameter count. CMA-ES and Powell methods fail to converge within realistic time constraints at higher resolutions, denoted by the hatched boxes.}
    \label{fig:appendix_scaling}
\end{figure}

For our adjoint gradient descent, we utilized a dynamic learning rate that scales proportionally with the loss magnitude, naturally taking smaller steps as the scene approaches the target comfort level. As illustrated in Figure \ref{fig:appendix_scaling}, our method maintains a near-constant optimization time regardless of dimensionality. Conversely, derivative-free methods exhibit exponential scaling, ultimately timing out or failing to converge entirely at higher resolutions.

It is worth noting that our current gradient-based implementation serves as a baseline proof-of-concept for the adjoint framework. We rely on a relatively simple proportional learning rate adjustment rather than highly optimized, adaptive scheduling algorithms or advanced line-search techniques. Consequently, the performance demonstrated by our method represents a lower bound; integrating more sophisticated solvers (such as L-BFGS or heavily tuned Adam schedulers) into our differentiable pipeline would likely yield even faster convergence, further widening the computational gap between our approach and black-box alternatives.

\subsection{Domain Generalization}

Table \ref{tab:domain_generalization} evaluates the robustness of our framework across fundamentally different physical material domains. For high-dimensional spatial mitigation tasks ($N=4096$), derivative-free baselines systematically fail to converge within the allotted time budget, regardless of whether the target is roughness, emitter intensity, or material blend weights. 

It is important to note that the absolute convergence time for our adjoint method varies between these spatial domains (ranging from 70.9s to 387.7s). This variance is not a failure of the optimizer, but rather a reflection of the underlying path tracing complexity; for instance, computing gradients for a blend weight requires evaluating multiple BSDF lobes simultaneously, increasing the per-iteration rendering cost compared to a single-lobe roughness evaluation. However, the optimization process itself remains stable and successfully converges in all cases.

To provide a complete baseline, we also include a scalar optimization task ($N=1$) targeting a global Plastic IOR. At this trivial dimensionality, derivative-free methods are able to succeed and converge slightly faster than our generalized pipeline. This highlights that while black-box methods are highly efficient for tuning isolated, global scalars, our adjoint framework is uniquely capable of scaling to the high-dimensional spatial maps required for complex architectural mitigation.

\begin{table}[h]
\centering
\resizebox{\columnwidth}{!}{
\begin{tabular}{@{}lccc@{}}
\toprule
\textbf{Parameter Domain} & \textbf{Ours} & \textbf{CMA-ES} & \textbf{Powell (SciPy)} \\ \midrule
Roughness ($\alpha$)      & 70.9 s                     & Timeout         & Timeout                  \\
Emitter Intensity         & 291.8 s                    & Timeout         & Timeout                  \\
Blend Weight              & 387.7 s                    & Timeout         & Timeout                  \\
Plastic IOR               & 91.5 s                     & 88.7 s          & 69.6 s                   \\ \bottomrule
\end{tabular}
}
\caption{Time to convergence (in seconds) across different material properties. All spatial domains (Roughness, Emitter Intensity, and Blend Weight) were optimized at a $64 \times 64$ resolution ($N=4096$), while Plastic IOR was a global scalar optimization ($N=1$). Derivative-free methods only succeed on the trivial scalar domain, timing out on all high-dimensional spatial maps.}
\label{tab:domain_generalization}
\end{table}

\subsection{Qualitative Results}

\begin{figure*}[t]
    \centering
    
    \begin{minipage}{0.32\textwidth}
        \centering
        \includegraphics[width=\linewidth]{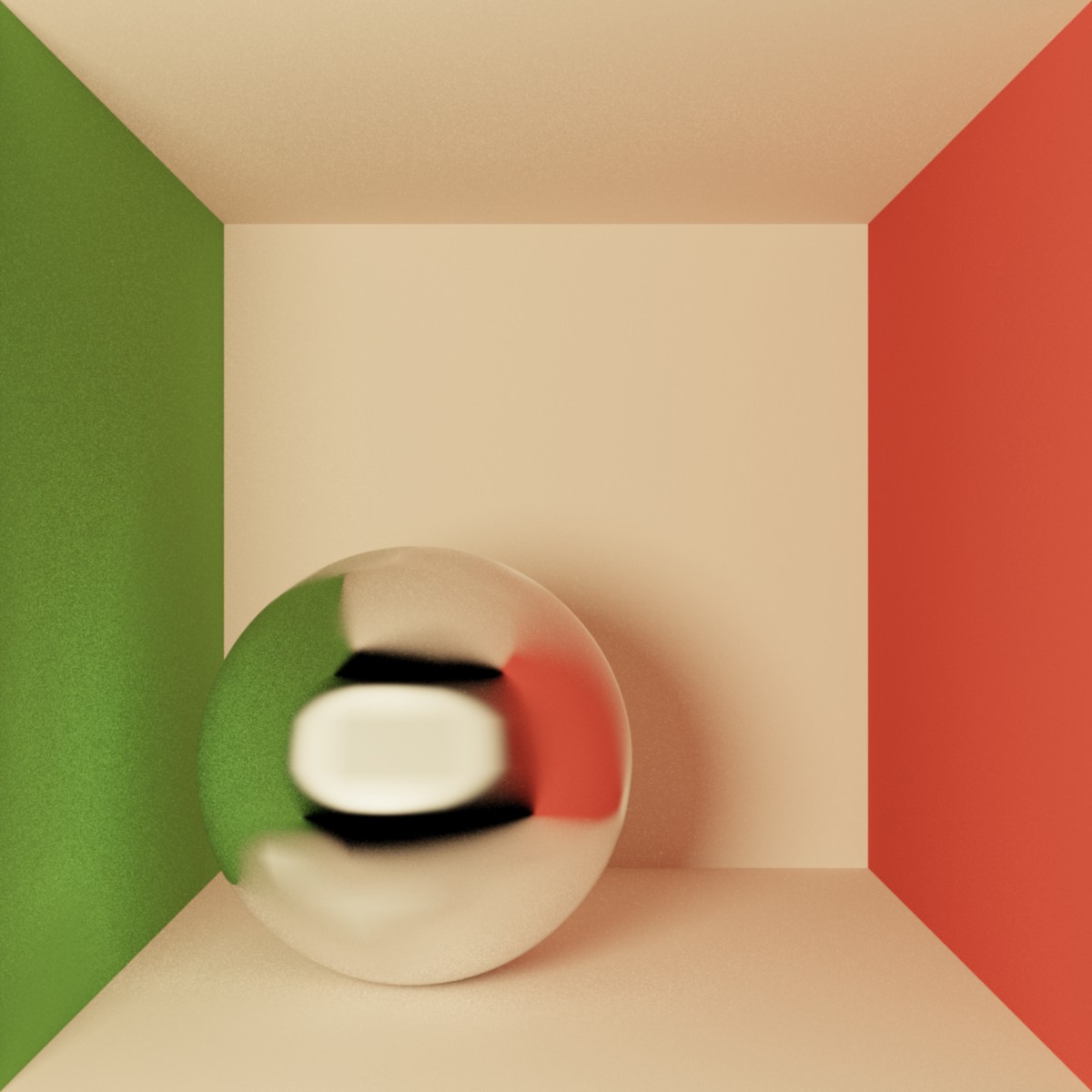}
        \vspace{-0.5cm}
        \caption*{Ours ($16 \times 16$)}
    \end{minipage}\hfill
    \begin{minipage}{0.32\textwidth}
        \centering
        \includegraphics[width=\linewidth]{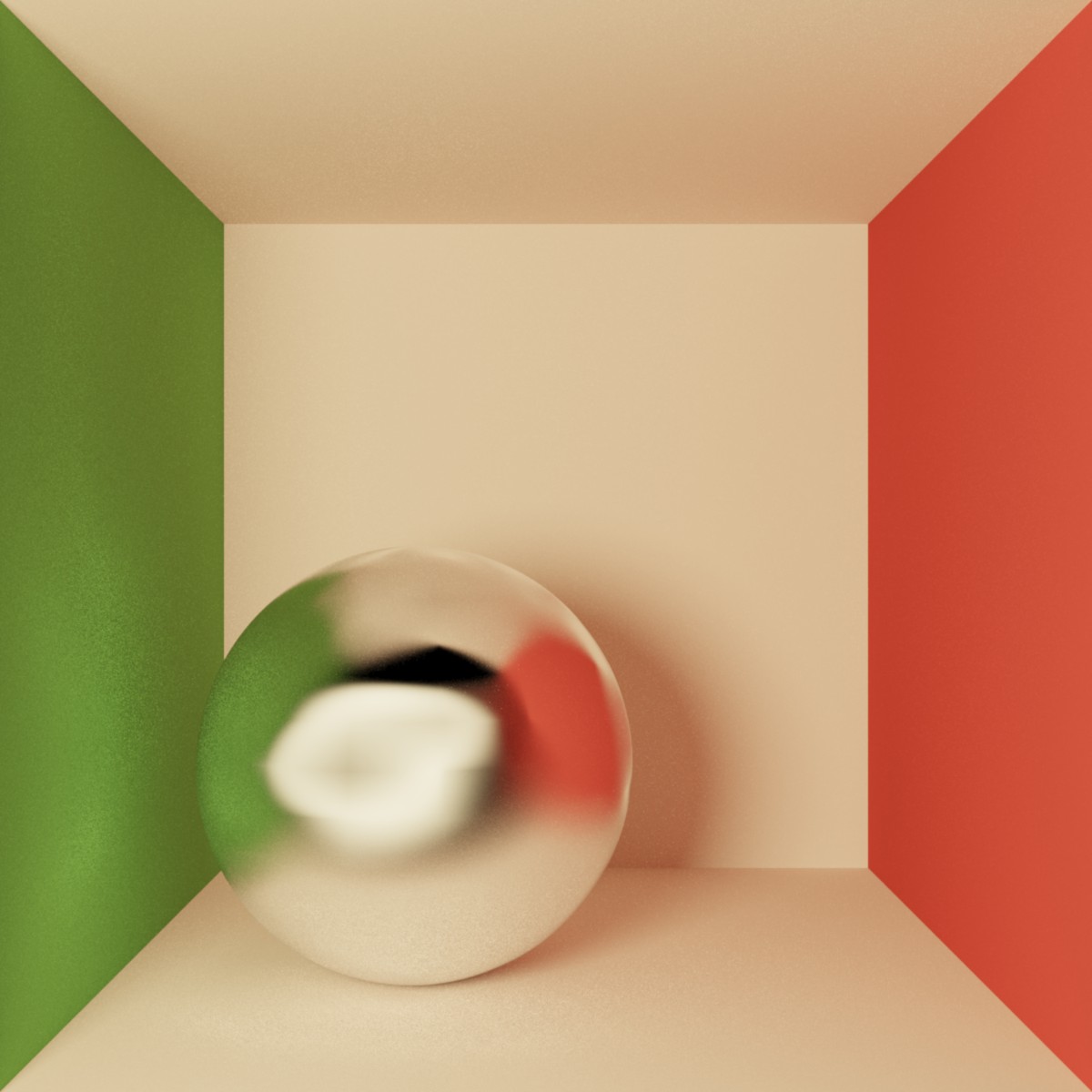}
        \vspace{-0.5cm}
        \caption*{CMA-ES ($16 \times 16$)}
    \end{minipage}\hfill
    \begin{minipage}{0.32\textwidth}
        \centering
        \includegraphics[width=\linewidth]{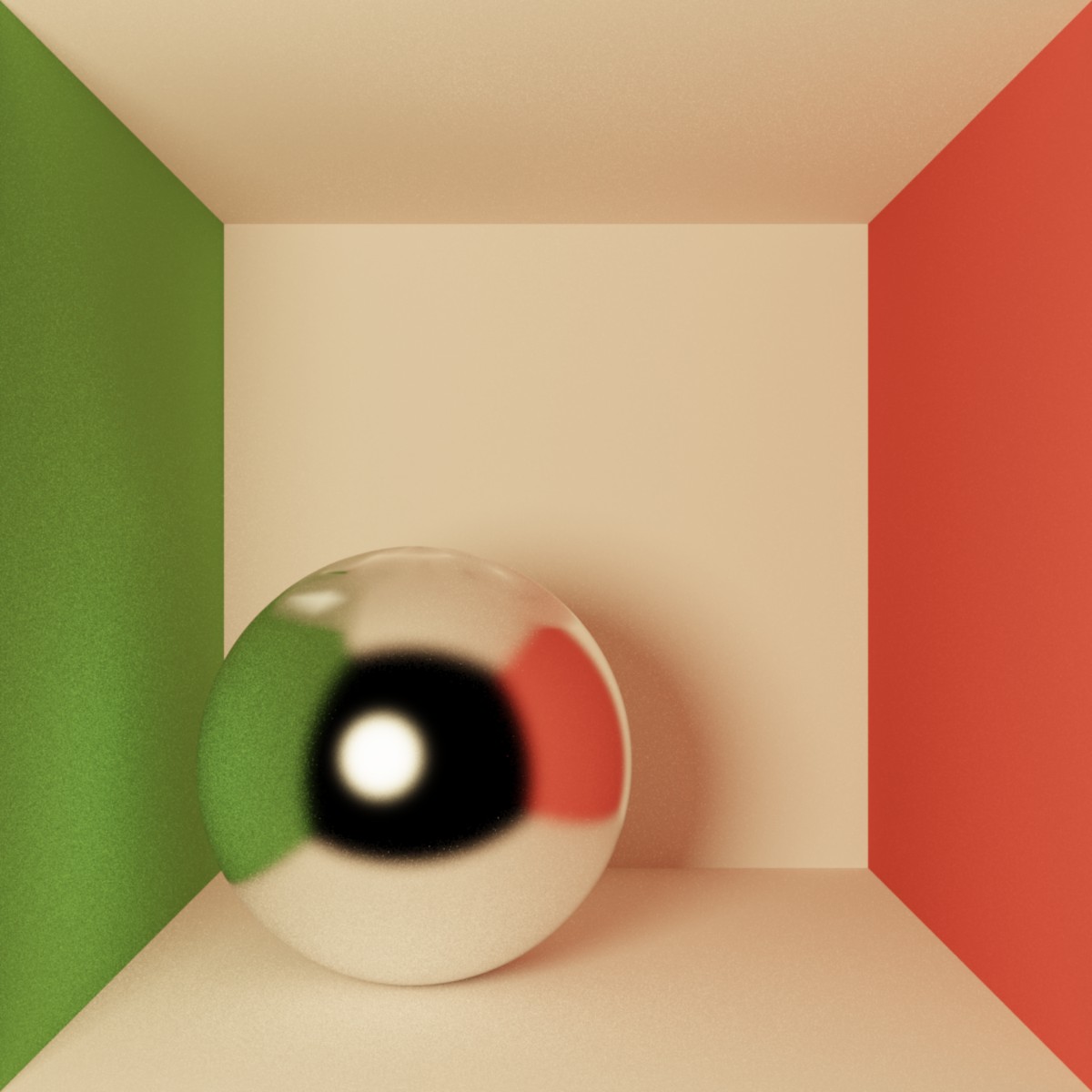}
        \vspace{-0.5cm}
        \caption*{Powell ($16 \times 16$ Timeout)}
    \end{minipage}
    
    \vspace{0.3cm} 
    
    \begin{minipage}{0.32\textwidth}
        \centering
        \includegraphics[width=\linewidth]{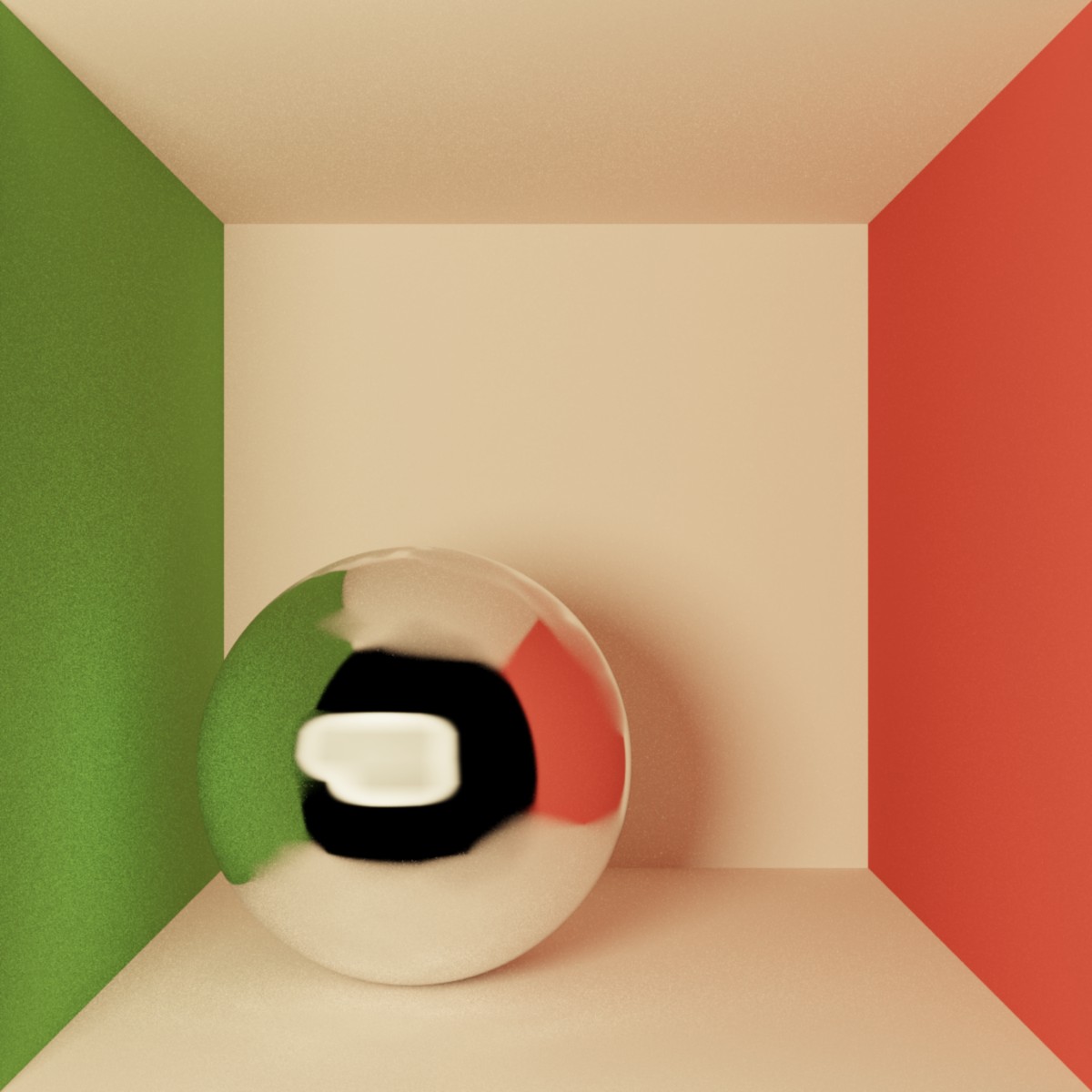}
        \vspace{-0.5cm}
        \caption*{Ours ($32 \times 32$)}
    \end{minipage}\hfill
    \begin{minipage}{0.32\textwidth}
        \centering
        \includegraphics[width=\linewidth]{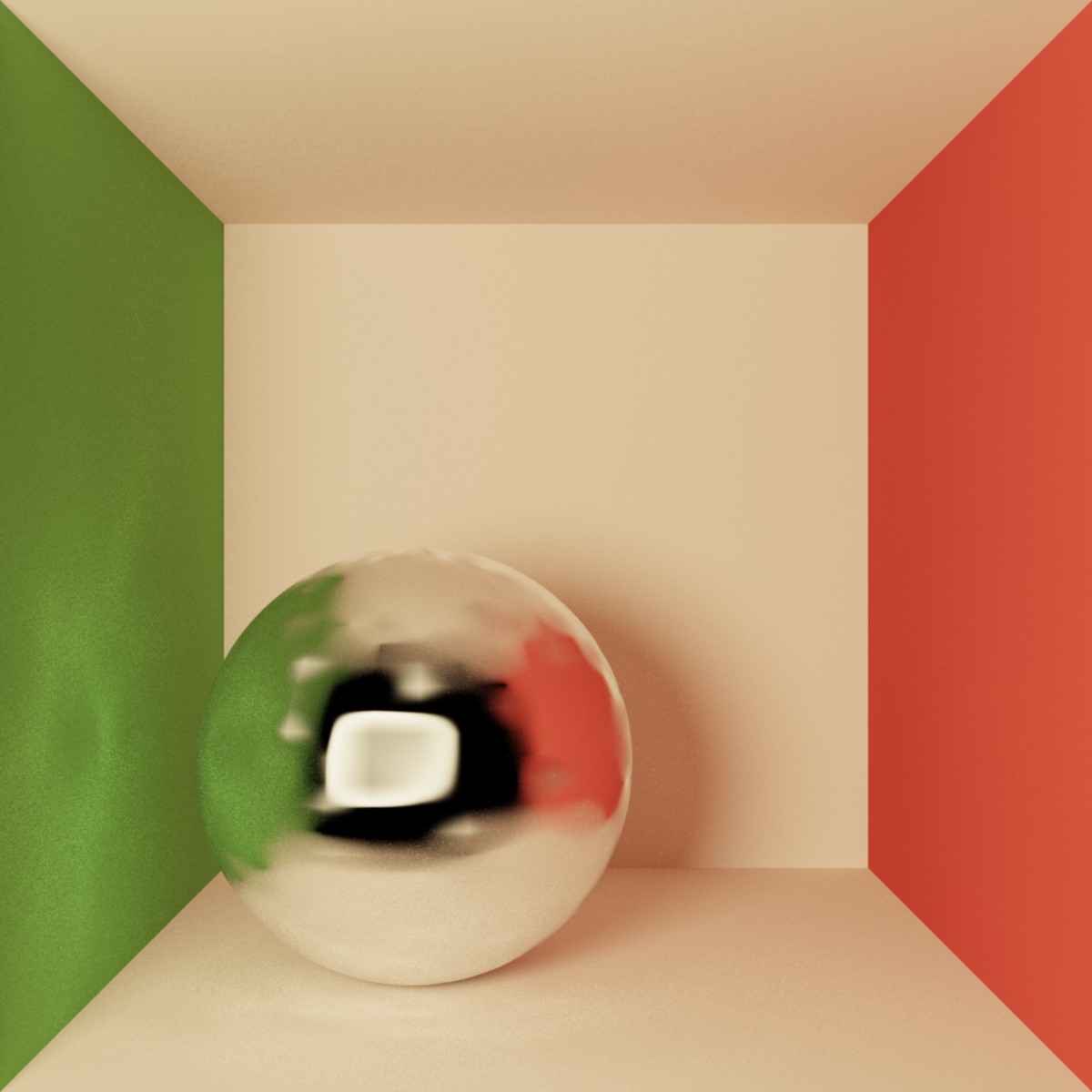}
        \vspace{-0.5cm}
        \caption*{CMA-ES ($32 \times 32$)}
    \end{minipage}\hfill
    \begin{minipage}{0.32\textwidth}
        \centering
        \includegraphics[width=\linewidth]{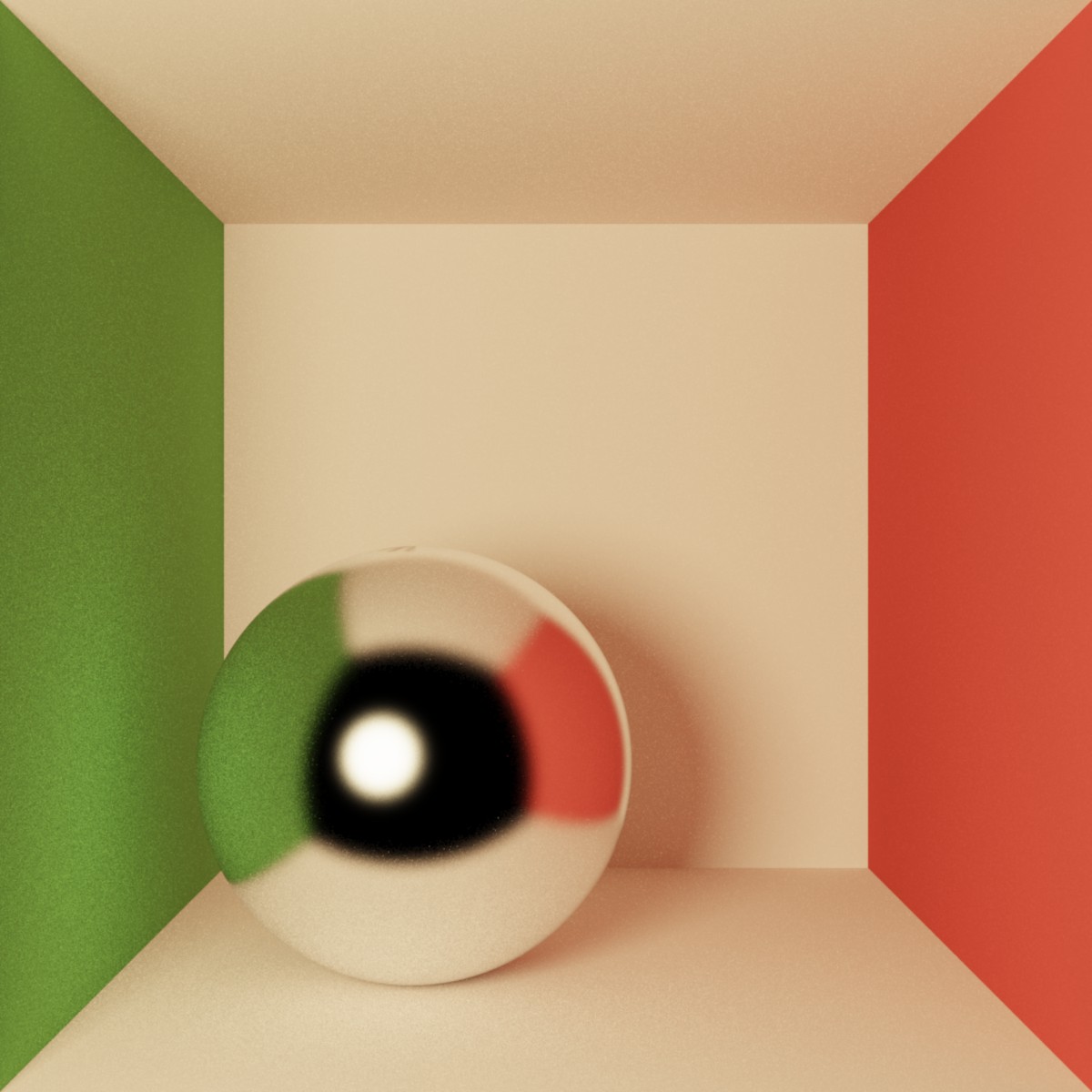}
        \vspace{-0.5cm}
        \caption*{Powell ($32 \times 32$ Timeout)}
    \end{minipage}
    
    \vspace{0.3cm} 
    
    \begin{minipage}{0.32\textwidth}
        \centering
        \includegraphics[width=\linewidth]{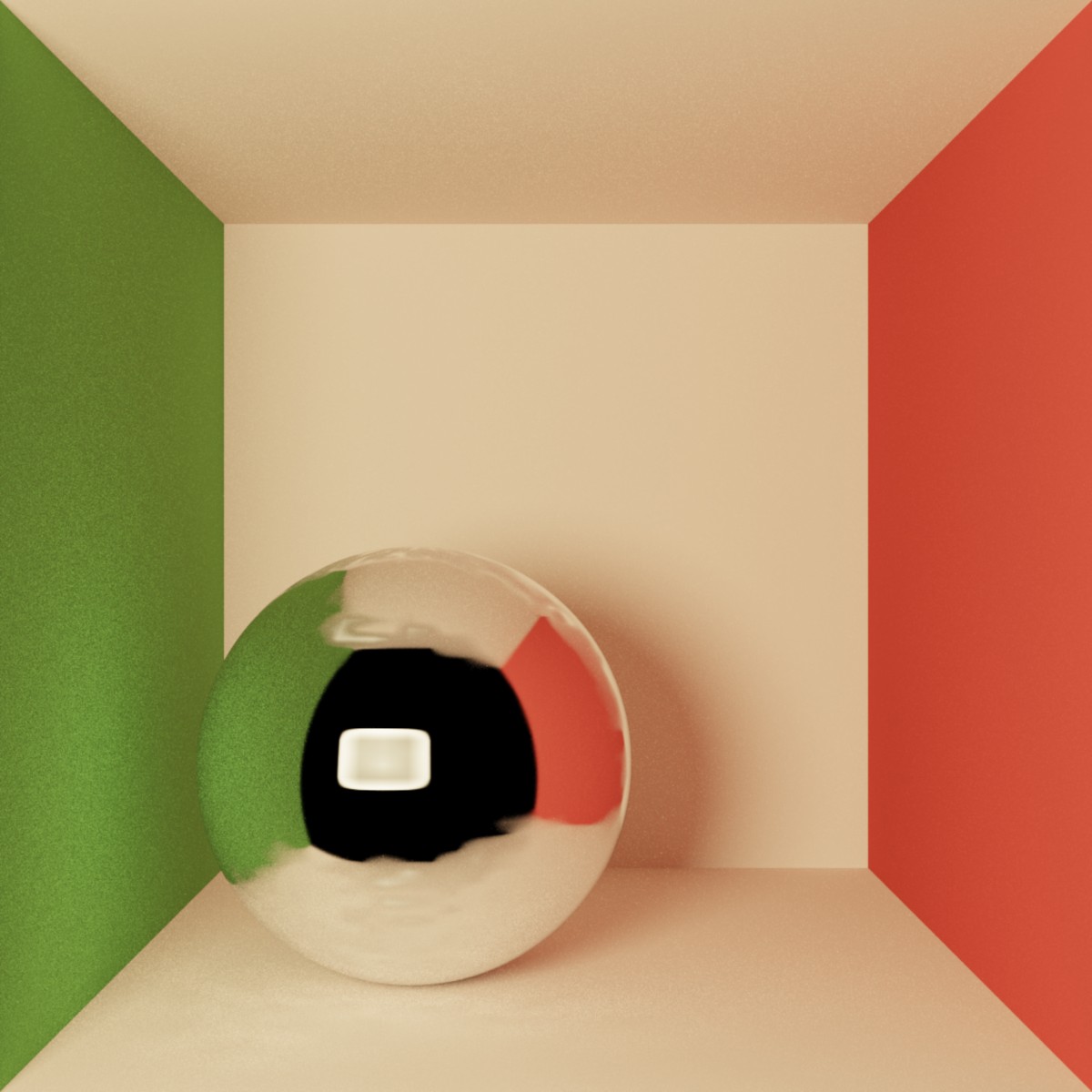}
        \vspace{-0.5cm}
        \caption*{Ours ($64 \times 64$)}
    \end{minipage}\hfill
    \begin{minipage}{0.32\textwidth}
        \centering
        \includegraphics[width=\linewidth]{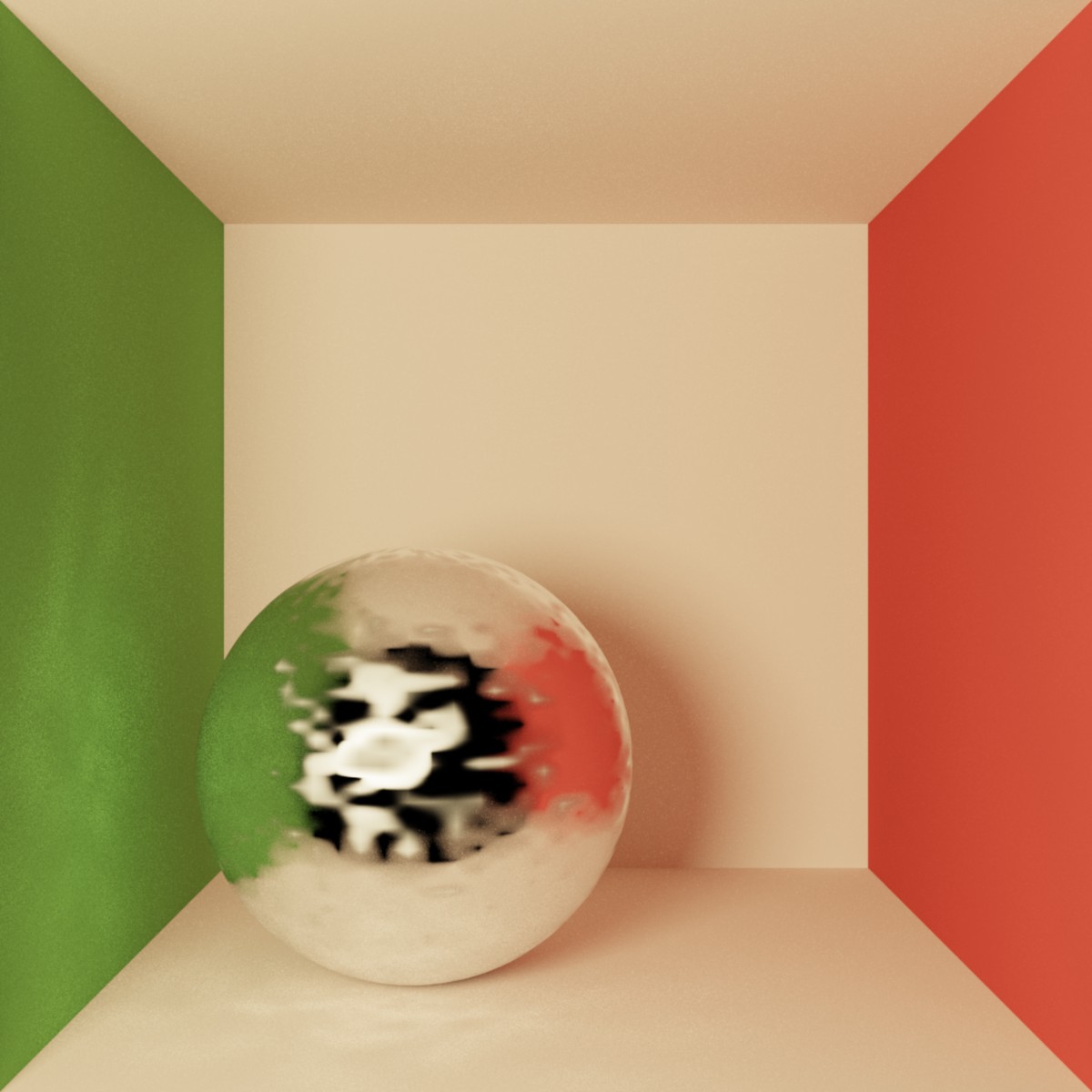}
        \vspace{-0.5cm}
        \caption*{CMA-ES ($64 \times 64$ Timeout)}
    \end{minipage}\hfill
    \begin{minipage}{0.32\textwidth}
        \centering
        \includegraphics[width=\linewidth]{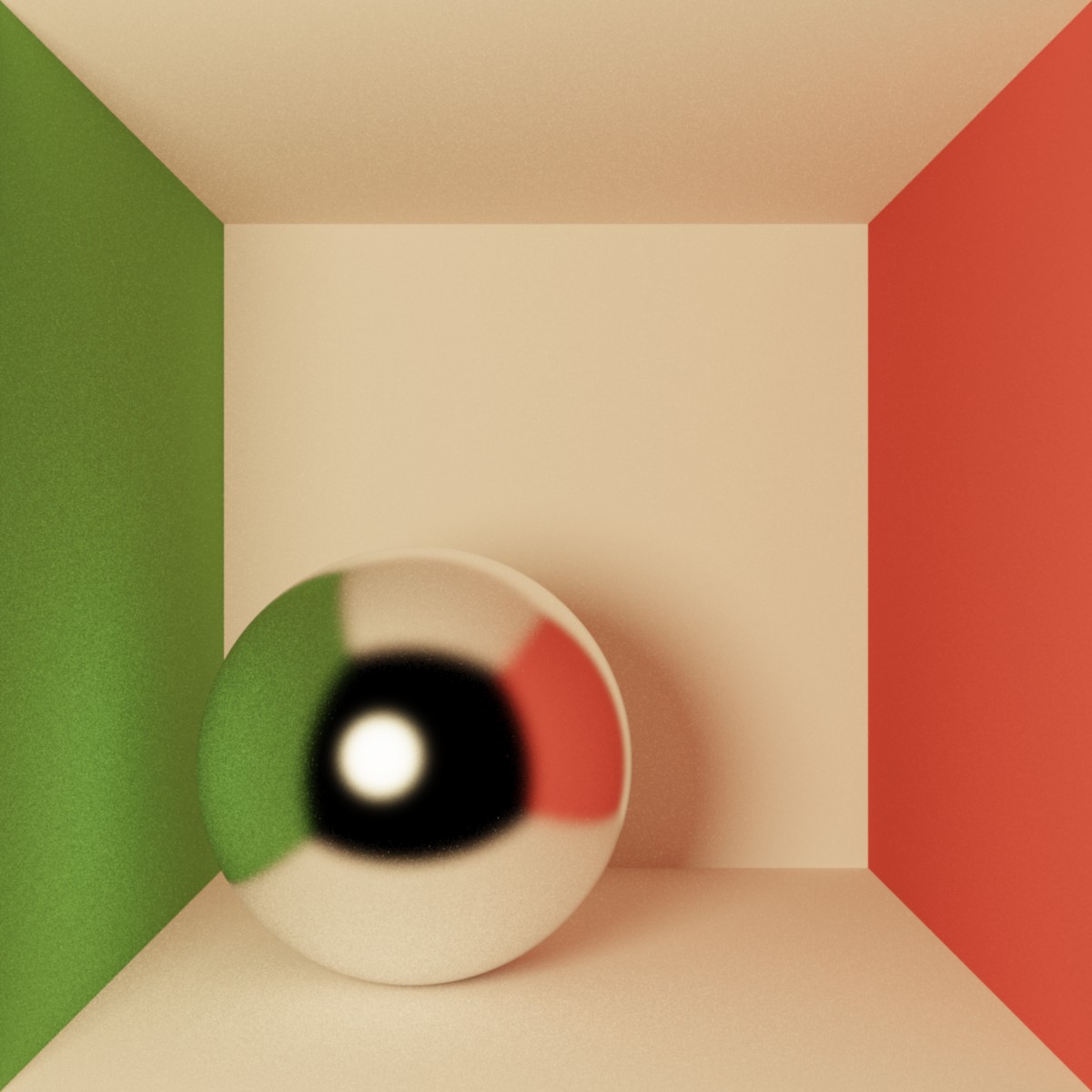}
        \vspace{-0.5cm}
        \caption*{Powell ($64 \times 64$ Timeout)}
    \end{minipage}
    
    \caption{Visual states after the allocated optimization budget. The initial, unoptimized scene exhibits extreme glare (UGR $\approx$ 32.3). Across all spatial resolutions, our gradient-based approach successfully suppresses targeted glare without destroying global aesthetics. Notably, this baseline implementation operates without explicit spatial regularization (e.g., Total Variation or sparsity constraints); consequently, minor, non-essential roughness alterations (blurring) can be observed in some non-glaring regions of the sphere. In contrast, as dimensionality increases (top to bottom), the derivative-free optimizers exhaust their evaluation budgets (indicated by \textit{Timeout}), failing to converge and leaving the severe glare largely unmitigated.}
    \label{fig:appendix_qualitative}
\end{figure*}

Figure \ref{fig:appendix_qualitative} provides visual validation of our optimization pipeline compared to derivative-free methods across escalating spatial resolutions. To isolate the effect of dimensionality, we compare our adjoint gradient descent against CMA-ES and Powell's method at grid resolutions of $16 \times 16$ ($N=256$), $32 \times 32$ ($N=1024$), and $64 \times 64$ ($N=4096$). 

While our adjoint method smoothly converges to a comfortable visual state across all dimensionalities, the derivative-free baselines exhibit progressive degradation. As the parameter count increases, the black-box optimizers exhaust their evaluation budgets or time out before discovering a viable spatial mitigation map, leaving the scene in an unresolved, highly glaring state.

\end{document}